République Algérienne Démocratique et Populaire
الجمهورية الجزائرية الديمقراطية الشعبية
Ministère de l'Enseignement Supérieur et de la Recherche Scientifique
وزارة التعليم العالي و البحث العلمي

ECOLE NATIONALE SUPÉRIEURE D'INFORMATIQUE

المدرسة الوطنية العليا للإعلام الآلي
(المعهد الوطني للتكوين في الإعلام الآلي سابقا)
Ecole nationale Supérieure d'Informatique
ex. INI (Institut National de formation en Informatique)

# Mémoire de fin d'études

Pour l'obtention du diplôme d'Ingénieur d'Etat en Informatique

**Option : Systèmes d'information**

**Thème**

## Protection de la vie privée à base d'agents dans un système d'e-learning

**Réalisé par :**

- BEKRAR Marwa

**Encadré par :**

- Mr. MENACER Djamel Eddine

Promotion : 2013/2014

# Résumé

Les systèmes d'e-learning visent à offrir un accès facile et permanent aux ressources pédagogiques mises en ligne. En effet, les systèmes d'e-learning sont dotés de capacités d'adaptation des contenus et des processus d'apprentissage selon le profil de l'apprenant. Les techniques d'adaptation utilisent des mécanismes avancés d'analyse de comportement, appelés «Learner Modeling » ou « Profiling ». Ces dernières nécessitent le traçage continu des activités de l'apprenant afin de détecter ses lacunes et ses points forts pour pouvoir adapter le contenu à son besoin ou le conseiller et l'accompagner durant son apprentissage. Cependant, l'inconvénient de ces systèmes est qu'ils causent le découragement des apprenants car l'apprenant, seul face à son écran, perd sa motivation à s'améliorer. L'ajout de l'extension sociale à l'apprentissage, pour éviter l'isolement des apprenants et stimuler l'entraide et les interactions entre les membres de la communauté d'apprentissage, a réussi à augmenter la motivation des apprenants. Cependant, les outils facilitant les interactions sociales intégrés aux plateformes d'e-learning peuvent être utilisés pour des besoins autres que l'apprentissage. Ces besoins, qui peuvent être éducatifs, professionnels ou personnels, créent un mélange de données issues de la vie privée et de la vie publique des apprenants. Avec l'intégration de ces outils aux systèmes d'e-learning et la croissance de la quantité de données personnelles stockée dans les bases de données de ces derniers**,** la protection de la vie privée des apprenants devient une préoccupation majeure. En effet, l'échange des profils entre les systèmes d'e-learning se fait sans le consentement explicite de leurs propriétaires. De plus, le profilage par l'analyse de comportements représente actuellement un moyen très rentable pour la génération des profits par la vente de ces profils aux sociétés publicitaires.

Aujourd'hui, le droit à la vie privée est menacé de toutes parts. En plus de la menace des pirates, la source de menaces la plus dangereuse est celle provenant des fournisseurs de services en ligne auxquels les utilisateurs vouent une confiance aveugle. Le contrôle et le stockage centralisé des données ainsi que les privilèges d'accès que possèdent les fournisseurs sont à l'origine de cette menace.

Notre travail se limite à la protection des données à caractère personnel dans les systèmes d'e-learning. Nous essayons de répondre à la question : Comment peut-on concevoir un système qui protège la vie privée des utilisateurs contre les menaces venant du fournisseur tout en bénéficiant de tous les services, y compris l'analyse de comportement?

Vu l'absence des solutions qui prennent en compte la protection et le respect de la vie privée dans les systèmes e-learning qui intègrent des outils d'apprentissage social, nous avons conçu notre propre solution. Notre système « ApprAide » utilise un ensemble de protocoles basés sur des techniques de sécurité afin de protéger la vie privée des utilisateurs. De plus, notre système intègre des outils qui favorisent les interactions sociales telles qu'un réseau social d'apprentissage, un outil de chat et un tableau virtuel. Notre solution autorise l'utilisation des techniques d'adaptation et de profilage afin d'aider les apprenants.

# Abstract

The e-learning systems are designed to provide an easy and constant access to educational resources online. Indeed, E-learning systems have capacity to adapt content and learning process according to the learner profile. Adaptation techniques using advanced behavioral analysis mechanisms, called "Learner Modeling" or "Profiling". The latter require continuous tracking of the activities of the learner to identify gaps and strengths in order to tailor content to their specific needs or advise and accompany him during his apprenticeship. However, the disadvantage of these systems is that they cause learners' discouragement, for learners, alone with his screen loses its motivation to improve. Adding social extension to learning, to avoid isolation of learners and boost support and interaction between members of the learning community, was able to increase learner's motivation. However, the tools to facilitate social interactions integrated to E-learning platforms can be used for purposes other than learning. These needs, which can be educational, professional or personal, create a mixture of data from the private life and public life of learners. With the integration of these tools for e-learning systems and the growth of the amount of personal data stored in the databases of these latter, protecting the privacy of students becomes a major concern. Indeed, the exchange of profiles between e-learning systems is done without the permission of their owners. Furthermore, the profiling behavior analysis currently represents a very cost-effective way to generate profits by selling these profiles advertising companies.

Today, the right to privacy is threatened from all sides. In addition to the threat from pirates, the source of the most dangerous threats is that from service providers online that users devote a blind trust. Control and centralized data storage and access privileges that have suppliers are responsible for the threat.

Our work is limited to the protection of personal data in e-learning systems. We try to answer the question: How can we design a system that protects the privacy of users against threats from the provider while benefiting from all the services, including analysis of behavior?

In the absence of solutions that take into account the protection and respect of privacy in e-learning systems that integrate social learning tools, we designed our own solution. Our "ApprAide" system uses a set of protocols based on security techniques to protect users' privacy. In addition, our system incorporates tools that promote social interactions as a social learning network, a chat tool and a virtual table. Our solution allows the use of adaptation techniques and profiling to assist learners.

À ma mère.

# Remerciement

Dieu merci, grâce à lui le tout puissant nous avons réalisé ce travail et le mener à terme.

Je remercie vivement monsieur Menacer Djamel Eddine pour ses précieux conseils et directives pédagogiques.

Je remercie Mme. Aimeur Esma, professeur à l’université de Montréal spécialiste dans le domaine de protection de la vie privée en ligne pour son aide.

Mes vifs remerciements à Mr. BELLAL Mohammed, Mr. AZOUAOU Faiçal, Mme YAKEK Zahida, Mr. GUEROUT Elhachemi, Mme. BENHAMIDA Fatima, Mme SAID EL HADJ Linda et Mme Ait Ali Yahia Dahbia pour leur aide.

Je remercie Mme DJOUDI Houria, enseignante de français pour la correction de ce travail.

Je remercie également ROBAI Amel et Meziane HADJADJ pour leur aide en développement.

Je remercie tous ceux qui ont participé de près ou de loin à la réussite de ce travail.

# Liste des figures

# Liste des tables

# Liste des algorithmes

# Table des matières

Partie 3.

Chapitre 5: Réalisation du système "ApprAide"

# Introduction générale

L'e-learning est une technique qui utilise les médias électroniques, les technologies éducatives ainsi que les technologies d'information et de communication pour faciliter l'apprentissage à distance. De nombreuses recherches avancées  ont été réalisées dans ce domaine, notamment dans le domaine des systèmes tutoriels intelligents (ITS). Les recherches dans ce domaine visent essentiellement à faire profiter les apprenants des avancées technologiques dans le domaine de l'intelligence artificielle. Cependant, ces systèmes qui basent leur enseignement sur des environnements isolés, capables d'offrir un tutorat personnalisé sans l'intervention d'un tuteur humain, ont tendance ces dernières années, à se retourner vers l'apprentissage à base de communautés d'apprentissage, afin d'éviter l'isolement et le découragement des apprenants au cours de la formation. L'ajout de la dimension sociale à l'apprentissage a créé de nouveaux problèmes car l'utilisation de ce type d'outils peut facilement être déviée de son objectif principal (l'apprentissage). Par exemple, les outils intégrés pour faciliter les interactions entre les membres peuvent être utilisés pour des raisons personnelles, ce qui augmente la quantité des données privées qui circulent sur le système et accroit le risque d'exposition aux menaces à la vie privée.

## 1. Contexte

L'apprentissage a deux formes : la première consiste en l'acquisition des connaissances de manière structurée et contrôlée par un tuteur au sein de l'école et qui aboutit à la fin de la formation à une reconnaissance (diplôme ou attestation). La deuxième forme consiste en l'apprentissage par interaction sociale où les connaissances se construisent à travers un processus d'interaction non structuré, de longue durée et non contrôlé par un tuteur. Dans les systèmes d'e-learning, la deuxième forme d'apprentissage se fait via des plateformes appelées «Social Learning portals». Le but de l'apprentissage social est de rassembler les connaissances des apprenants afin que d'autres apprenants puissent en bénéficier. Les «Social Learning portals » fournissent l'accès au contenu d'enseignement et permettent aux apprenants de produire et de partager différents types de contenus: liens, cours, notes, discussions, connaissances, etc. Ces portails permettent également aux utilisateurs d'intégrer facilement des contenus d'autres médias sociaux externes, comme des liens vers des vidéos de YouTube. Ce type de systèmes tisse de véritables réseaux sociaux et permet un échange entre les élèves qui peut déborder. Par conséquent, les apprenants de ces systèmes se trouvent aujourd'hui face aux risques qui menacent leur vie privée et qui peuvent avoir des conséquences très graves sur leur vie professionnelle. La ruine de la réputation des apprenants en ligne est due aux profils professionnels utilisés sur ce type de systèmes et qui facilitent l'identification des utilisateurs et la violation de leur vie privée.

## 2. Problématique

Le besoin à la sécurité et à la vie privée a été largement ignoré dans les systèmes d'e-learning. Malgré l'apparition des « social learning portals », qui sont actuellement exposés aux mêmes risques sur la vie privée que les autres réseaux sociaux, tel que Facebook et Linkedin, la protection de la vie privée des apprenants continue d'être ignorée. Les « social Learning portals » n'offrent pas des mécanismes de protection de la vie privée aux membres de leur communauté. Par exemple, lors de la création et la diffusion des contenus sur un forum, il se peut que les utilisateurs veulent restreindre les droits d'accès pour éviter les commentaires de certains pairs. Les paramètres de confidentialité sont indispensables pour restreindre la visibilité des contenus partagés ; quoi que ce moyen ne protège pas les utilisateurs contre la divulgation et la vente de leurs données par les fournisseurs du service.

Pour ces raisons nous soulevons la problématique suivante: Comment peut-on concevoir un système qui protège la vie privée des utilisateurs contre les menaces venant du fournisseur tout en bénéficiant de tous les services y compris l'analyse de comportement?

# 3. Objectifs

Notre travail a principalement deux objectifs

- Le premier est de concevoir un système de « e-learning» permettant aux apprenants et aux enseignants d'interagir, de s'entraider et de partager leurs connaissances. Le système tel qu'il est conçu, vise à profiter de l'enthousiasme des jeunes envers les réseaux sociaux pour l'apprentissage.
- Le deuxième objectif est de proposer un modèle de sécurité pour la protection de la vie privée des utilisateurs de notre système.

# 4. Organisation du mémoire

Le mémoire est organisé en trois parties :

**La première partie : Etat de l'art**

Cette partie comprend trois chapitres consacrés respectivement aux systèmes d'apprentissage en ligne, à la protection de la vie privée et à l'utilisation des agents dans les systèmes d'apprentissage.

**La deuxième partie : Conception**

Cette partie est composée de deux chapitres, le premier chapitre décrit l'aspect fonctionnel de notre système « ApprAide». Le deuxième chapitre présente notre proposition pour la protection de la vie privée des apprenants.

**La troisième partie : Implémentation**

Dans cette partie, nous présenterons l'architecture technique de notre solution, l'ensemble des outils de développement utilisés ainsi que les fonctionnalités et les outils implémentés du système « ApprAide».

Nous achèverons notre mémoire par une conclusion générale.

# Partie 1.
# Etat de l'art

# Chapitre1.
# Apprentissage en ligne

## Avant-propos

Ce chapitre présente des généralités sur les systèmes d'e-learning, les théories d'apprentissage et quelques standards dans le domaine. Nous nous intéressons particulièrement aux systèmes et outils utilisés pour l'apprentissage social.

## Introduction

Quand l'e-Learning est apparu il y a 20 ans, il consistait en un texte affiché sur l'écran permettant un accès distant aux ressources pédagogiques et présentant un environnement d'apprentissage isolé similaire à la lecture d'un livre. L'e-Learning n'était pas assez efficace ni populaire chez les apprenants.

Grâce aux technologies du Web 2.0 l'e-Learning est devenu plus interactif et plus riche en contenu multimédia. Ces nouvelles technologies ont donné une vision différente de l'enseignement, en offrant des outils permettant des apprentissages différents selon les élèves. En effet, c'est l'apprenant qui choisit le moment, le lieu et le rythme de son apprentissage, contrairement aux méthodes classiques d'enseignement, les élèves apprennent tous de la même manière. Ainsi, avec le passage à des méthodes d'enseignement centrées sur l'apprenant, ce dernier n'est plus vu comme un simple récipiendaire de contenu mais un membre qui peut jouer un rôle très actif dans l'enrichissement des contenus pédagogiques et l'avancement de sa communauté d'apprentissage.

Ce chapitre a pour objectif de présenter les concepts d'e-Learning, de présenter le social Learning et l'impact des outils du Web 2.0 qu'il utilise sur la vie privée des apprenants.

## 1. Qu’est-ce que le e-learning ?

Cerner le mot « e-Learning » dans une définition serait une tâche difficile vu la pluralité de ses aspects ainsi la richesse de ses concepts et technologies. Ce néologisme qui a vu le jour avec l’émergence du web2.0, ses origines remontent aux années 1960 mais son apparition réelle était au milieu des années 1990**[ACA, 2012]**. Le terme « e-Learning » ou « electronic Learning » est un terme anglais qu’on peut traduire littéralement par « apprentissage électronique ». Nous avons remarqué une grande diversité entre les définitions, chacune insiste sur un aspect ; soit elles mettent l’accent sur le support technologique, soit sur l’aspect pédagogique tandis que d’autres proposent une synthèse. Par conséquent, les libellés donnés au terme « e-Learning » étaient diversifiés. Parmi les termes utilisés pour traduire « e-Learning » on trouve : « formation en ligne » (recommandé en France par la délégation générale à la langue française et aux langues de France), « apprentissage en ligne » (recommandé au canada) et bien d’autre**. [AWT, 2008]**

**Définition 1 :** la définition donnée par la commission européenne : « *utilisation des nouvelles technologies multimédias et de l’Internet pour améliorer la qualité de l’apprentissage en facilitant l’accès à des ressources et des services, ainsi que les échanges et la collaboration à distance* ». **[AWT, 2008]**

Cette définition qui semble être claire, synthétique et complète, met l’accent sur la qualité d’apprentissage. Cette définition tente d’influer les usages et améliorer les pratiques**. [AWT, 2008]**

**Définition 2 :** la définition proposée par le LabSET**:** *« E-learning (ou electronic learning): apprentissage en ligne centré sur le développement de compétences par l'apprenant et structuré par les interactions avec le tuteur et les pairs».* **[AWT, 2008]**

Cette définition restreint le e-learning au dispositif en ligne et exclut les autoformations sur CD-ROM. Elle fait référence aux « interactions avec le tuteur et les pairs » ce qui ancre la définition dans un contexte pédagogique. **[AWT, 2008]**

## 2. Les théories d’apprentissage

Les pratiques et les méthodes pédagogiques appliquées dans les systèmes d’e-Learning sont basées sur des théories d’apprentissage visant à définir des méthodes de transmission des connaissances et des méthodes pour la bonne maitrise des connaissances par les apprenants.

Dans ce qui suit nous présenterons les théories d’apprentissage les plus connues, spécialement celles qui sont appliquées dans les systèmes d’e-Learning.

### 2.1 Le behaviorisme (comportementaliste)

L'apprentissage se fait en suivant les méthodes d'enseignement « étape par étape » traditionnelles, qui s'appuient sur la création du comportement souhaité chez l'apprenant. Le behaviorisme se concentre sur un nouveau modèle de comportement et demande à l’apprenant de le répéter jusqu'à ce qu'il devienne automatique. Une fois l'action ou le comportement répliqué par l'apprenant, l'apprentissage est supposé réalisé. **[Hathaway, 2007]**

### 2.2 Le constructivisme

Le « constructivisme » se concentre sur la préparation de l’apprenant à résoudre un problème dans des situations ambigües **[Mergel, 1998].** La signification et la compréhension sont créées et construites par l’apprenant à travers ses activités de résolution des problèmes ou par ses interactions et ses expériences dans le monde que ce soit individuellement (cognitive constructivisme) **[Bruner, 1966, Piaget, 1972]** ou dans des contextes sociaux (social constructivisme) **[Bandura, 1977][Vygotsky, 1978]**. L’apprentissage par les problèmes ouverts, étude de cas est le produit de cette école. **[Kozanitis, 2005] [Hathaway, 2007]**

### 2.3 Le Socio-constructivisme

Issu en partie du constructivisme, le socioconstructivisme ajoute la dimension du contact avec les autres afin de construire les connaissances. Vygotsky **[Vygotsky, 1978]** est le concepteur de cette théorie interactionniste. Il considère que l’apprentissage est renforcé et élargi par la présence d’autres personnes bien informées ou plus expérimentées (Théorie de la zone proximale de développement). L’apprentissage par projets, discussions sont les produits de cette école. **[Kozanitis, 2005]**

### 2.4 Le Social Learning

Une théorie alternative qui a été élaborée par Albert Bandura[1] **[Hathaway, 2007]**. Selon Bandura, le représentant actuel de cette théorie, « tout apprentissage est social » et l’apprentissage passe par trois procédures : **[Bandura, 1977**]

1. l’apprentissage vicariant : résulte de l’imitation par l’observation d’un pair qui exécute le comportement à acquérir (formateur par exemple).
2. la facilitation sociale : la personne améliore sa performance sous l’effet de la présence d’autres observateurs.

[1] **Albert Bandura** : psychologue canadien connu pour sa théorie de l'apprentissage social.

3. l’anticipation cognitive : chercher une solution à une nouvelle situation par raisonnement à partir des situations similaires.

La principale différence entre cette théorie et la théorie du « socio-constructivisme » est que Bandura considère l’apprentissage à 100% social. Selon Bandura, même si la personne est seule dans un laboratoire, les résultats dont il aboutira ont toujours relation avec les anciens acquis issus de son environnement social. A notre époque, le terme « social learning » est le plus répandu en ligne et dans les entreprises pour désigner l’apprentissage par les interactions sociales. Dans les entreprises, les outils déployés pour le social learning visent essentiellement à collecter les connaissances informelles des employés.

## 3. Les types d’interaction en e-Learning

L’interaction en e-Learning peut avoir lieu entre: **[Hathaway, 2007]**

- **Enseignant-apprenant :** le rôle de l’enseignant dans ce type d’interaction est, soit de transmettre directement les connaissances à l’apprenant, soit le guider pour arriver à la connaissance.
- **Apprenant-apprenant:** Ce type d’interaction crée de la connaissance.
- **Apprenant-contenu:** L’apprenant peut lire les documents de la formation définis par le système d’e-Learning ou définis par son enseignant, comme il peut chercher et lire des contenus publiés sur le Web.

## 4. Les modes d’apprentissage dans les systèmes d’e-learning [ORZ, 2010]

### 4.1 La formation synchrone (synchronous education)

Dans une formation synchrone, l'échange s'effectue en temps réel avec les autres apprenants ou avec les tuteurs de la classe virtuelle avec possibilités d'échanges oraux et écrits (par chat, par web-conférence ou par visioconférence). Les formations synchrones permettent également de partager des applications et d'interagir sur celles-ci au moment où le tuteur leur donne la main sur le document partagé. **[EDU, 2009] [CAL, --]**

### 4.2 La formation asynchrone (asynchronous education)

C'est une formation en ligne sur Internet pendant laquelle l'apprenant n'a pas de contact simultané (en temps réel) avec son formateur ou les autres apprenants. Il peut s'agir de forums de discussion (pour poser des questions, réaliser des activités pédagogiques collaboratives, résoudre des exercices en groupe), de la messagerie électronique ou bien encore par l'échange des cours de formation multimédia que l’apprenant peut utiliser de manière autonome. L’utilisateur peut se former à son rythme, en fonction de ses besoins et de ses disponibilités. **[BOD, 2005] [EDU, 2009]**

### 4.3 Le Blended-learning

Nous pouvons également distinguer le Blended[2]-Learning appelé aussi la formation mixte **(présentiel-distantiel)** ou hybride. Le Blended-Learning permet d'introduire, de prolonger et compléter une formation en présentiel. Généralement, le blended learning concentre les apprentissages liés à la partie théorique de la formation dans les contenus e-learning et organise la formation présentielle sur la partie pratique.

## 5. Les systèmes de gestion d'e-learning

Les cours en ligne sont créés, gérés et utilisés grâce à des systèmes de gestion d'e-learning appartenant aux catégories suivantes:

### 5.1 LMS - Learning Management System [ORZ, 2010]

LMS est une famille de systèmes, qui permet la gestion de toutes les activités de la formation. Les systèmes LMS catégorisent les utilisateurs, leur donnent certaines autorisations à des modules de formation et affectent les utilisateurs à des groupes spécifiques de formation.

### 5.2 CMS - Content Management System [ORZ, 2010]

CMS est une famille de systèmes d'e-learning dédiés à la création, le stockage, la gestion et la présentation des contenus. L'important dans ce modèle est de fournir des possibilités de réutilisation des objets d'apprentissage, appelés RLG (Reusable LO[3]) ou des composants de contenu.

### 5.3 LCMS - Learning Content Management System [ORZ, 2010]

LCMS est une famille de systèmes d'e-learning les plus avancés technologiquement. Les systèmes LCMS intègrent des fonctionnalités des LMS et des CMS, assurant à la fois la création et la gestion des contenus éducatifs. Ils offrent la possibilité d'évaluer les connaissances assimilées par les utilisateurs.

Parfois, la catégorie suivante est aussi distinguée **[ORZ, 2010]:**

### 5.4 VCS –Virtual Classroom System [Finkelstein, 2006]

Les VCS sont des systèmes qui incluent les fonctionnalités suivantes:

- Transmission de la voix et de la vidéo en temps réel entre tous les participants.
- Tableau blanc partagé (shared whiteboard).
- Espace intégré pour la projection de diapositives ou autres supports visuels.
- Capacité d'interaction textuelle, y compris les conversations ou « note-passing».
- Des moyens permettant aux apprenants d'indiquer qu'ils ont des questions.

---

[2]Blend : signifie mélange

[3]L'unité de base de contenu électronique est appelée Learning object – LO ou objet pédagogique en français, qui peut être soit un contenu élémentaire tel que: un seul fichier image, vidéo ou texte.

- Outils pour évaluer l'humeur actuelle, les opinions et la compréhension ainsi que pour solliciter des questions ou des commentaires.

### 5.5 ITS- Intelligent Tutoring Systems

Les ITS sont similaires aux LMS. Ils peuvent donner des feedback intelligents à l'utilisateur. Les ITS emploient les techniques de l'intelligence artificielle, pour pouvoir comprendre, informer et diriger l'apprenant quand il termine ses exercices ou ses tests. Ils visent à répliquer le rôle du tuteur qui guide et forme les apprenants d'une manière efficace. Le tuteur humain est souvent remplacé par des entités intelligentes, appelées « agents » capables de suivre et guider l'apprenant durant son apprentissage. **[Dal, 2009]**

## 6. Le social learning, Un nouvel air du e-Learning

*« Where once content was king, interconnectivity has taken the crown »***[Hathaway, 2007]**

Bien que le terme social learning fût utilisé bien avant l'apparition du e-Learning, le terme *social learning* représente un nouveau phénomène du Web 2.0[4]. Actuellement, l'expression de « social learning » s'est répandue sur le web comme un concept qui caractérise l'utilisation des médias électroniques synchrones ou asynchrones pour le développement des savoirs, par le biais de connexion avec des collègues, des mentors ou des experts dans une optique collaborative **[Domon, 2009].**

### 6.1 Pourquoi le social learning?

Le découragement des apprenants et l'arrêt en cours de formation est l'un des grands problèmes des formations dans les systèmes d'e-Learning, car à travers le temps, l'apprenant, seul face à son écran, perd sa motivation pour s'améliorer. Le « social Learning » permet aux apprenants de discuter, échanger et progresser ensemble. De plus, La théorie d'apprentissage socio- constructiviste affirme que les interactions avec l'environnement est une exigence pour la création des connaissances car la connaissance est le produit de l'interaction sociale et culturelle. Un rapport important a été publié par Palloff et Pratt en 2001 **[Hathaway, 2007]** sur les milieux d'apprentissage efficaces. Ces rapports ont montré que l'une des clés de succès de l'apprentissage est la communauté.

Selon **[Hathaway, 2007]**, l'élève veut avoir dans un système d'e-learning:

- ✓ La communication et le feedback
- ✓ L'interactivité
- ✓ Un sens de la communauté
- ✓ Une orientation adéquate
- ✓ L'habilité d'effectuer les tâches demandées pour le cours.

---

[4] **Le Web 2.0 [Hage, 2011] :** Le terme web2.0 représente la deuxième génération du Web à base de communauté (Comme les blogs, Wikis, etc.). Il vise à faciliter la créativité, promouvoir la collaboration et le partage entre les utilisateurs. Avec le web1.0, les internautes étaient juste des récipiendaires de l'information publiée par l'auteur. Cependant, avec le Web2.0, les utilisateurs sont également des participants actifs à la création des contenus.

L'élève est un être humain apprenant continuellement et l'utilisation du social Learning facilite le partage des connaissances et des expériences, la recherche des informations, la résolution des problèmes et les interactions avec les autres. Que ce soit pour l'entreprise ou pour les établissements d'éducation, le social Learning est un mode de formation rapide, peu couteux et efficace.

### 6.2 Les outils du social Learning

Récemment, l'e-Learning a bénéficié des développements avancés d'outils du Web 2.0, spécialement les outils collaboratifs. Les moyens d'apprentissage social se basent essentiellement sur ce type d'outils.

Dans ce qui suit, quelques exemples d'outils du Web2.0 conçus pour l'apprentissage ou utilisés pour faciliter les interactions entre l'enseignant et ses élèves ou entre les élèves eux-mêmes:

- **Les forums**: Un espace où les membres laissent leurs messages en attente d'une réponse.
- **Les outils de conversation:** Sont des outils de communication synchrone comme le chat ou le tableau blanc. **Le tableau blanc (Figure 1)** est un écran interactif qui permet de dessiner des images et d'écrire sur un espace dédié. Il permet également le partage de documents et la communication écrite entre le professeur et l'apprenant pendant les cours à distance. L'enseignant et l'élève peuvent communiquer en utilisant la vidéo, les appels audio ou la messagerie instantanée. **[GFD, --]**

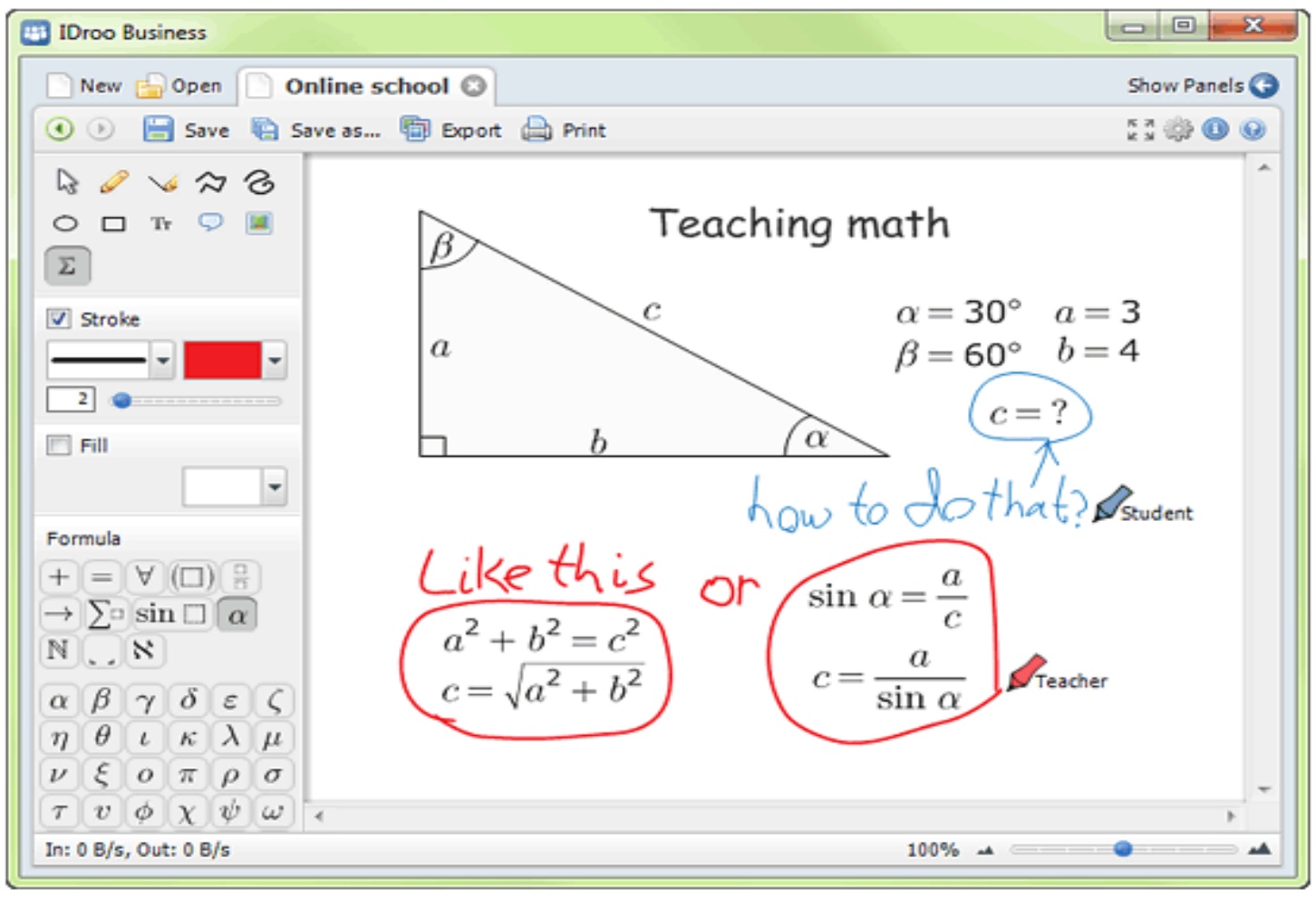

**Figure 1: Le tableau blanc d'IDroo Business [IDroo, --]**

- **Les réseaux sociaux**

Les réseaux sociaux, tel que Facebook et Twitter sont parmi les outils les plus utilisés dans les entreprises et dans les établissements d'éducation pour l'apprentissage social et le travail collaboratif. Actuellement, cet outil est très utilisé pour l'apprentissage collaboratif des langues étrangères[56]. Souvent, le partage de documents, de liens, d'informations ou la création

5 http://eduscol.education.fr/numerique/actualites/veille-education-numerique/archives/fevrier-2010/apprentissage-des-langues-tic-nouveaux-medias-

de projets par groupes, se font sur les groupes de Facebook. La création de groupe sur Facebook facilite les rencontres en ligne pour des séances de travail en groupe. Ainsi, les réseaux sociaux en ligne sont un moyen gratuit pour ce type d'apprentissage. Cependant, les groupes de Facebook ne sont pas conçus pour le travail collaboratif et ne supportent pas le suivi des activités des apprenants par le tuteur (par les outils dédiés à l'apprentissage collaboratif). La figure 2 montre une de nos expériences en utilisation des groupes de Facebook pour l'apprentissage par projet.

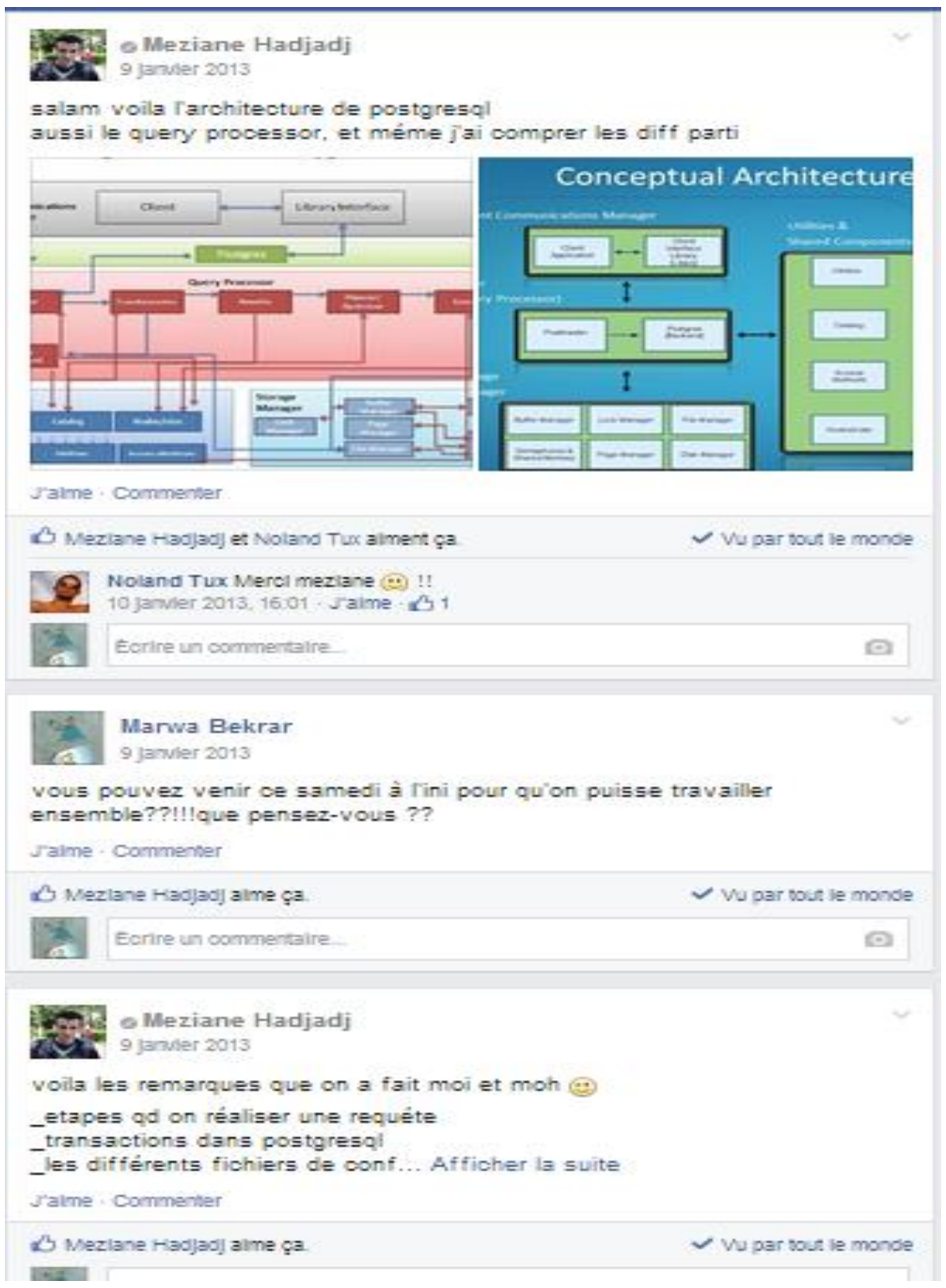

**Figure 2: Un groupe sur Facebook**

## 7. Les réseaux sociaux et l'apprentissage

Depuis les années 2000, la transmission de savoir, savoir-faire et comportement a réellement évolué et a été considérablement accéléré avec le développement des réseaux sociaux **[Vaufrey, 2010]**. L'usage croissant des outils numériques de réseautage[7] social est devenu aujourd'hui un moyen d'apprentissage et d'acquisition de compétences pour l'élève alors que depuis longtemps ils étaient considérés comme des espaces de vie sociale à des personnes mal adaptées dans la vie réelle. Les établissements d'éducation se trouvent souvent face à des situations où ils doivent décider de permettre ou non l'utilisation des nouvelles technologies, comme les RS, pour renouveler ou améliorer l'enseignement. Une décision qui dépend essentiellement de l'apport que ces technologies peuvent ajouter à l'enseignement. Dans ce qui suit nous essayerons de montrer l'apport des réseaux sociaux à l'apprentissage

[6] http://alsic.revues.org/2513
[7] Réseautage: nom masculine, Action de « réseauter ». Néologisme inventé pour traduire le terme anglais « networking »

ainsi que les conséquences de l'utilisation inappropriée de ces derniers. Les avantages et les inconvénients listés ont étaient démontrés par des expériences jeunes documentées dans les articles **[Grégoire, 1996] [PIN, 2011] [AUR, 2012] [EDU, 2011] [EDU, 2012]**

## 7.1 Les avantages

Les réseaux peuvent être des outils d'apprentissage performants, Les expériences ont montré que :

- ✓ Les élèves apprennent dans les réseaux sociaux à communiquer, à reformuler pour bien expliquer et pour bien être compris. En effet, cela rend les élèves plus aptes à défendre leurs idées. Ainsi, l'utilisation des RS augmente l'apprentissage des langues, la maîtrise de la syntaxe, la compréhension de textes et la capacité à rédiger. Elle permet aux élèves de s'engager dans des activités qui les invitent à créer et à partager avec les autres, et même les intéresser à l'écriture collective.
- ✓ Les réseaux sociaux suscitent l'intérêt des élèves pour des activités d'apprentissage et les amènent à y consacrer plus de temps et d'attention que dans les classes habituelles.
- ✓ Ils représentent un moyen de motivation des élèves car les réseaux sociaux proposent un environnement et présentent des contenus d'une manière qui est plus stimulante et sollicite plus directement leur participation que ne le font les manuels et le matériel d'enseignement plus traditionnel.
- ✓ L'élève apprend à utiliser ce type de technologie pour obtenir de l'information, de développer son esprit de recherche, de demander l'aide des autres, de dire ce qu'il pense et ce qu'il ne comprend pas.
- ✓ L'utilisation des réseaux sociaux favorise la collaboration entre les élèves et le partage des liens d'accès à des connaissances pertinentes non strictement définies à l'avance par l'enseignant.
- ✓ Dans les réseaux sociaux, l'enseignant s'occupe davantage des élèves qui ont besoin d'aide, qui sont généralement les élèves les moins avancés de la classe, alors que, dans la classe traditionnelle, il a tendance à s'adresser en priorité aux plus forts. Les RS lui facilitent aussi la détection des points forts de ses élèves, de même les difficultés qu'ils rencontrent ou de leurs apprentissages préalables erronés ou mal assimilés
- ✓ Ce moyen peut créer des conflits entre les élèves, quand leurs idées s'affrontent, mais de telles interactions sociales jouent un rôle important dans le développement de la capacité de penser de l'élève et d'avoir des retours et de revenir sur son apprentissage. Les élèves apprennent aussi à coopérer et transformer la structure sociale de compétition en une structure sociale de coopération.
- ✓ L'utilisation des réseaux sociaux pour l'apprentissage éduque les élèves à l'usage approprié des réseaux sociaux et les amène à réfléchir sur leurs identités numériques et la visibilité de leurs informations et ce qu'ils publient. **[CAN, 2012]**

## 7.2 Les inconvénients

- Les réseaux sociaux sont considérés comme source de distraction, les étudiants qui se rendent régulièrement sur les réseaux sociaux passent moins de temps à étudier et obtiennent des notes faibles aux examens.
- L'utilisation des réseaux sociaux crée des résistances idéologiques du fait que les réseaux sociaux d'apprentissage viennent modifier la pédagogie transmissive où le professeur n'est plus au centre et il n'est plus le seul détenteur du savoir. De plus, les élèves de la génération Y[8] ou Digital natives, sont en avance sur leurs professeurs ce qui oblige les enseignants à se former pour pouvoir utiliser ces réseaux.
- L'usage des réseaux sociaux à des fins éducatives peut facilement détourner de son objectif primaire et l'usage de celle-ci est parfois particulièrement inapproprié, voire illégal.
- Le mélange dangereux de la vie privée et la vie publique que crée ce type d'outils d'apprentissage (vu que les profils utilisés sont des profils professionnels), et les dérives possibles d'une traçabilité ineffaçable qui pourra poursuivre l'élève durant toute sa vie et nuire à son identité numérique dans l'avenir.

Les initiatives d'utilisation des réseaux sociaux par les enseignants pour l'apprentissage restent isolées et ne semblent pas être encouragées par les établissements d'éducation à cause de leur inquiétude sur le mélange de la vie privée et la vie publique des élèves que crée ce type d'outils. Cependant, l'intégration d'un réseau social d'apprentissage dans une plateforme d'e-Learning représente une excellente manière pour rendre les échanges autour de l'apprentissage plus dynamiques et initier les élèves à l'utilisation des réseaux sociaux, connaitre ses conséquences afin de mieux gérer leurs attitudes sur les réseaux sociaux du Web et sensibiliser les élèves à la visibilité de leur information personnelle et de leur vie privée et à l'importance d'avoir un certain contrôle de son identité numérique.

# 8. Les profils dans les systèmes e-learning

Un profil utilisateur est une représentation explicite de ses caractéristiques. Les principaux profils (acteurs) d'inscrits sur une plateforme d'apprentissage en ligne (e-Learning) sont les suivants **[Madjarov, 2005]** :

## 8.1 Les types de profil

### 8.1.1 L'apprenant

L'apprenant est un individu qui s'engage à suivre les activités d'une e-formation afin d'acquérir des connaissances. Il est l'acteur central pour lequel la formation est conçue.

[8] **La génération Y** (1970-2000) a connu la **« révolution technologique ».** Cette génération veut être tout le temps connectée ; ils contribuent, coopèrent et partagent des informations régulièrement. Ce groupe est **très performant dans l'usage des réseaux sociaux.**

### 8.1.2 L’enseignant

Avant le passage à la méthode d’enseignement centrée sur l’apprenant, le professeur était la pièce maîtresse dans le processus d’apprentissage, c’était lui qui détenait le savoir, mais avec la nouvelle méthode, l’accent est mis sur l’apprenant afin de le rendre plus actif. Le rôle de l’enseignant est beaucoup plus de guider et d’orienter les apprenants. Actuellement, l’enseignant peut prendre le rôle d’un :

- **Auteur (concepteur) de cours :** Son rôle est la création du contenu.

- **Tuteur:** Il a un rôle d’accompagnement des groupes d’apprenants, surtout au cours des phases d’action à distance (suivi et motivation…) en tant qu’animateur de la formation.

- **Evaluateur :** Son rôle essentiel est la création des activités de validation de connaissances par la création des tests, le suivi et l’évaluation de l'apprenant.

### 8.1.3 L’administrateur:

L’administrateur d'un système e-Learning s’occupe de l’installation et la maintenance de la plate-forme, la gestion des droits d’accès et la suppression des acteurs et des contenus en cas d’abus et la création de liens avec les systèmes. **[GRE, 13].**

## 8.2 Le contenu du profil

Les systèmes d’e-Learning utilisent les informations des apprenants pour adapter l’activité d’apprentissage et optimiser le comportement du système. Ces informations représentent le profil ou le modèle de l’apprenant. Plusieurs systèmes d’e-Learning utilisent leur propre présentation interne du modèle de l’apprenant. Cependant, les informations sur l’apprenant doivent être bien décrites afin de permettre leur réutilisation par différentes plateformes d'apprentissage à distance avec lesquelles l'apprenant est censé travailler. Il existe quelques standards et spécifications pour représenter le modèle de l’apprenant ; les efforts de normalisation les plus importants pour la modélisation de l’apprenant sont les suivants [**Oubahssi, 2005]**:

### 8.2.1 Le modèle IMS LIP

Le modèle « IMS LIP », qui désigne « IMS Learner Information Package », définit une structure XML pour l'échange des données de l’apprenant entre systèmes coopérants utilisant les processus d'apprentissage. Il représente les informations de l'apprenant dans un package défini et structure les données en onze éléments (appelés « segments ») :

1. **Identification** : Ce segment contient des éléments qui aident à identifier la personne tels que son nom, son âge, son adresse, son email, etc.
2. **Accessibilité** : Ce segment décrit l'accessibilité générale comme les capacités linguistiques, les handicaps et les préférences d’apprentissage.
3. **Qualifications, Certifications et Licences (QCL):** Ce segment décrit l’ensemble des qualifications, certifications et diplômes de la personne.
4. **Activités** : Ce segment regroupe les données sur les activités liées à l'apprentissage dans n'importe quelle étape de son cursus (formation, expérience professionnelle, etc.)

5. **Objectif** : Ce segment définit l'objectif de la tâche d'apprentissage, la carrière envisagée et les objectifs personnels etc.
6. **Compétence** : Ce segment décrit les compétences, l'expérience et les connaissances acquises associées à la formation.
7. **Intérêt** : Ce segment regroupe les informations décrivant les loisirs de l'apprenant.
8. **Relevé de notes**: c'est un dossier qui permet de décrire les données sur les bulletins de notes de la personne (résultats scolaires).
9. **Affiliation** : Ce segment présente des informations sur l'adhésion aux organisations professionnelles.
10. **Clé de sécurité** : Ce segment regroupe les données de sécurité d'une personne : L'ensemble des mots de passe et clés de sécurité liés à l'apprenant.
11. **Relation** : Ce segment permet de décrire les relations entre les structures de données utilisées pour stocker les données de la personne employées dans ce modèle.

### 8.2.2 Le modèle PAPI

**PAPI** (Public And Private Information for Learners) est un standard développé au sein du groupe « Learner Model Working Group ». Il s'est donné comme objectif de spécifier la sémantique et la syntaxe des informations sur l'apprenant. Ces informations peuvent être de diverses natures : acquisition de connaissances, préférences de l'apprenant, ses performances, ses compétences, ses relations avec d'autres apprenants, etc.

Le profil d'apprenant PAPI comprend donc les éléments suivants :

- informations personnelles sur l'apprenant (PAPI Learner Personal) : nom, adresse et numéro de téléphone. Ces informations ne sont pas directement liées à l'apprentissage, mais principalement à l'administration.
- informations relationnelles (PAPI Learner Relations) : relatives aux relations entretenues par l'apprenant avec les autres utilisateurs - professeurs, autres étudiants, etc.
- informations sur la sécurité (PAPI Learner Security): Mot de passe, clés, etc.
- informations sur la performance de l'apprenant (PAPI Learner Performance) : Le système mémorise les activités actuelles de l'apprenant ou ses objectifs futurs afin d'optimiser son parcours d'apprentissage.
- informations « portfolio » qui constituent une collection représentative des travaux de l'apprenant utilisées en tant qu'illustrations de ses capacités.

Chacun des standards présentés se focalise sur une catégorie spécifique de données. Le standard PAPI est plus orienté vers les données administratives et relatives aux préférences de l'apprenant plutôt que celles qui décrivent ses connaissances et compétences. Le standard IMS-LIP permet de représenter les différentes compétences et connaissances acquises par un apprenant particulier. De plus, les deux standards ne prennent pas en compte les publications partagées par l'utilisateur dans le réseau social ou le forum. **[HAGE, 2011]**

## 9. Tracking (Traçabilité)

Le « tracking » est le traçage des activités effectuées par l'apprenant pendant sa connexion au système e-Learning. Il vise à récupérer les informations pédagogiques relatives à l'avancement des apprenants dans leurs parcours e-Learning ; comme le temps passé sur la formation à distance, le nombre de connexions, les scores obtenus aux évaluations en ligne, etc. La récupération de données de traçage est un élément indispensable pour le travail d'un tuteur pour lui permettre de réaliser un tutorat en ligne rigoureux et individualisé (suivi de chaque apprenant, étude des résultats enregistrés, accompagnement, etc.).

## Conclusion

Depuis l'émergence du e-Learning, plusieurs modèles d'apprentissage à distance ont vu le jour. Chacun utilise des technologies et des médias différents et chacun se repose sur un modèle pédagogique différent mais ils visent tous le même objectif de transmission des connaissances et d'accessibilité.

Dès leur apparition, les plateformes d'e-Learning, se sont concentrées sur le développement d'outils permettant la création et le partage des contenus pédagogiques et facilitant le travail de l'enseignant. Mais, avec le passage à des méthodes d'enseignement centrées sur l'apprenant de nouvelles exigences sont apparues, ce qui a poussé les recherches vers le développement d'outils plus interactives et plus intelligents.

L'intégration des techniques de l'intelligence artificielle dans les plateformes d'e-Learning (ITS), a offert la possibilité de suivre et comprendre les besoins de l'apprenant, de connaitre ses points faibles et ses points forts, de le conseiller et l'accompagner tout au long de son processus d'apprentissage grâce à des entités logicielles appelées « agents intelligents » qui permettent de simuler le rôle de l'enseignant ou un camarade qui suit le cours avec l'apprenant afin de créer des situations d'apprentissage similaires à celles vécues dans une classe réelle. Ces solutions essayent de remplacer les êtres humains (qui ne peuvent pas être tout le temps avec l'apprenant pendant son apprentissage) par les agents intelligents ont tendance à isoler les apprenants. Cependant, les systèmes d'ITS doivent éviter l'isolement de l'apprenant (par des environnements d'apprentissage isolés) et passer vers des environnements d'apprentissage à base de communautés sociales créant des échanges riches, des interactions motivantes au travail, une ambiance et une entraide qui encouragent l'apprenant à continuer son parcours.

Les systèmes d'apprentissage social en ligne se distinguent par rapport aux autres systèmes d'e-learning classiques par la participation active des apprenants et des enseignants, qui sont considérés à la fois comme créateurs et consommateurs des contenus partagés dans cet environnement. C'est un environnement où tout le monde enseigne et apprend en interagissant les uns avec les autres. Cette participation active des individus pendant le processus d'apprentissage dans les systèmes de social Learning a généralement pour résultat l'aboutissement de l'apprenant à des compétences de pensée créatives, un esprit critique et un raisonnement logique.

Une solution combinant les deux approches (social Learning et agents intelligents) héritera des bénéfices que peuvent apporter ces dernières mais elle héritera aussi de l'inconvénient le plus dangereux qui est l'exposition des apprenants aux menaces à la vie privée.

# Chapitre2.
# La vie privée

## Avant-propos

Vu l'absence des solutions de protection de la vie privée dans les systèmes d'apprentissage social, nous présentons dans ce chapitre une étude sur la protection les réseaux sociaux. Nous terminerons par une présentation des problèmes de la vie privée dans les systèmes d'e-Learning ainsi qu'une présentation de la seule solution existante en littérature pour la protection de la vie privée dans les systèmes d'e-learning.

## I- Problématique de la vie privée dans les réseaux sociaux

Les réseaux sociaux en ligne sont devenus le portail de fait pour l'accès au Web pour des millions d'utilisateurs, ce qui a impliqué un changement fondamental sur le Web. Auparavant, le contenu du Web était principalement créé par un groupe d'éditeurs restreint (les entreprises, les universités et les gouvernements). Ce contenu était, en général, à accès public, avec comme objectifs, un partage ouvert et un accès universel à l'information. Aujourd'hui, une grande partie du contenu partagé sur le Web est en cours de création par les utilisateurs finaux, et la nature personnelle de ces informations est à l'origine de la problématique de la vie privée. Avec la motivation de communiquer et de maintenir les relations d'amitié en ligne, la quantité d'informations révélée volontairement par les utilisateurs sur les RS est beaucoup plus grande que ce que les utilisateurs généralement partagent sur les autres médias. Actuellement, les mécanismes de contrôle d'accès appropriés sont indispensables pour protéger la vie privée mais les menaces à la vie privée vont bien au-delà des problèmes de simples paramètres de confidentialité proposés par les réseaux sociaux. Les informations échangées dans les RS sont généralement à caractère personnel et de nature numérique; ce qui les rend faciles à être copiées, dévoilées voire même vendues à des tiers sans l'autorisation explicite de leur propriétaire. L'émergence des menaces à la vie privée causées par le partage d'informations sur les utilisateurs par les propriétaires des réseaux sociaux avec les sociétés publicitaires (pour générer des profits) ou avec les laboratoires de recherche (pour permettre la recherche scientifiques) présentent de nouveaux défis aux chercheurs car ces dernières utilisent des techniques très avancées pour l'extorsion des données privées. En effet, une fois l'instance de la base publiée, ni l'utilisateur, ni les fournisseurs n'ont le contrôle sur l'utilisation de ces informations par ces sociétés. La vente d'informations représente une source importante de financement pour les fournisseurs des RS et constitue en quelque sorte le prix payé par les utilisateurs pour bénéficier de leurs services gratuits.

Depuis quelques années, les chercheurs essaient de trouver de nouveaux instruments de protection contre les menaces de la vie privée venant des fournisseurs en les considérant comme la source de menaces la plus dangereuse. Comme les fournisseurs de RS résistent aux réclamations des groupes militants pour la vie privée et comme la conception de leurs RS ne respecte pas les lois et les principes de protection de la vie privée, les chercheurs s'efforcent de trouver de nouvelles solutions libres et indépendantes pour faire face à ces menaces. Les solutions proposées se scindent en deux grandes catégories : les solutions basées sur la sécurité informatique et les efforts juridiques des militants pour définir des lois imposant une reconnaissance générale du droit à la vie privée.

# 1. Les réseaux sociaux

## 1.1 Qu’est-ce qu’un réseau social ?

Les progrès récents en technologies de l'information ont apporté des changements importants à la nature de la communication et de socialisation. En effet, tout au long de ces dernières années, les blogs, les forums, la messagerie instantanée, les albums photos en ligne ont fleuri partout sur Internet. Aujourd'hui, tous ces médias sont rassemblés dans des sites de réseaux sociaux.

Bien que le concept de réseau social semble être évident, chaque chercheur le décrit d'une manière assez différente. Les réseaux sociaux en ligne ont été étudiés dans de nombreux contextes et différentes définitions ont été données. Il n'y a aucun modèle cohérent et reconnu des réseaux sociaux, ce qui explique pourquoi les chercheurs les nomment différemment: les réseaux sociaux assistés par ordinateur (computer supported social networks), les réseaux sociaux en ligne (online social networks), les sites de réseautage social(social Networking sites), les réseaux sociaux basés Web (web-based social networks), les communautés web (web communities), ou des communautés virtuelles (virtual communities ou Online communities), les portails des réseaux sociaux (social networking portals)…etc.

Au niveau le plus élémentaire, les RS consistent en la représentation de chaque utilisateur (souvent un profil). La page de profil agit comme la page d'accueil de l'utilisateur et peut inclure des informations personnelles tels que les photos, la date de naissance, le sexe, la religion, la musique préférée, les citations des livres,…etc. De plus, les utilisateurs sont en mesure de créer un réseau de contacts avec lesquels ils peuvent se connecter. Ces contacts sont appelés parfois «Amis ». Les utilisateurs peuvent afficher et parcourir la liste de leurs amis et les listes d’amis des autres utilisateurs du même RS [**Boyd, 2007**]. Pour définir le concept de réseau social en ligne, nous avons choisi la définition de **[Boyd, 2007]** qui est la plus citée en littérature:

**Définition**

"Les réseaux sociaux sur Internet sont des services basés web qui permettent aux individus de construire un profil public ou semi-public dans un système fermé, de créer une liste utilisateurs avec lesquels ils partagent une connexion (lien entre deux utilisateurs), de visualiser et d’explorer leur liste de contacts (ensemble de connexions) et celles des autres au sein du système. La nature et la nomenclature de ces connexions peuvent varier d'un site à un autre."

Cette définition est loin de décrire les RS actuels tels que Twitter ou les sites webs du partage des ressources tels que Youtube et Picasa. Par ailleurs, cette définition ne comprend pas l’aspect de la confidentialité des données qui apparait dans une communauté en ligne. La seule définition qui considère cette dimension est celle proposée par Ai Ho **[Ho, 2012]** :

Un réseau social (Social Networking Site) est un site Web qui permet aux utilisateurs de:

- Etre en contact avec d'autres utilisateurs par les demandes d'ajout(Facebook), par l'abonnement (Twitter, Youtube)...etc.
- Interagir avec les contenus publiés par d'autres utilisateurs, par exemple en commentant ou en évaluent les publications et les commentaires.
- Restreindre l'accès à leur propre contenu qu'aux utilisateurs autorisés.

## 1.2 Classification des réseaux sociaux

Plusieurs critères existent pour classifier les réseaux sociaux. L'un des critères de classification le plus intéressants est « le profil » des utilisateurs. Sur cette base, Musiał **[Musiał, 2013]** classifie les sites de réseaux sociaux comme suit**:**

| | |
|---|---|
| **Général** | Facebook, Friendster, Orkut, et de nombreux services locaux habituellement limités à une seule langue. |
| **Rencontres** | Yahoo! Personals, OkCupid, Fubar, Match.com, eHarmony, Plentyoffish, Zoosk, Christian Mingle, JDate. |
| **Anciens scolaires** | Classmates.com, Friends Reunited, Nasza Klasa College Tonight, StudiVZ |
| **Professionnels** | LinkedIn |
| **Chercheurs** | SciSpace.net, Epernicus, ResearchGate |
| **Artistes** | DeviantArt, Quarterlife, Taltopia |
| **Militants** | Care2, WiserEarth |
| **Intéressés par la politique** | dol2day |
| **Les fans de fantasy** | Elftown |
| **Adolescents** | Piczo, Faces.com, Habbo |
| **Les communautés mobiles** | itsmy, MocoSpace, mobikade |
| **Religieux** | MyChurch, Xt3, Muxlim |
| **Hommes d'affaires -** | Talkbiznow, XING |
| **Les clients** | Yelp, Inc., Epinions.com; |
| **Les athletes** | Athlinks |

**Tableau 1:** Classification des réseaux sociaux selon les profils **[Musiał, 2013]**

## 1.3 Les caractéristiques des réseaux sociaux

Les points détaillés ci-dessous sont les composantes principales des réseaux sociaux :

### 1) Profil

Les profils peuvent être considérés comme les briques de base du RS. Les profils contiennent généralement des informations démographiques basic sur l'utilisateur tels que son nom, son sexe, sa ville natale et son emplacement actuel…etc. Parallèlement à ces informations personnelles considérées essentielles pour chaque profil, la plupart des RS encouragent également les utilisateurs à écrire une courte biographie sur eux-mêmes et de partager leurs goûts et leurs intérêts. Pourtant ces types d'informations ne sont pas obligatoires pour pouvoir s'inscrire sur les RS, de nombreux utilisateurs mettent beaucoup de détails facultatifs sur leurs profils. Les figures suivantes montrent deux profils, un profil d'un utilisateur sur Facebook  (Figure3) et un autre sur Twitter (Figure4).

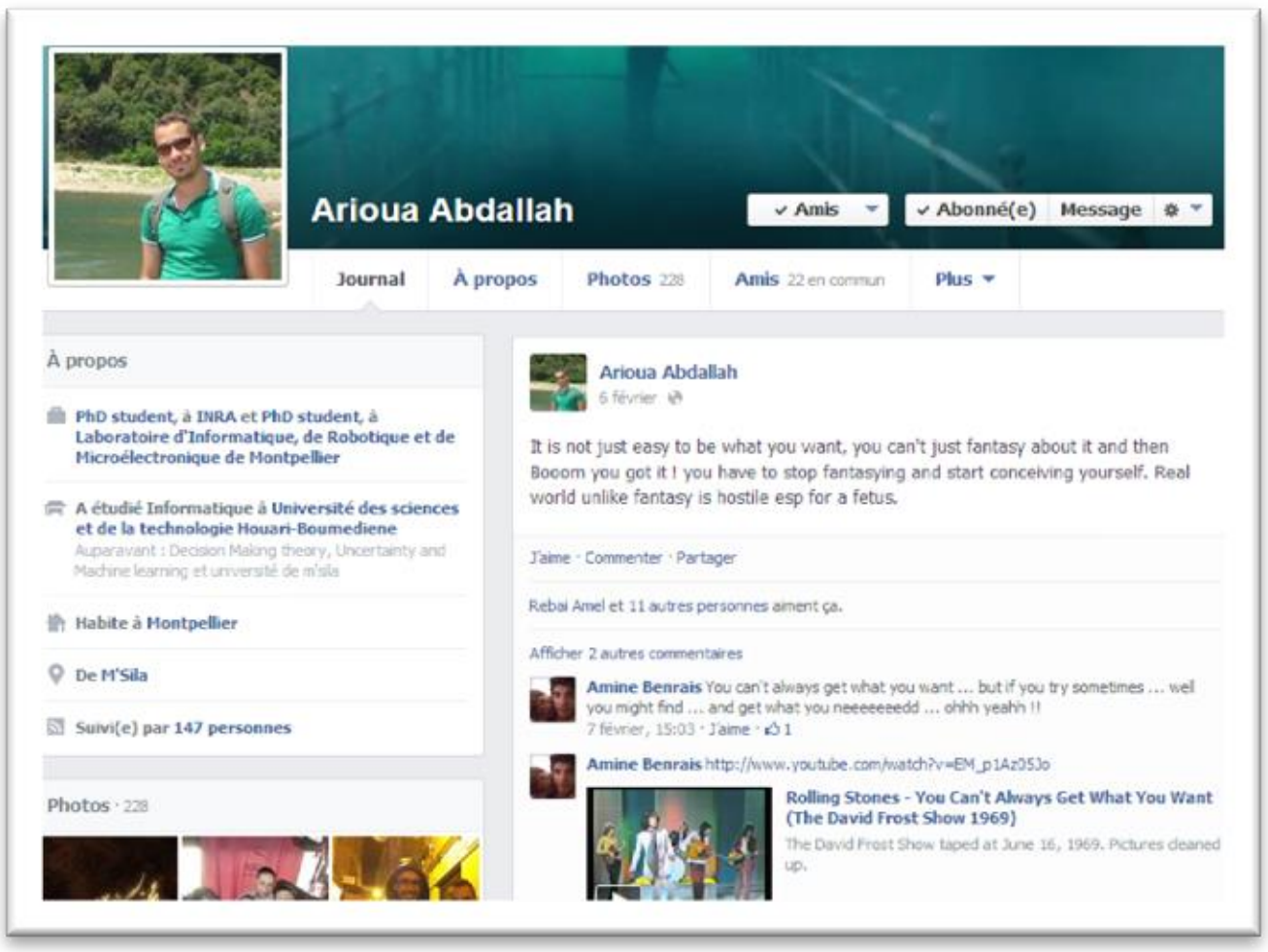

**Figure 3: profile Facebook avec une vue de son Timeline**

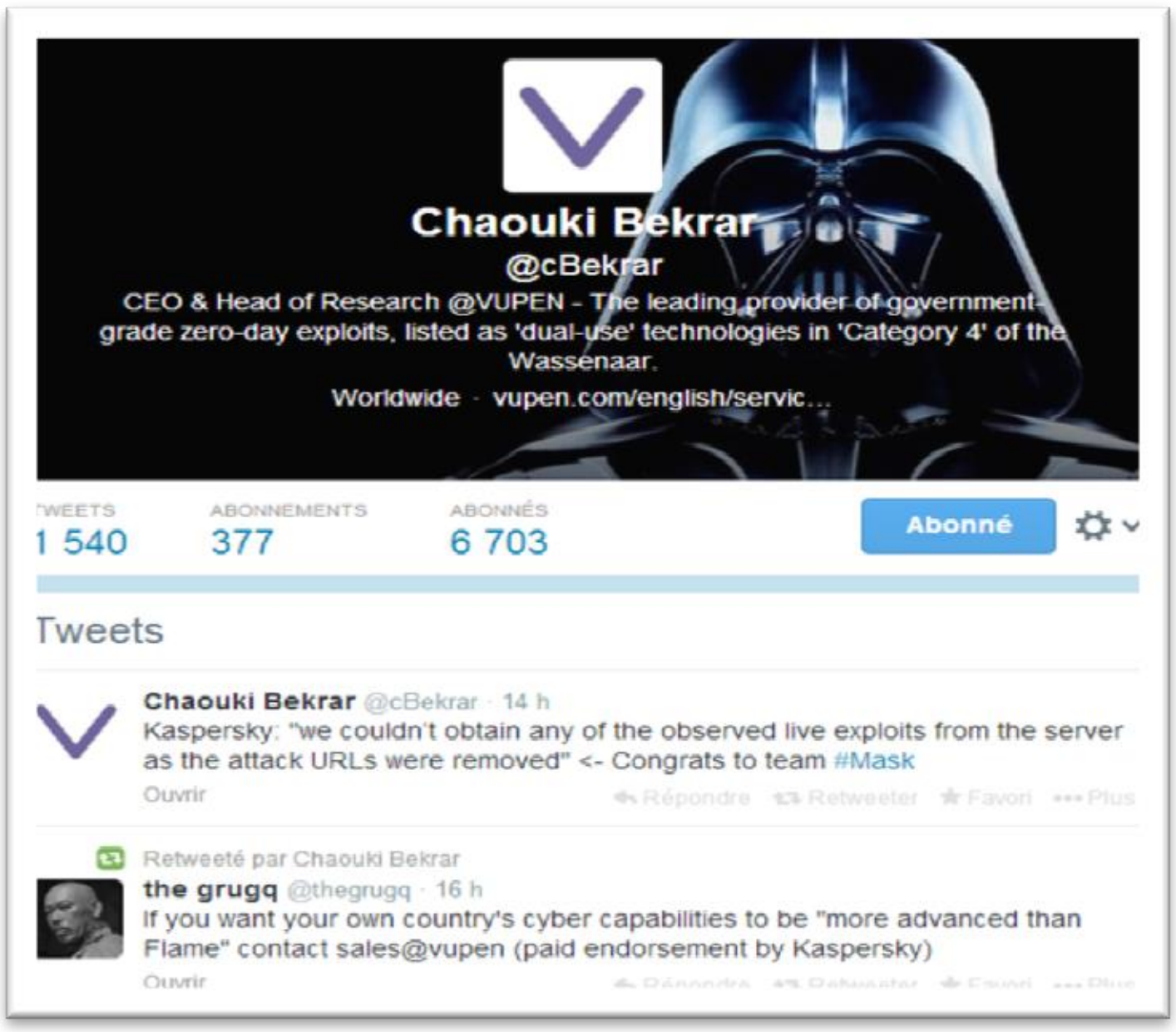

**Figure 4: profile sur twitter**

Les profils des utilisateurs Facebook sont très détaillés contrairement à Twitter qui montre seulement les informations de base.

## 2) Les amis

La plupart des RS sont conçus et construits autour du concept d'« Amis » ou « Friends ». Sur un RS, un « Ami » peut être un ami, un membre de la famille, une connaissance, un ami d'un ami, ou même quelqu'un que l'utilisateur n'a jamais rencontré auparavant, sauf en ligne. En 2013, le nombre moyen d'amis pour chaque utilisateur sur Facebook est de 175. Le RS permet à l'utilisateur de garder la trace  des activités de ses amis:

par exemple, quand ils publient une nouvelle photo, mettent à jour leurs profils, changent leurs statuts ou lorsqu'ils achètent quelque chose de nouveau en ligne. Le RS a généralement une fonctionnalité de recherche qui peut aider l'utilisateur à trouver de nouveaux amis. Par exemple, les utilisateurs peuvent rechercher des amis partageant les mêmes centres d'intérêts, qui appartiennent à un certain groupe d'âge, ou qui vivent dans la même région. Le RS peut aussi lui suggérer des amis qui pourrait les connaitre parce qu'ils ont des amis en commun. La relation « ami » peut être symétrique (Amitié) ou asymétrique (abonnement). Par exemple, sur Facebook, l'utilisateur peut ajouter des amis et voir également les contacts qui lui ont ajouté ou accepté dans leurs liste d' « Amis ».

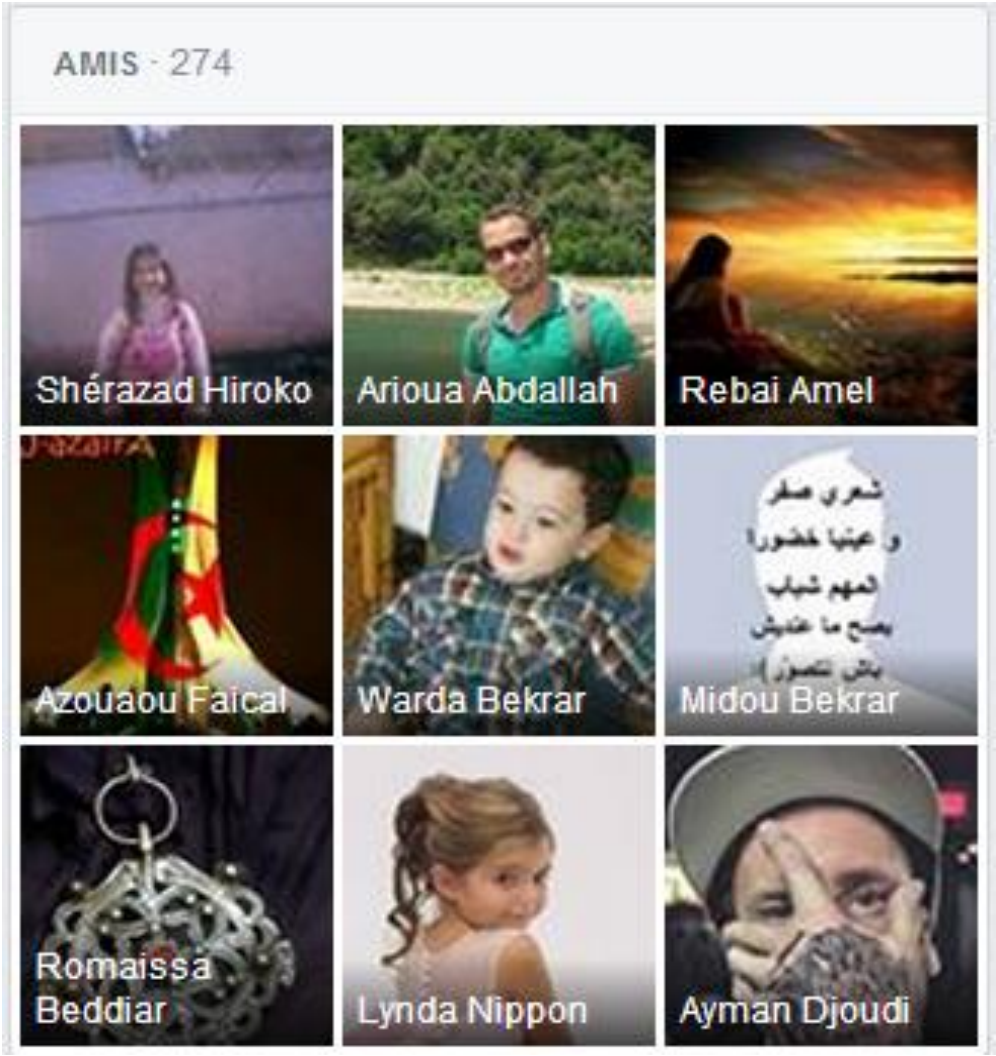

**Figure 5: Les « Amis » sur Facebook**

### 3) Fonctionnalités de réseautage

En plus des relations d'amitié, certains RS proposent également des fonctionnalités de réseautage pour faciliter l'interaction entre les utilisateurs, tels que les groupes, chat rooms et la messagerie instantanée. Chaque RS a aussi des fonctionnalités particulières propres à lui tels que l'envoie des « pokes » sur Facebook ou « High five » sur Hi5[9].

- **Les groupes**

La plupart des RS s'appuient sur la notion de groupe pour aider les utilisateurs à trouver des personnes ayant des intérêts similaires ou à s'engager dans des discussions sur certains sujets. Parfois, les groupes sont appelés par d'autres noms, tel que «les réseaux» sur Linkedin.

- **Les évènements**

C'est une fonctionnalité de réseautage permettant aux « Amis » de savoir les événements à venir dans leur communauté ainsi que d'organiser des rassemblements sociaux. Par exemple, sur MySpace, il est possible de publier un questionnaire ou de décorer la page de l'événement.

[9] http://www.hi5.com

- **Les « Tags »**

Un tag est un mot-clé ou terme assigné à un élément d'information. Par exemple, un tag peut être un bookmark en ligne, une photo numérique ou un fichier. Ce type de métadonnées décrit un objet et permet de le trouver par la recherche ou en navigant. Facebook et Friendster permettent aux utilisateurs d'associer un tag à une zone spécifique dans l'image. Par exemple, l'utilisateur peut taguer les personnes figurant dans une image d'une famille dans une place particulière par leurs noms et mettre un tag pour spécifier le nom de la place où la photo a été prise. Si le nom utilisé pour le tag est associé à un membre de Facebook ou à une page (région connue par exemple), le tag se transforme en un lien hypertexte vers le profil ou la page.

- **Flux d'actualité (News Feeds)**

Les flux d'actualité sont des outils utiles pour rester en contact avec les « Amis ». Par exemple, les mises à jour de profil, les messages sur le blog, les photos et vidéos publiées sont souvent diffusées sous forme de « news feeds » (voir la figure 6).

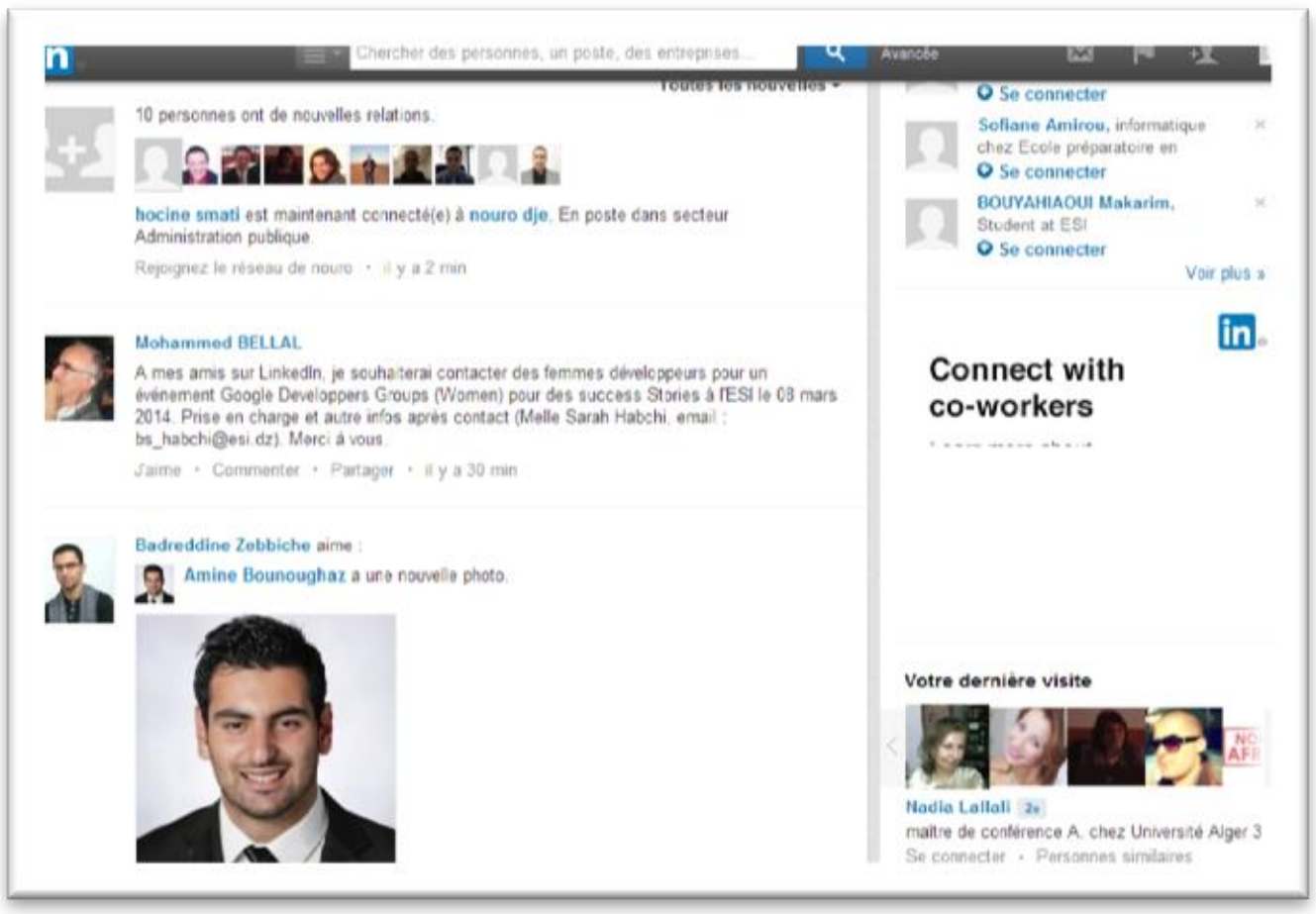

**Figure 6: Exemple des « New Feeds » de LinkedIn.**

## 4) Les applications sociales

Les RS comprennent un grand nombre d'applications sociales que les utilisateurs peuvent ajouter à leur profil. Ces applications peuvent être programmées à travers des interfaces ouvertes aux développeurs tiers pour concevoir et mettre en œuvre des applications ou des jeux sur la plateforme. Les applications créées à l'aide de ces API[10] posent des problèmes pour la vie privée car elles demandent aux utilisateurs d'accéder à leurs informations personnelles (situation familiale...etc.), à leurs listes d'amis, aux informations personnelles de leurs amis…etc. Voire même de publier sur leur profil et avoir un accès permanent à ces derniers. Beaucoup de recherches dans les systèmes d'e-learning travaillent sur le déploiement des jeux éducatifs « **Serious games**» dans les sites de réseaux sociaux tels que Facebook et Google+. Ces travaux visent essentiellement à profiter de l'enthousiasme des jeunes adolescents envers les réseaux sociaux pour les encourager à consacrer un peu du temps passé sur les RS pour l'apprentissage **[Allognon, 2012]**. Malheureusement, les applications ne sont pas accessibles au grand public**.** La figure 7 montre l'application « **Saif Almarifa**», un jeu éducatif très connu sur la plateforme de Facebook.

[10] Application Programming Interface : traduit par« interface de programmation » ou « interface pour l'accès programmé aux applications). Une API permet de fournir un certain niveau d'abstraction au développeur, c'est-à-dire qu'elle lui masque la complexité de l'accès à un système ou à une application en proposant un jeu de fonctions standard dont seuls les paramètres et les valeurs retournées sont connus.

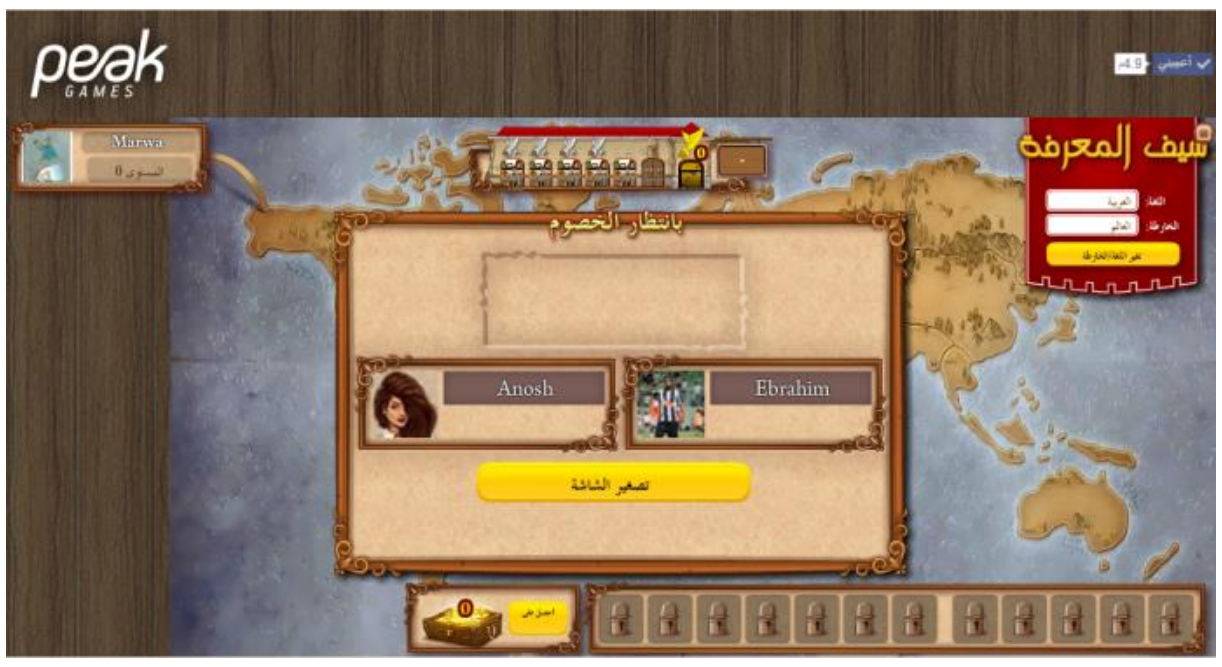

**Figure 7:** Application "Saif Almarifa"

## 1.4 Les architectures des réseaux sociaux

Les architectures des réseaux sociaux sont de 4 types : **[Marlier, --]**

### 1.4.1 Architecture centralisée

Les systèmes centralisés (Figure 8) sont généralement présenté par le système client-serveur qui se base sur un serveur central auquel se connectent les utilisateurs. Ce serveur est chargé de les mettre en relation directe et de stocker toutes leurs données. L'intérêt de cette technique réside dans la recherche rapide des ressources partagées par les utilisateurs du système. Cependant, cette architecture pose des problèmes de sécurité, de robustesse et de limitation de la bande passante .

Les réseaux sociaux les plus connues actuellement, tels que Facebook, Twitter, Linkedin et Google+ sont des systèmes centralisés.

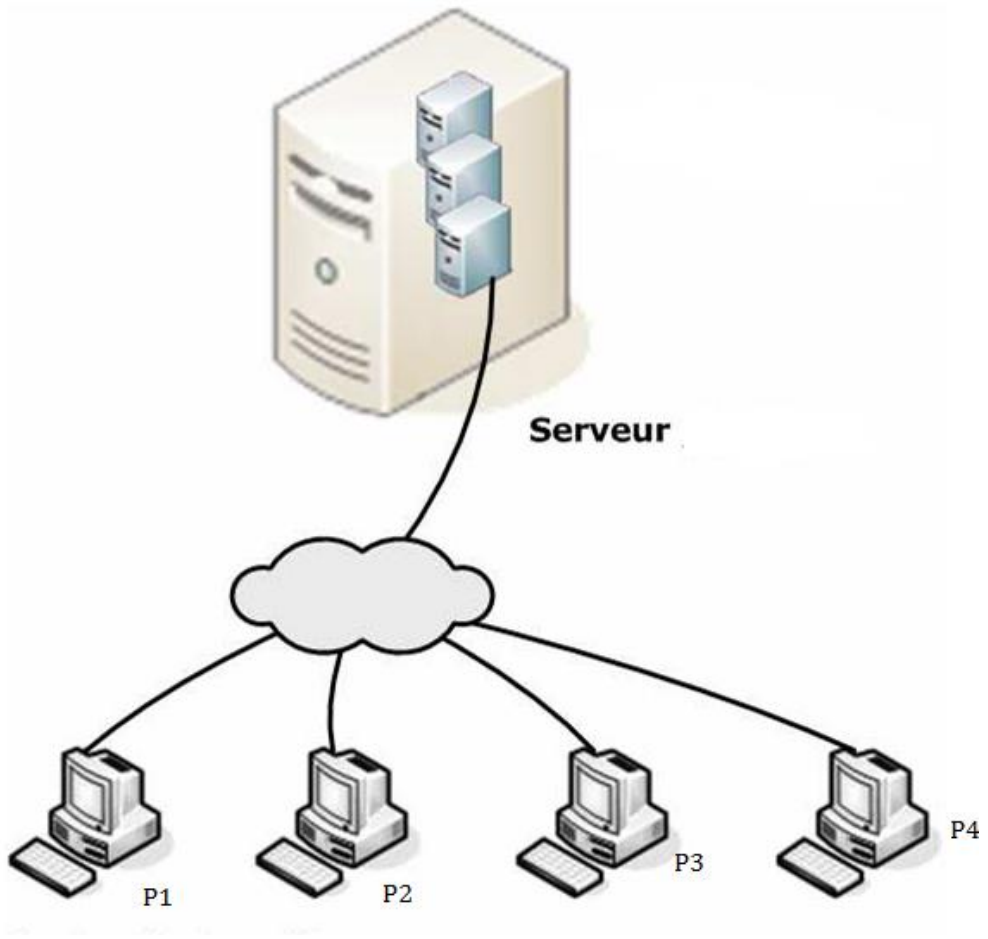

**Figure 8: Architecture centralisée [MARLIER,--]**

### 1.4.2 Les architectures décentralisées

Les systèmes P2P sont des systèmes répartis de nœuds interconnectés. Ils forment un réseau logique construit au-dessus de la structure physique. Cette structure est souvent appelée réseau de « recouvrement » ou « Overlay » en anglais. Le but de ces réseaux est de construire un réseau de pairs capables de s'auto-organiser afin d'échanger ou partager des services et des ressources tels que des données, des programmes, des capacités de stockage ou de calcul. Les types de ces réseaux se distinguent par le degré de centralisation des données. Il existe trois types d'architecture de systèmes P2P :

#### 1.4.2.1 Architecture purement décentralisée

Dans cette architecture chaque nœud du réseau est serveur et client au même temps, il n'y a pas de coordination centrale de leurs activités. Les systèmes P2P pure sont évolutifs, tolérant aux pannes et ont un plus grand degré de contrôle d'autonomie sur leurs données et leurs ressources. Ces systèmes présentent une découverte lente de l'information et La localisation d'une ressource est une opération critique qui demande a priori le parcours d'une partie plus ou moins importante du réseau. Les solutions consistent, en générale, en la constitution d'un annuaire incluant chaque client, puis de les faire communiquer. C'est sur ce mécanisme que se basent les réseaux « Peer to Peer » décentralisés. Cependant, il n'y a aucune garantie sur la qualité des services, en raison du manque de vision globale au niveau du système.

Parmi les réseaux sociaux purement décentralisés on trouve le projet Diaspora[11] qui a été résumé par ses créateurs comme un "réseau décentralisé, où les ordinateurs totalement indépendants (non liés) se connectent les uns aux autres directement, permettant ainsi aux utilisateurs de se connecter sans dévoiler leur vie privée".

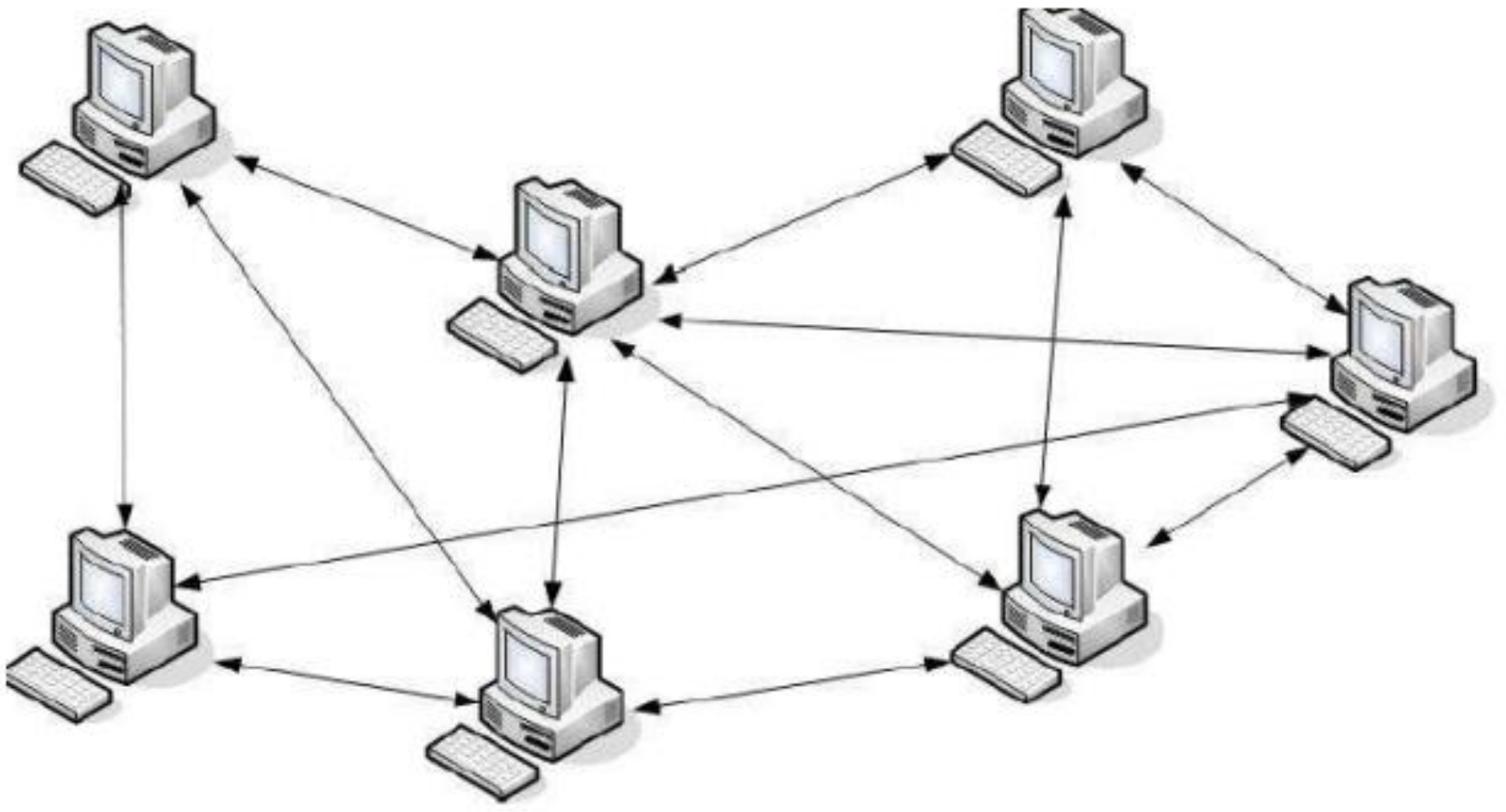

**Figure 9:**Architecture d'un P2P purement décentralisé **[MARLIER,--]**

[11] https://joindiaspora.com/

### 1.4.2.2 Architecture partiellement décentralisée

Le principe est similaire aux systèmes purement décentralisées. Ces systèmes se reposent sur l'élection des nœuds qui ont pour rôle d'agir comme des indices locaux centraux pour les fichiers partagés par les pairs connectés à ce nœud. Ces nœuds sont appelés « super-peers ». Chaque système a sa façon pour élire les super-nœuds et en cas de défaillance d'un super nœud le réseau prendra automatiquement des mesures pour le remplacer par un autre.

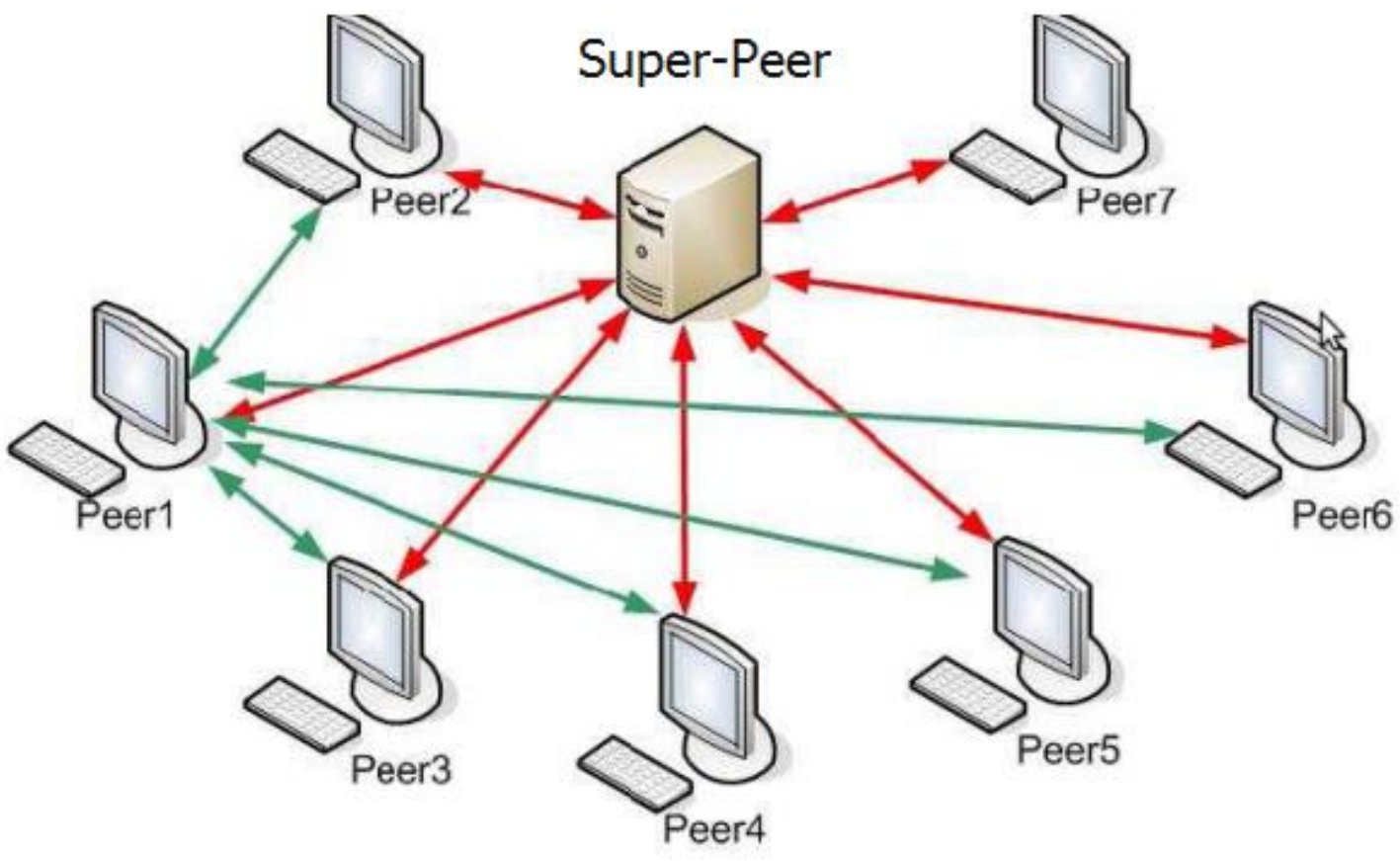

**Figure 10** Architecture partiellement décentralisée **[MARLIER,--]**

Skype est un réseau social partiellement décentralisé. Cependant, Skype possède un serveur central qui sert à authentifier les utilisateurs. (Figure11)

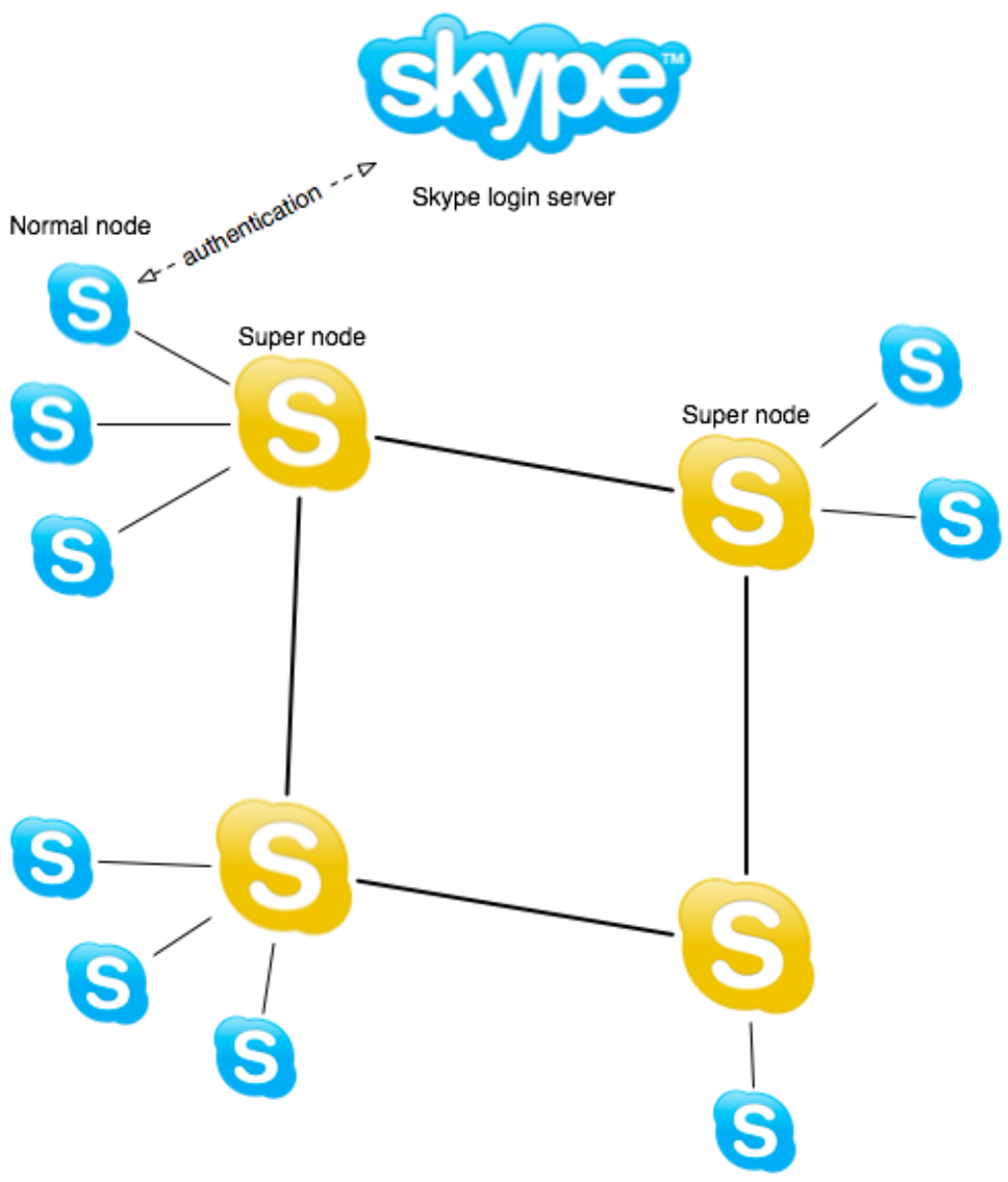

**Figure 11: Architecture de Skype**

### 1.4.2.3 Les architectures hybrides

Il s'agit d'un modèle semi-décentralisés entre le modèle centralisé (client-serveur) et le modèle purement décentralisé. Ces architectures utilisent un serveur de localisation central qui facilite l'interaction entre les pairs grâce à des répertoires de métadonnées, décrivant les données ou les fichiers partagés et stockés par les nœuds. Ces réseaux peuvent utiliser un ensemble de serveurs répartis afin de minimiser les pannes.

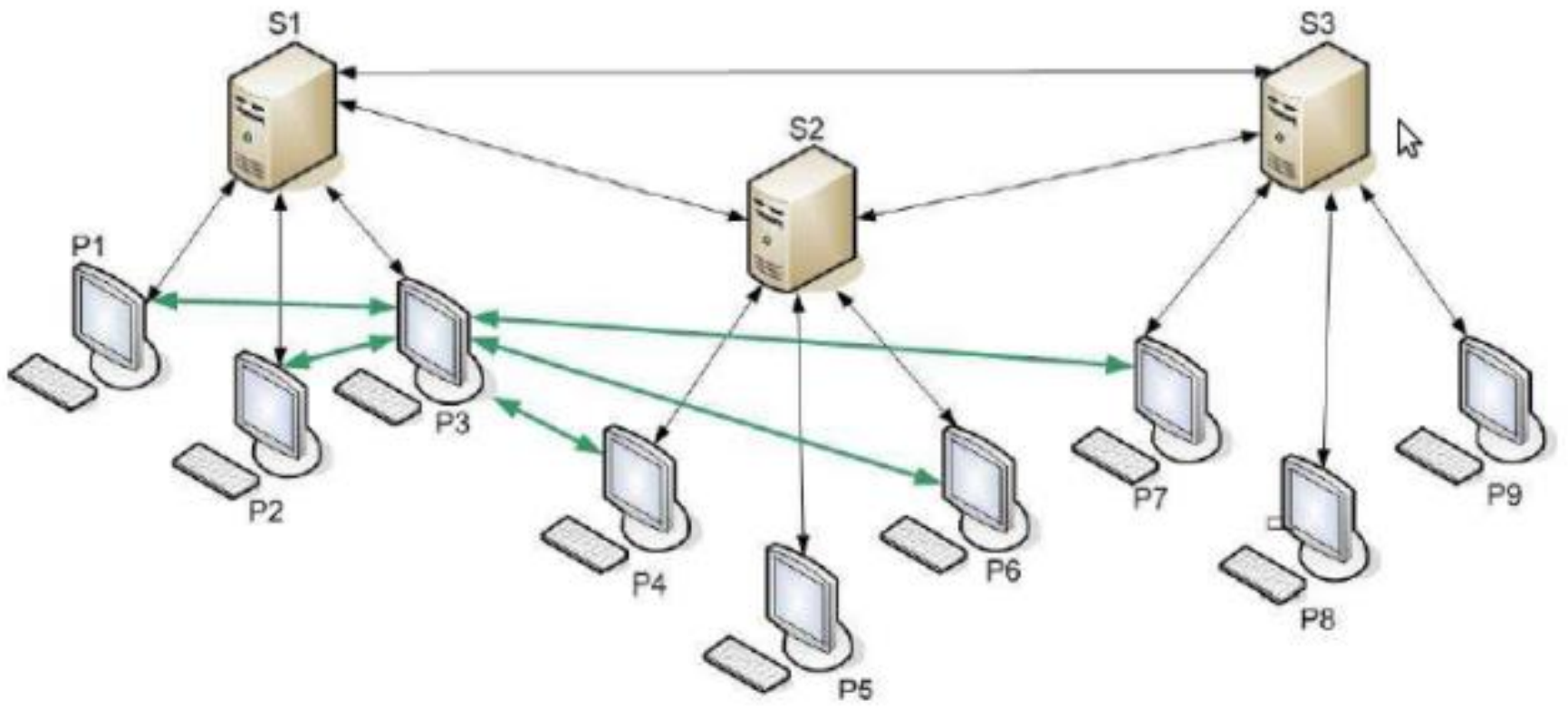

**Figure 12**: Architecture Hybride **[MARLIER,--].**

Les réseaux sociaux de cette architecture telle que « PrivacyWatch » **[Ho, 2012]**, sont des réseaux sociaux centralisés qui utilise Elgg[12], une plate-forme de réseautage social open source pour implémenter les fonctionnalités. Elgg implémente aussi des mécanismes pour contrôler les accès. La protection de la vie privée des utilisateurs, en utilisant le cryptage des données, est gérée par un plugin Firefox (Coté client).

## 2. La vie privée, un droit fondamental de l'homme

La vie privée, est tout ce qui est strictement intime, personnel et qui n'est pas ouvert à tout public**.** Le droit au respect de la vie privée est proclamé par la loi. **[Kolsi, 2007]** Chacun a le droit au respect de sa vie privée. Le droit au respect de la vie privée est défini par " le droit pour une personne d'être libre de mener sa propre existence avec le minimum d'ingérences extérieures ", ce droit comportant " la protection contre toute atteinte portée au droit au nom, à l'image, à la voix, à l'intimité, à l'honneur et à la réputation, à l'oubli, à sa propre biographie ". **[SENAT, 2014]**

Le contenu de la vie privée est variable selon les circonstances, les personnes concernées et les valeurs d'une société ou d'une communauté. Généralement, les domaines inclus dans la protection de la vie privée comprennent essentiellement l'état de santé, la vie sentimentale, l'image, la pratique religieuse, les relations familiales et, plus généralement, tout ce qui relève du comportement intime. **[SENAT, 2014]**

[12] http://www.elgg.org

## 2.1 Les lois de protection de la vie privée

La vie privée est un droit fondamental de chaque individu mais c'est également une notion difficile à définir et à formaliser. En effet, les lois de protection de la vie privée ne sont pas unifiées dans le monde entier et chaque loi ne peut protéger que les citoyens du pays qui l'a définie. L'Internet est très peu réglementé et les lois relatives à la vie privée en ligne sont en général en cours de développement **[Clearinghouse, 2011]**. Dans cette section, nous présentons certaines règles et lois de protection de la vie privée. **[Ho, 2012]**

### 2.1.1 Les lignes directrices de l'OCDE

Les lignes directrices de la vie privée ont été émises par l'Organisation pour la Coopération et le Développement Economiques (OCDE) **[OCDE, 1980]** en 1980. Elles sont devenues la base des lois de la vie privée et les politiques connexes dans de nombreux pays, dont les États- Unis, Canada, la Suède, l'Australie et Nouvelle-Zélande, ainsi que l'Union Européenne. Ces lignes directrices comportent huit principes, qui sont souvent désignés comme **« Les pratiques d'information équitables** » : **[Ho, 2012]**

1) **Principe de la Limitation de la Collection** *:* La collecte de données personnelles doit être limitée et obtenue par des moyens légaux et équitables et seulement au besoin avec la connaissance ou le consentement du propriétaire des données. Pour l'implémenter, la collection des données personnelles doit être limitée à un minimum strict dans la conception d'une application. De plus, la conception doit assurer qu'il n'y a pas des données identifiables qui sont collectées par un tiers. Ce processus n'empêche pas l'utilisation ni la manipulation des données, mais il empêche la collection des données inutiles.
2) **Principe de la qualité des données** : Les données personnelles doivent être pertinentes, exactes, complètes et mises à jour avant l'opération de la collecte de données.
3) **Principe de spécification des buts** : Les buts de la collecte de données doivent être précisés au moment de la collecte et au moment d'apport des changements à ces buts. L'utilisation de données à caractère personnel devrait suivre des fins similaires.
4) **Principe de la limitation de l'utilisation** : Les données personnelles ne doivent pas être divulgués, mises à la disposition ou utilisées à des fins autres que celles spécifiées conformément au Principe 3, sans le consentement du propriétaire des données.
5) **Principe des mesures de sécurité** : des mesures de sécurité raisonnables devraient être utilisées pour protéger les données personnelles contre la perte ou l'accès non autorisé, la destruction, l'utilisation inappropriée, la modification ou la divulgation de données.
6) **Principe de la transparence** : Le public devrait être informé sur les politiques et les pratiques de confidentialité. Aussi, les individus doivent avoir un accès facile à l'utilisation et la collecte de l'information personnelle.
7) **principe de la participation des individus**: l'individu devrait avoir le droit de savoir sur la collecte de données, de récupérer les données recueillies, de modifier ou de supprimer les données recueillies de manière raisonnable et de contester le refus de ces droits .
8) **Principe de la responsabilité** : l'entreprise qui gère les données devrait être responsable sur l'application de ces principes.

Les lois de la vie privée qui seront présentées dans les paragraphes suivants ont été directement influencées par les principes et les lignes directrices équitables de l'OCDE.

### 2.1.2 Les lois de protection de la vie privée en Europe

Au début de l'année 2009, L'union Européenne (UE) a investi des efforts considérables pour encourager tous les fournisseurs des services de réseautage social à rédiger et à adapter leur réglementation interne afin d'améliorer le niveau de sécurité des jeunes utilisateurs sur ces réseaux. Dix-huit (18) fournisseurs des réseaux sociaux (y compris Facebook, Bebo, Dailymotion et MySpace) ont uni leurs efforts dans un accord appelé « the Safer Social Networking Principles for the EU » **[EUROPA, 2009]**, à travers lequel les participants acceptent de se conformer aux principes de base d'autorégulation suivants:

1. Sensibiliser les utilisateurs aux risques de ces réseaux.
2. S'assurer que les services sont appropriés à l'âge pour le public visé.
3. Fournir des mécanismes faciles à utiliser pour signaler les comportements illicites ou contenu inapproprié.
4. Répondre rapidement aux notifications de contenus illicites.
5. Permettre et encourager les utilisateurs à employer une approche sûre pour protéger les informations personnelles.
6. Évaluer les moyens de vérification des contenus / comportements illégaux ou interdits.

En Mars 2011, la Commission européenne a déclaré que les médias sociaux en ligne doivent se conformer aux lois européennes s'ils ont des informations sur les citoyens européens et en particulier, les quatre piliers de la législation de la vie privée dans l'UE **[Bisaerts, 2011]** :

1. **Droit à l'oubli** : les entreprises doivent démontrer la nécessité de collecter les données à caractère personnel et les utilisateurs ont le droit de se retirer ou de ne pas participer à tous les efforts de collecte de données. Le droit à l'oubli, qui est appelé actuellement le droit à la suppression **[DataGuidance, 2013]**, oblige le contrôleur des données à prendre toutes les étapes pour avoir les données de l'utilisateur effacées (i.e. l'individus demandant la suppression de ses données), y compris les données retenues par les tiers «sans tarder» et les données personnelles que le contrôleur a rendu publiques « sans justification légale ». Actuellement, une grande partie du débat sur le « droit à l'oubli» a mis l'accent sur la faisabilité de l'implémentation et le respect de ce droit surtout avec l'avènement du cloud computing et du Big Data.
2. **Transparence**: Les entreprises doivent dévoiler entièrement aux utilisateurs toutes les informations concernant le processus de collecte de données. La logique pour la transparence est d'établir la confiance.
3. **Protection des données par défaut** : Les paramètres de confidentialité par défaut devraient refléter le niveau de la " vrai vie privée " pour les utilisateurs.
4. **Protection indépendamment de l'emplacement des données**: Les normes de vie privée pour les citoyens européens doivent s'appliquer indépendamment de la région du monde où leurs données sont traitées.

## 3. Les menaces à la vie privée dans les réseaux sociaux :

Les menaces à la vie privée sont essentiellement de trois types:

- **Les accès non autorisés** : ils sont généralement des attaques réalisées par les pirates (attaques externes) ou par des utilisateurs malveillants du même réseau. On trouve aussi dans cette catégorie les accès aux messages et contenus de l'utilisateur (sans son consentement) par le fournisseur du système pour des buts d'espionnage.
- **La vente des données** : effectuée par les fournisseurs ou bien par ceux qui collectent les informations publiques sur les réseaux sociaux, pour détecter par exemple les membres les plus influents d'un réseau, ou les préférences et les gouts des utilisateurs à partir des « j'aime » qu'ils mentionnent sur les pages et les publications. La détection des membres influents ou les préférences sert à détecter les clients potentiels pour la publicité pour les entreprises. Cela se fait par les techniques de « Profilage » qui enregistrent et classifient les comportements. Il s'agit d'une véritable industrie, parfois appelée Client Relations Management (CRM) ou tout simplement « Personnalisation ». Les informations collectées servent à construire des profils compréhensibles des individus, afin de vendre leurs produits. Ceci est souvent effectué sans le consentement explicite de la personne. Il est à noter que la vente des données publiques est légale. Cependant, la vente des données privées qui est illégale.
- **La révélation des informations personnelles** : La révélation des informations personnelles sans la connaissance et sans l'autorisation de la personne concernée est un grand problème qui n'a pas encore de solution. Les amis de l'utilisateur peuvent révéler des informations ou publier des photos qui peuvent ruiner sa réputation ou l'exposer à des menaces telles que « Cyberbullies » ou les prédateurs en ligne. Les tags associés aux photos et qui pointent directement vers le profil de l'utilisateur sont un moyen très dangereux qui peut ruiner la vie professionnelle et sociale de l'utilisateur.

Les applications accessibles via Facebook peuvent aussi causer trois types de violations de la vie privée :

- **Les attaques coercitives** : Les réseaux sociaux sont très vulnérables aux attaques dites « attaques coercitives » par lesquelles les applications obligent l'utilisateur à fournir son mot de passe pour pouvoir bénéficier de ses services (exemple, les jeux sur Facebook). Ce type d'attaques représente un grand danger sur la sécurité des comptes utilisateurs et facilite la divulgation des informations confidentielles. Sur Facebook, la plupart des utilisateurs ne savent pas qu'ils se connectent aux applications à travers leurs mots de passe associés leurs comptes Facebook **(Figure 13).**

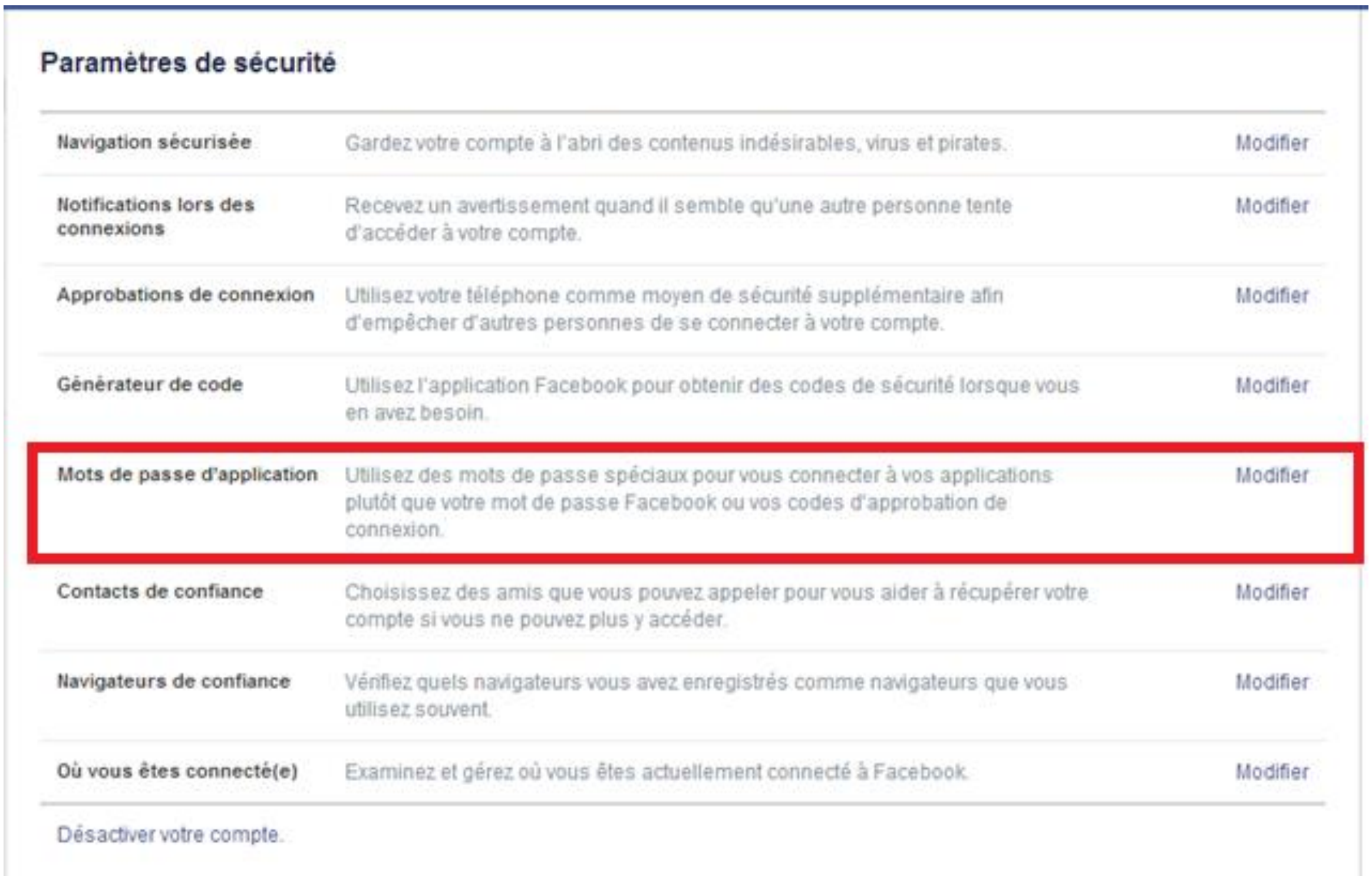

**Figure 13: les mots de passe pour les applications sur Facebook**

- **La vente des données personnelles** : Comme ces applications accèdent et récupèrent les données privées des utilisateurs, elles peuvent aussi vendre les données récupérées. Il existe même des applications qui sont allées plus loin. Il existe une application sur Facebook qui demande l'accès aux messages privés de l'utilisateur afin de compter le nombre des messages échangés avec chaque personne et lui dire avec quelle personne il a échangé le plus grand nombre de messages. Une telle application qui prend l'image d'un simple jeu, peut cacher dernière elle des objectifs très dangereux tels que l'espionnage.
- **La vulnérabilité aux menaces à la vie privée** : Même si l'application ne dévoile pas les données privées, elle peut être elle-même vulnérable aux menaces à la vie privée.

## 4. Les modèles de l'attaquant

Les utilisateurs des réseaux sociaux sont exposés aux attaquants suivants:

- **Le fournisseur du réseau social:** il possède tous les droits et les privilèges d'accès aux données des utilisateurs.
- **Les utilisateurs du réseau social:** ceux sont les utilisateurs inscrits dans le réseau social mais qui ne font pas partie de la liste d'amis de l'utilisateur.
- **Les utilisateurs externes**: ceux sont les utilisateurs qui ne sont pas connectés au réseau social. C'est la responsabilité du fournisseur du RS d'implémenter les mécanismes nécessaires pour la protection de ses utilisateurs contre les accès externes non autorisés.

- **Les applications et les parties tierces :** Ceux sont les applications et les parties tierces autorisées (par le fournisseur du RS) à récupérer les informations des utilisateurs.

## 5. Etude des solutions existantes

Avant de présenter les solutions, il serait préférable de définir quelques concepts pour mieux comprendre les solutions présentées dans ce chapitre.

## Quelques définitions

### 1) Certificats anonymes de pseudonymes

Les autorités de certification (CA) sont des entités de confiance dont la responsabilité centrale est de certifier l'authenticité des entités (personnes ou organisations) et leurs clés publiques. Plus précisément, un certificat d'une entité délivré et signé par une autorité de certification agit comme une preuve que la clé publique est associée à l'entité en question. En 1985, Chaum **[Chaum, 1985]** a introduit le concept de certification de pseudonyme pour protéger la vie privée. Plus précisément, le système résultant permet aux utilisateurs de participer à plusieurs transactions (avec les systèmes) électroniques, anonymes et qui ne peuvent pas être tracées.

### 2) Le chiffrement

Le chiffrement consiste à effectuer une opération (dite de chiffrement) par une clé (suite d'octets) pour chiffrer les données. Il existe deux sortes de chiffrement **:**

- ✓ Chiffrement asymétrique ou à clé publique
- ✓ Chiffrement symétrique ou à clé symétrique

### 2.1) Chiffrement asymétrique [Hage, 2011]

Les systèmes à clé publique ou Public Key Cryptosystems (PKCs) étaient introduits indépendamment par Merkle **[Merkle, 1978]** par Diffie et Hellman. **[Diffie, 1976]** Formellement, un PKC se compose de trois algorithmes efficaces: un algorithme de génération de clé qui génère des paires de clé secrète (SK) et clé publique (PK), un algorithme de cryptage, E, qui calcule le cryptogramme d'un message en utilisant la clé publique, et un algorithme de déchiffrement, D, qui calcule le message en clair à partir du message chiffré, en utilisant la clé secrète.

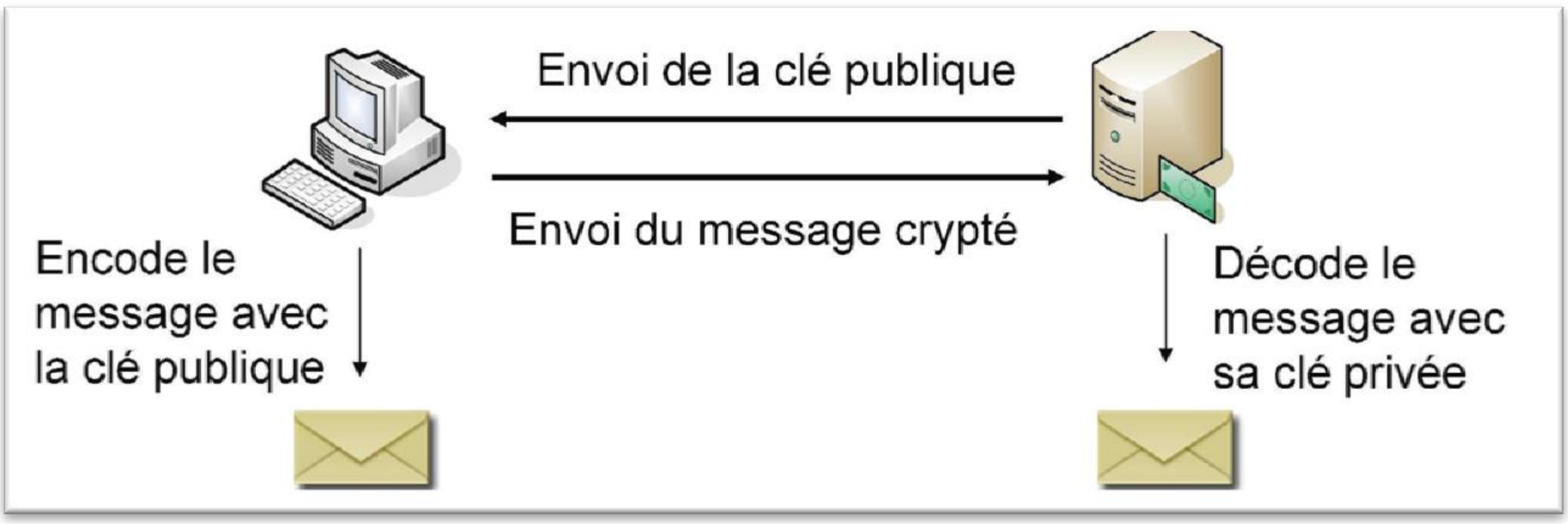

**Figure 14: chiffrement à clé asymétrique [wikileaks, 2010]**

### 2.2) Le chiffrement à clé symétrique [Hage, 2011]

La clé de chiffrement est la même que la clé de déchiffrement (clé secrète, privée).

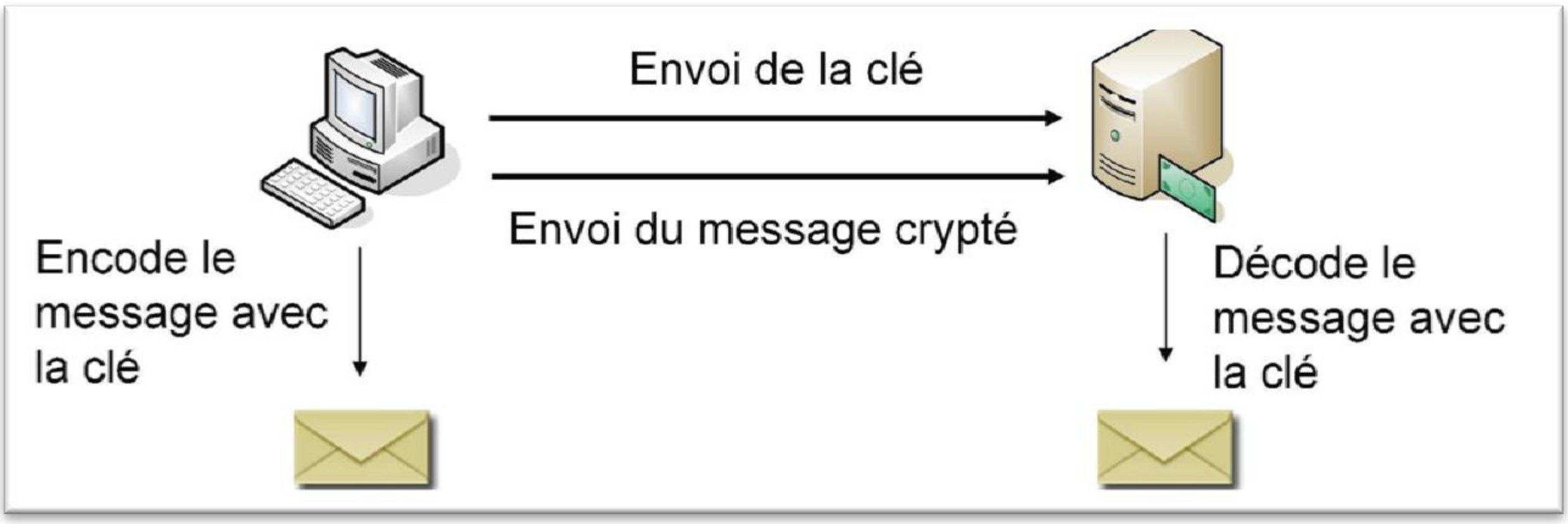

**Figure 15: Chiffrement symétrique [wikileaks, 2010]**

L'objectif de la cryptographie est de permettre à deux personnes de communiquer au travers d'un canal non sécurisé (téléphone, réseau informatique ou autre), sans qu'un espion puisse comprendre ce qui est échangé.

La cryptographie a essentiellement quatre objectifs principaux qui sont d'assurer :

- **La confidentialité**: garantir que le contenu d'une communication (ou d'un fichier) n'est pas accessible à tout autre personne que le destinataire légitime par l'utilisation des méthodes de chiffrement.
- **L'authenticité**: Assurer que l'identité de l'interlocuteur et bien celle qu'il prétend. Généralement, elle est vérifiée par les moyens d'identification ou de signature numérique.
- **L'intégrité**: s'assurer que le contenu d'une communication (ou d'un fichier) n'a pas été modifié de façon malveillante, généralement la vérification se fait par les fonctions de hachage.
- **La non-répudiation**: mécanisme pour enregistrer un acte ou un engagement d'une personne ou d'une entité de telle sorte que celle-ci ne puisse pas nier avoir accompli cet acte ou pris cet engagement. Les signatures numériques sont le moyen le plus utilisé pour assurer la non-répudiation.

### 3) RSA [Wikipedia, --]

RSA est un algorithme de chiffrement nommé par les initiales de ses trois inventeurs en 1977 par Ronald **R**ivest, Adi **S**hamir et Leonard **A**dleman**. RSA** se base sur la théorie d'Euler définit comme suit:

| |
|---|
| Soit $n = p*q$ (p et q premiers) |
| et quel que soit m, Nous avons $m^{(p-1)(q-1)} \bmod n = 1$ |

Ce qui peut aussi donner :

$$m*m^{(p-1)(q-1)} \bmod n = m$$

$$m^{(p-1)(q-1)+1} \bmod n = m$$

Cela veut dire que si on élève m à une certaine puissance, on retombe sur m.

Si on applique l'algorithme pour la cryptographie, on obtient:

Soient:

- **M** le message en clair
- **C** le message chiffré (Cryptogramme)
- **(e, n)** constitue la clé publique
- **(d, n)** constitue la clé privée
- **n** le produit de 2 nombres premiers
- ^ l'opération de mise à la puissance (a^b : a puissance b)
- **mod** l'opération de modulo (reste de la division entière)

Pour chiffrer un message M, on fait: C = M^e mod n.

Pour déchiffrer un cryptogramme C : M = C^d mod n.

**4) Digital Signature (signature numérique)**

La **signature numérique** est un mécanisme permettant de garantir l'intégrité d'un document électronique et d'en authentifier l'auteur. L'algorithme consiste en la préparation d'un condensat du message M par l'émetteur en utilisant la fonction de hachage choisie **H**(**M**). Puis il chiffre ce condensat grâce à une fonction de chiffrement **C** en utilisant sa **clé privé** $K_{pr}$: $\mathbf{S_M} = \mathbf{C}(K_{pr},\mathbf{H}(\mathbf{M}))$ Le résultat obtenu est la signature du message. Il prépare le message signé en plaçant le message en clair **M** et la signature $\mathbf{S_M}$ dans un conteneur quelconque : $\mathbf{M_{signé}}=(\mathbf{S_M},\mathbf{M})$. Il transmet le message signé au destinataire par un canal non sécurisé.

A la réception du message signé, le récepteur, pour vérifier l'authenticité du message, produit un condensat du texte clair en utilisant la fonction de hachage convenue : **H**(**M**). Ensuite, il **déchiffre** la signature en utilisant la fonction de déchiffrement **D** avec la **clé publique** $K_{pb}$ soit : $\mathbf{D_{Sm}}=\mathbf{D}(K_{pb},\mathbf{S_M})$. Puis, Il compare $\mathbf{D_{Sm}}$ avec **H**(**M**).Si la signature est authentique, $\mathbf{D_{Sm}}$ avec **H**(**M**) sont égaux, le message est alors authentifié.

**5) Blind signature [Hage, 2011]**

Un schéma d'une signature aveugle est un schéma d'une signature numérique qui permet à une entité centrale de signer un message u, sans rien connaître sur u. Les signatures aveugles sont utiles pour les applications où l'entité qui demande la signature (l'utilisateur par exemple) a besoin de protéger sa vie privée. Dans le travail de Hage, il utilise le schéma de signature aveugle basée sur la signature numérique RSA. Pour utiliser le RSA, il faut générer un module public N = p * q (où p et q sont deux grands nombres premiers), un exposant public e, et un exposant secret d. **[Hage, 2011]**

## 5.1 Les solutions centralisées

La plupart des RS connus, tels que Facebook, LinkedIn, Twitter ou Google+, sont des RS centralisés dans lesquels les données des utilisateurs sont sous le contrôle d'une entité centrale qui les stocke sur ses serveurs. La majorité des RS sont basés sur une architecture centralisée, dans laquelle les données des utilisateurs sont stockées sur un serveur contrôlé par les fournisseurs des RS. Facebook est souvent attaqué en justice à cause de la vente de profils, voire même les messages privées entre les utilisateurs **[JAU, 201A]**. En fait, Facebook est considéré comme la plus terrible machine à espionner **[Bor, 2012]**. Cependant, Facebook et les autres réseaux sociaux centralisés proposent des solutions pour protéger la vie privée des utilisateurs. Nous prenons Facebook comme un réseau social de référence. Dans ce qui suit nous allons présenter les solutions proposées par Facebook pour protéger la vie privée de ses utilisateurs. Nous discutons aussi les problèmes de ces solutions centralisées.

### 5.1.1 Protection de la vie privée des utilisateurs sur Facebook

Facebook[13] est un réseau social qui a été lancé en Février 2004. Depuis sa création, Facebook a connu une transformation remarquable. Au début, il était un espace privé pour la communication avec un groupe choisi. Récemment, il a été transformé en une plateforme dans laquelle presque toutes les informations de l'utilisateur sont par défaut publiques. Aujourd'hui, il est devenu une plateforme où les utilisateurs n'ont pas le choix de mettre certains renseignements publics tels que le nom, la photo de profil et les commentaires sur les murs des autres. Ces informations publiques sont partagées par Facebook avec ses sites partenaires et utilisées pour la publicité ciblée (les annonces sur Facebook). Facebook offre des options puissantes pour protéger les utilisateurs en ligne, mais c'est aux utilisateurs de les utiliser de façon proactive. Facebook propose les paramètres de confidentialité suivants :

1) **Les paramètres de confidentialité du partage des publications** :

L'utilisateur peut utiliser le menu sélecteur d'audience en ligne pour contrôler qui peut voir le contenu posté (comme les mises à jour de statut, photos et vidéos). L'utilisateur peut aussi contrôler la visibilité de ses informations personnelles de son profil (par exemple, l'anniversaire et les informations de contact) et même les contenus que les autres partagent à propos de lui (par exemple, des commentaires sur ses publications et ses photos). Facebook fournit un vocabulaire standard des paramètres de confidentialité (moi uniquement, amis, amis-des-amis, tout le monde) à partir de laquelle les propriétaires des ressources peuvent choisir **(Voir Figure 16).**

Les utilisateurs peuvent partager leurs publications et leurs informations à des audiences spécifiques en listant les noms d'amis ou les groupes d'amis. Facebook permet également aux utilisateurs de décider qui peut rechercher leur profil par nom ou par des informations de contact (exemple : seulement les amis de mes amis qui peuvent me trouver). De plus, Facebook supporte la révision des tags dans le sens où les utilisateurs peuvent approuver ou supprimer des tags pointant vers leur profil. Les utilisateurs de Facebook peuvent utiliser un menu en ligne, c'est-à-dire : les paramètres de confidentialité qui sont intégrés à chaque publication, pour spécifier qui peut accéder à leurs publications. Sur Facebook, la politique de contrôle d'accès d'un élément de données est directement enlacée avec la publication ou l'application **(Voir Figure 17)**.

[13] www.facebook.com

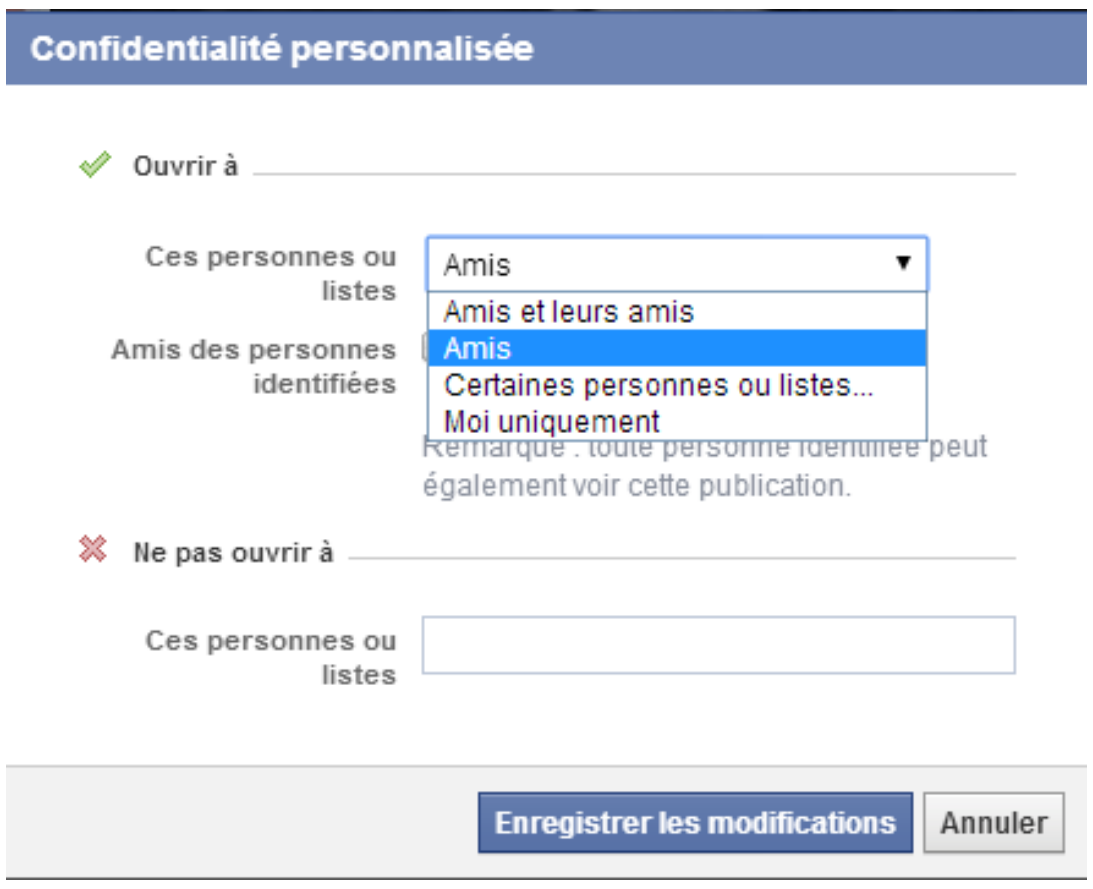

**Figure 16: Paramètres de confidentialité de Facebook**

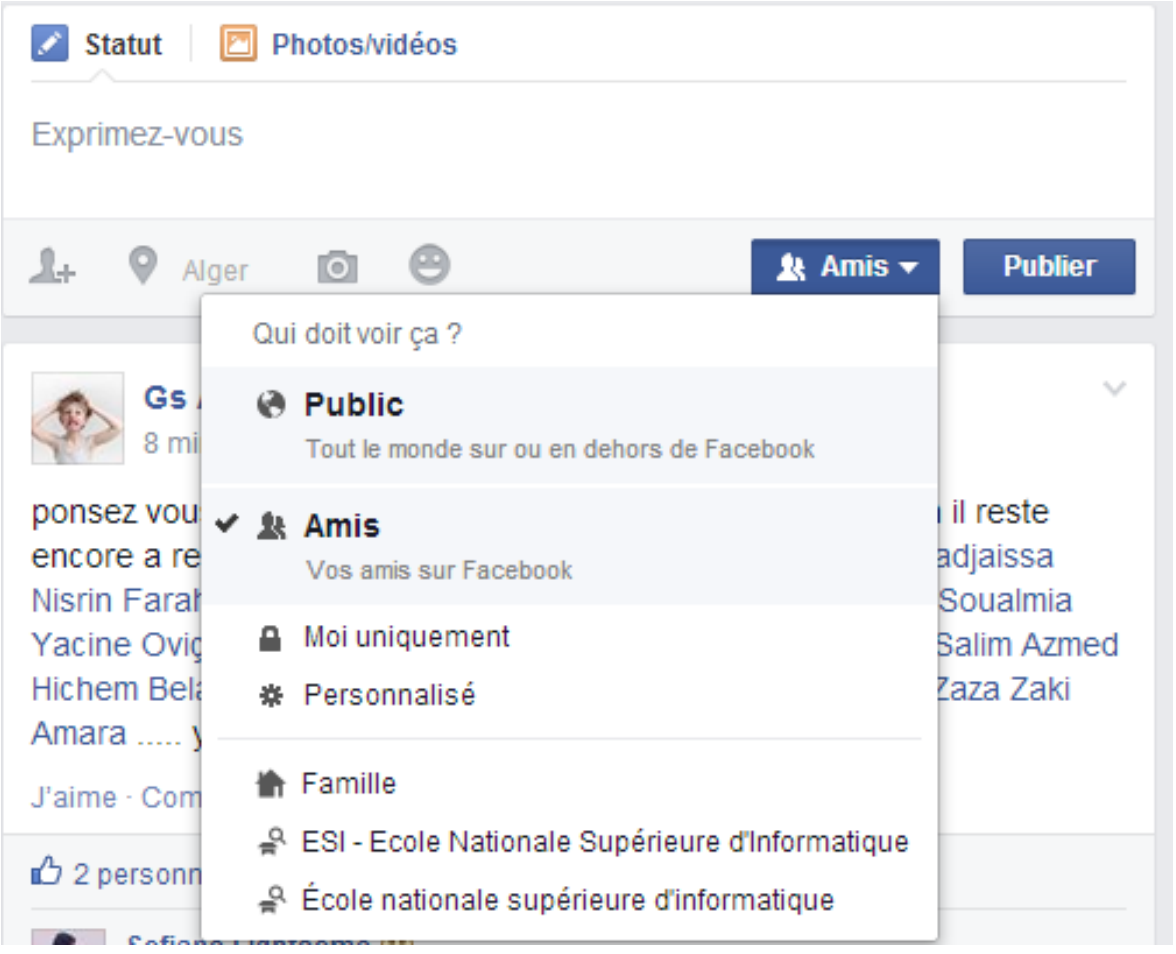

**Figure 17:** Les contrôles d'accès joints à chaque publication

Cependant, les autres RS tels que Linkedin permettent à l'utilisateur de modifier ses paramètres à partir de la page des configurations associée à son profil. La configuration se fait d'une manière à généraliser le paramètre de confidentialité pour toutes les publications du même type (**Figure 18**), ce qui rend la navigation à travers ses paramètres de confidentialité très difficile. Souvent les utilisateurs oublient de mettre à jour leurs paramètres de confidentialité ou bien ils ne trouvent pas l'endroit où ils peuvent les modifier.

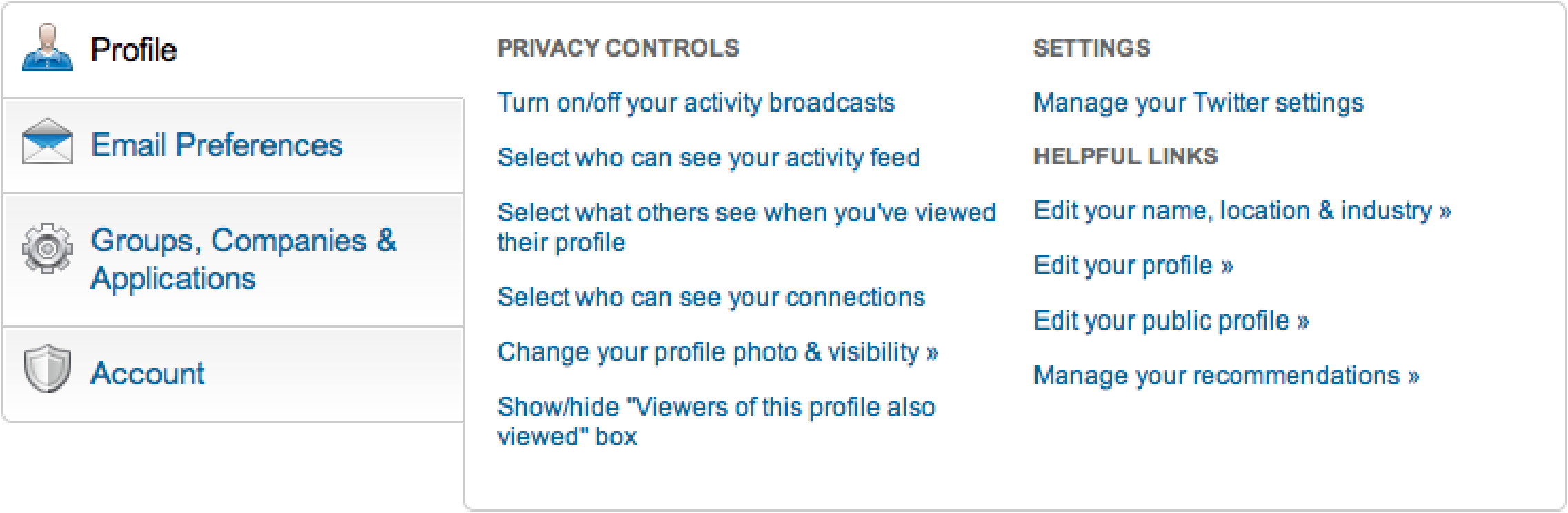

**Figure 18:** Page "paramètres du compte" sur Linkedin

2) **Les contrôles d'accès pour les applications et les sites web**

L'utilisateur peut contrôler les informations accessibles aux applications ou aux sites Web (y compris les moteurs de recherche) (Figure 19). L'utilisateur peut choisir de permettre de publier ou non sur son mur par les applications qu'il utilise, de supprimer les applications ou les bloquer, il peut même désactiver complètement la plateforme des applications de Facebook.

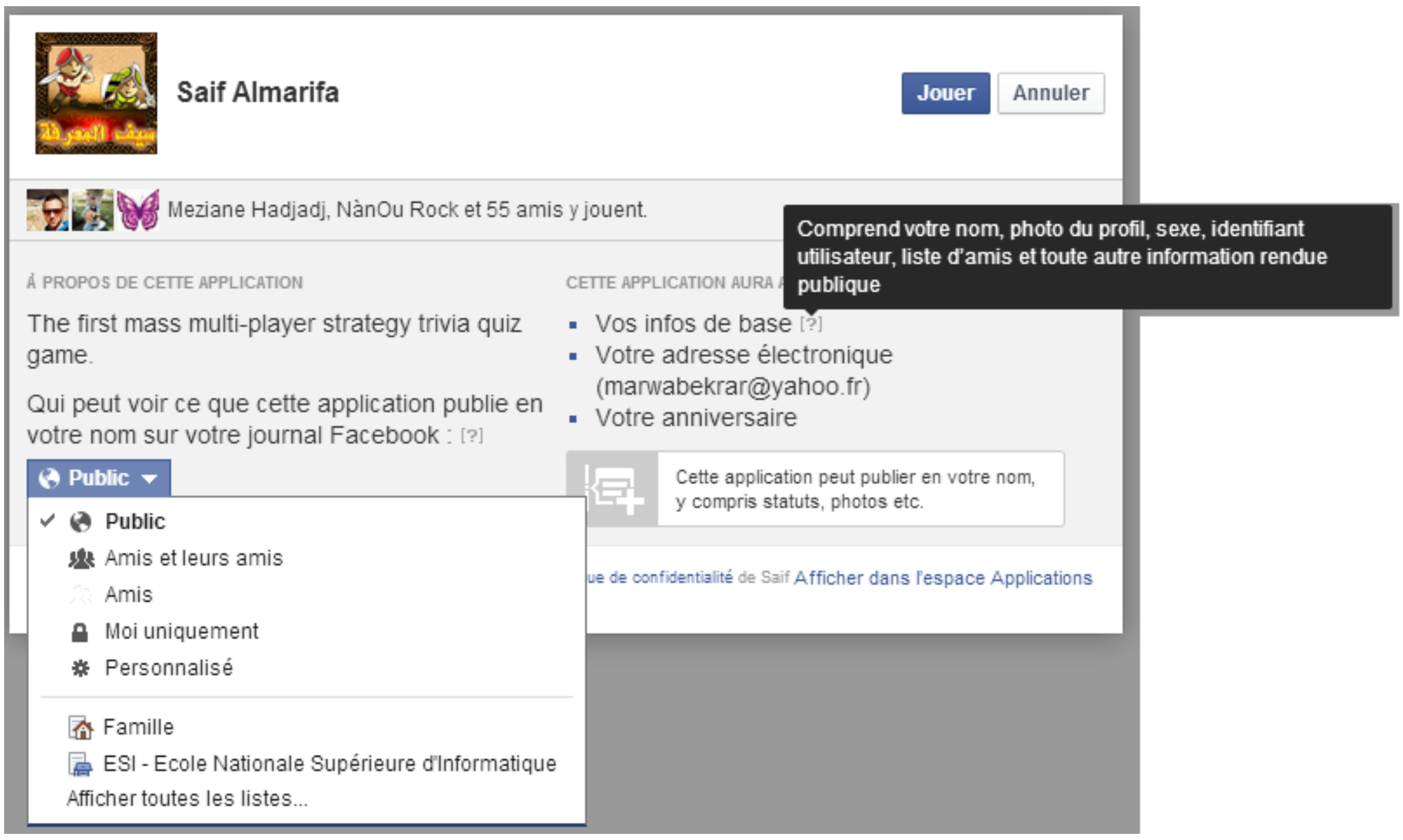

**Figure 19:** Paramétrage des accès avant l'installation des applications sur le profil

Facebook offre à ses utilisateurs des solutions qui permettent de personnaliser le contrôle d'accès par groupes d'utilisateurs et par types d'informations. Ainsi, Facebook offre un accès rapide et une utilisation facile des paramètres d'accès (contrairement à Linkedin par exemple). Facebook offre aussi à ses utilisateurs la possibilité de personnaliser la recherche de leurs profils sur le web ou sur le réseau en précisant à Facebook qui peut rechercher leurs profils. De plus, Facebook offre même la possibilité de réviser les paramètres de confidentialité choisis par une option appelée « La loupe de vie privée» (**Figure 20**).

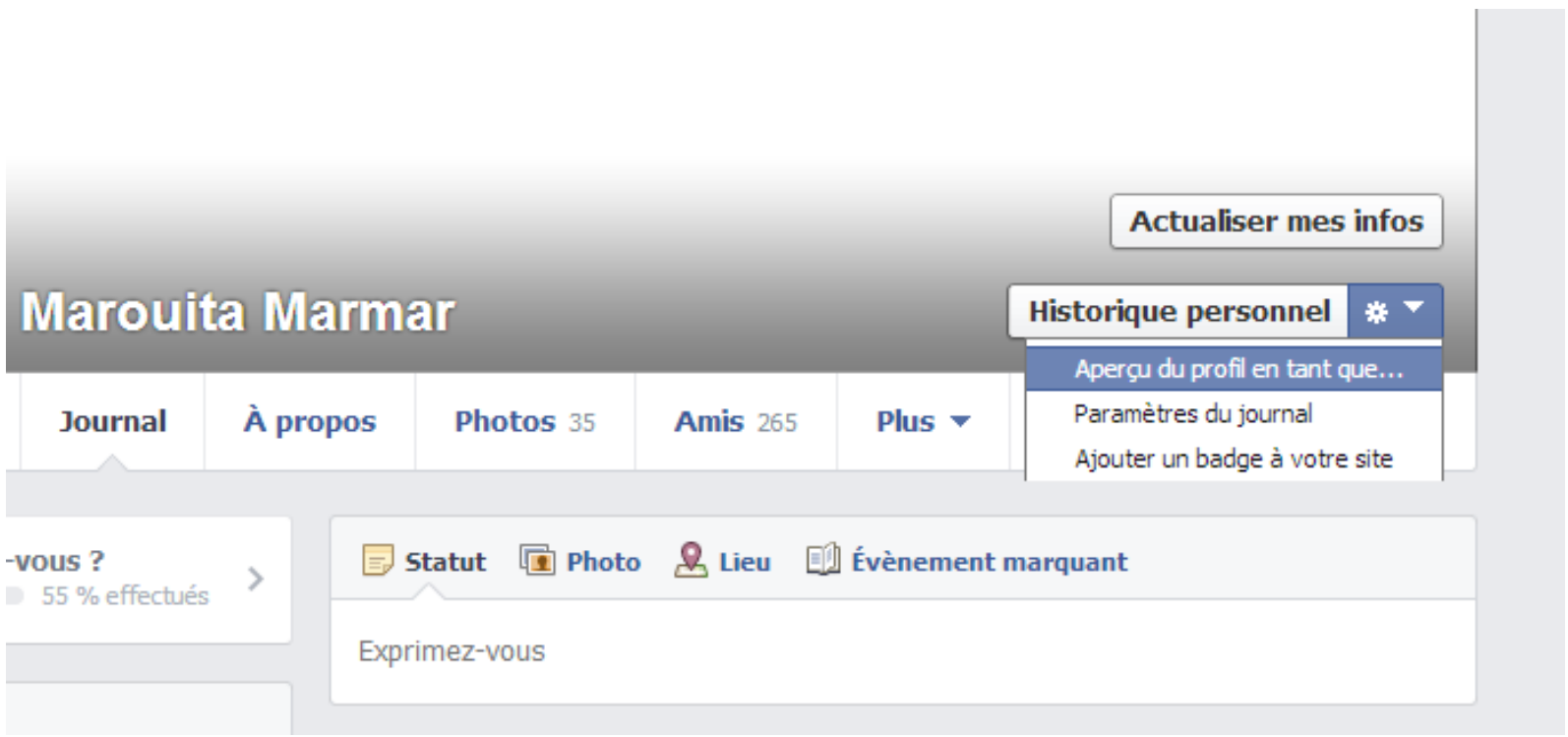

**Figure 20: La loupe de vie privée, Facebook**

En réalité, Facebook ne donne à ses utilisateurs que la possibilité de contrôler la visibilité de leurs informations. Cependant, les utilisateurs ne sont pas les propriétaires de leurs données publiées sur Facebook. En effet, Facebook exige à ses utilisateurs, dans ses condition d'utilisation, de donner une licence " non - exclusive, transférable, sous-licenciable, libre de droits dans le monde entier» sur tous les contenus qu'ils publient sur le réseau social. Ce transfert de propriété viole le principe de la souveraineté des données. De plus, Facebook protège la confidentialité des données pour ses propres intérêts car s'il est vulnérable aux attaques externes qui visent à récupérer massivement les données privées, il perd beaucoup de profits qu'il génère à partir de la vente des profils. Ainsi, Facebook encourage activement ses utilisateurs à partager leurs données privées afin d'enrichir les profils qu'il possède et bénéficier des techniques de profilage utilisées pour la publicité. Malheureusement, par manque de conscience, les utilisateurs remplissent toutes les informations qu'il demande (pourtant elles sont facultatives) en ignorant qu'ils ne peuvent pas exercer leur droit à l'oubli plus tard. En fait, Facebook cache seulement les données et en aucun cas il les supprime. Cela est déclaré dans sa politique de confidentialité (conformément au principe de la transparence) mais malheureusement les politiques de confidentialité sont rarement lues par les utilisateurs. La problématique de la vie privée est essentiellement un problème comportemental car si l'utilisateur est conscient aux risques, il sera moins exposé aux violations de sa vie privée.

## 5.2 Les solutions décentralisées

### 5.2.1 Les solutions hybrides

Ceux sont des solutions basées sur l'observation du fonctionnement de certains RS sur de "fausses" données. La plupart des solutions sont implémentées comme des plugins Firefox. Ce type de solutions était initialement introduit par **NOYB** (qui signifie "None Of Your Business" ou "Ce n'est pas vos Affaires") **[Guha, 2008]**. En **NOYB**, les informations personnelles de l'utilisateur sont d'abord cryptées, ensuite le texte chiffré est remplacé par un pseudo sélectionné d'une manière aléatoire à partir d'un dictionnaire public afin de le faire ressembler à des données légitimes. Le service en ligne peut fonctionner sur les données cryptées, mais seuls les utilisateurs autorisés peuvent décoder et décrypter le résultat. En effet, dans NOYB, les informations sont remplacées par des termes du dictionnaire car le cryptage des données est contre les conditions d'utilisation de Facebook et si Facebook détecte ce type d'activité il suspendera directement le compte. Une version « proof-of-concept » de NOYB a été implémentée sous la forme d'un plugin Firefox. Malheureusement, cette solution peut seulement crypter des petits textes telles que les informations personnelles du profil et ne permet pas de crypter des textes longs. Le canal d'échange de clé, nécessaire pour envoyer les clés de décryptage correspondantes à ses amis, est indépendant du RS et est implémenté sous la forme de courrier électronique, par l'utilisation d'une partie tierce ou d'un réseau pair à pair (peer-to-peer).

**FaceCloak [Luo, 2009]** suit les étapes de NOYB**.** FaceCloak stocke de fausses données du profil sur Facebook et les données cryptées correspondantes sont stockées sur un serveur tiers. Le problème de cette solution est lorsque ce serveur tombe en panne, il est impossible de récupérer les données réelles. FaceCloak est aussi une extension Firefox. (Figures 21,22 ;23)

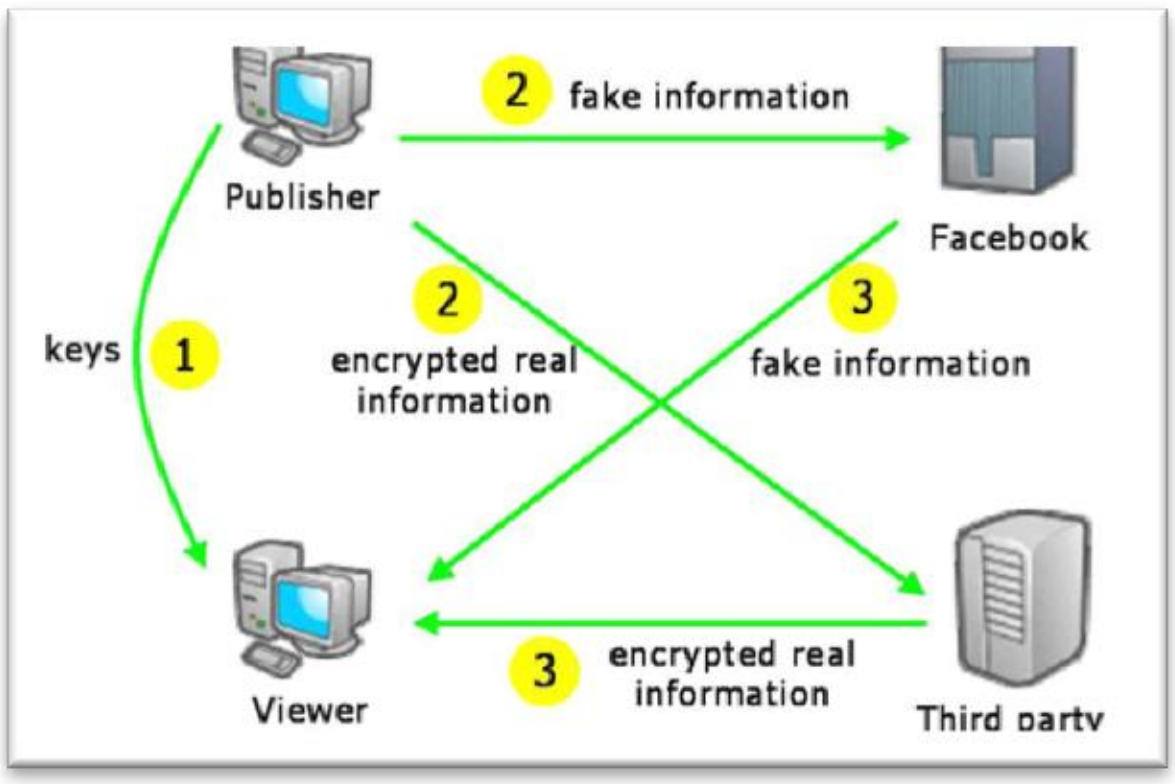

**Figure 21:** Architecture de FaceCloak **[Luo, 2009]**

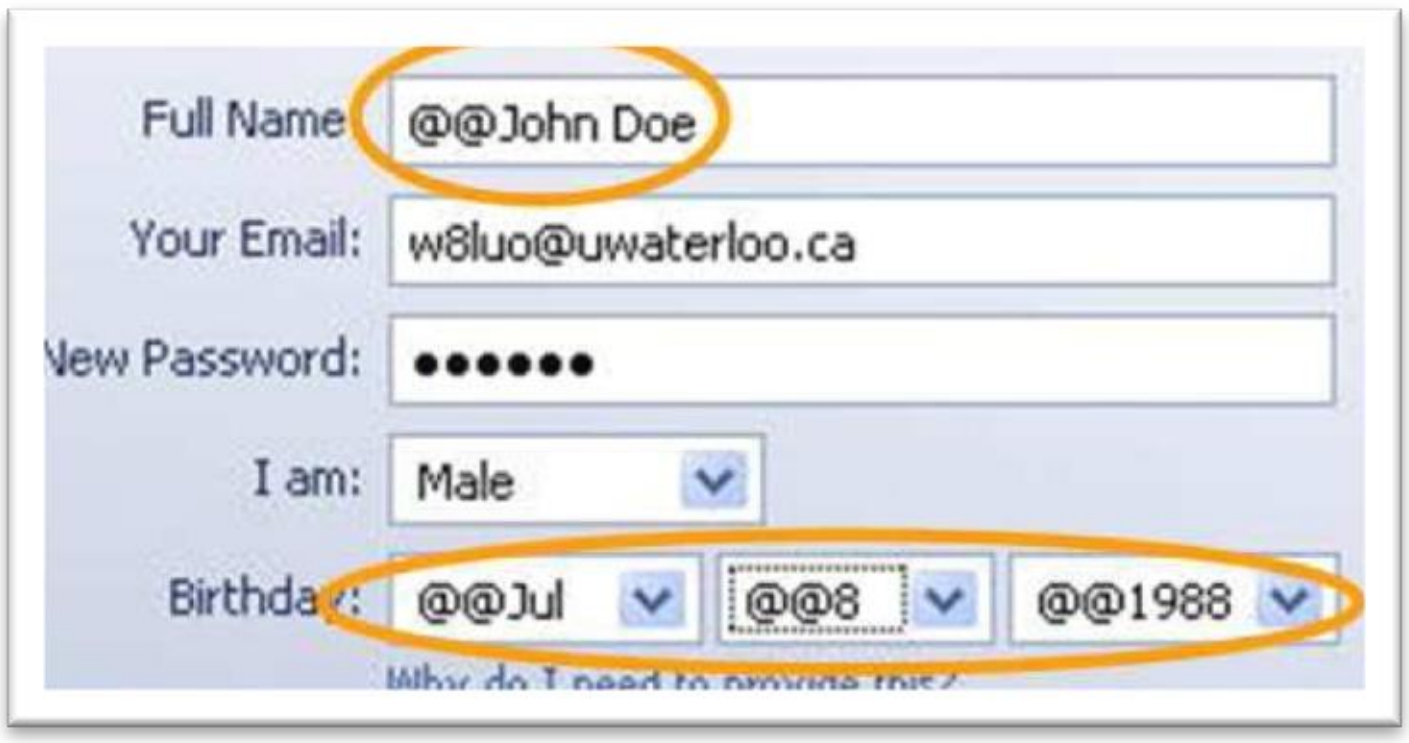

**Figure 22:** Les vraies informations avant de cliquer sur « submit » **[Luo, 2009]**

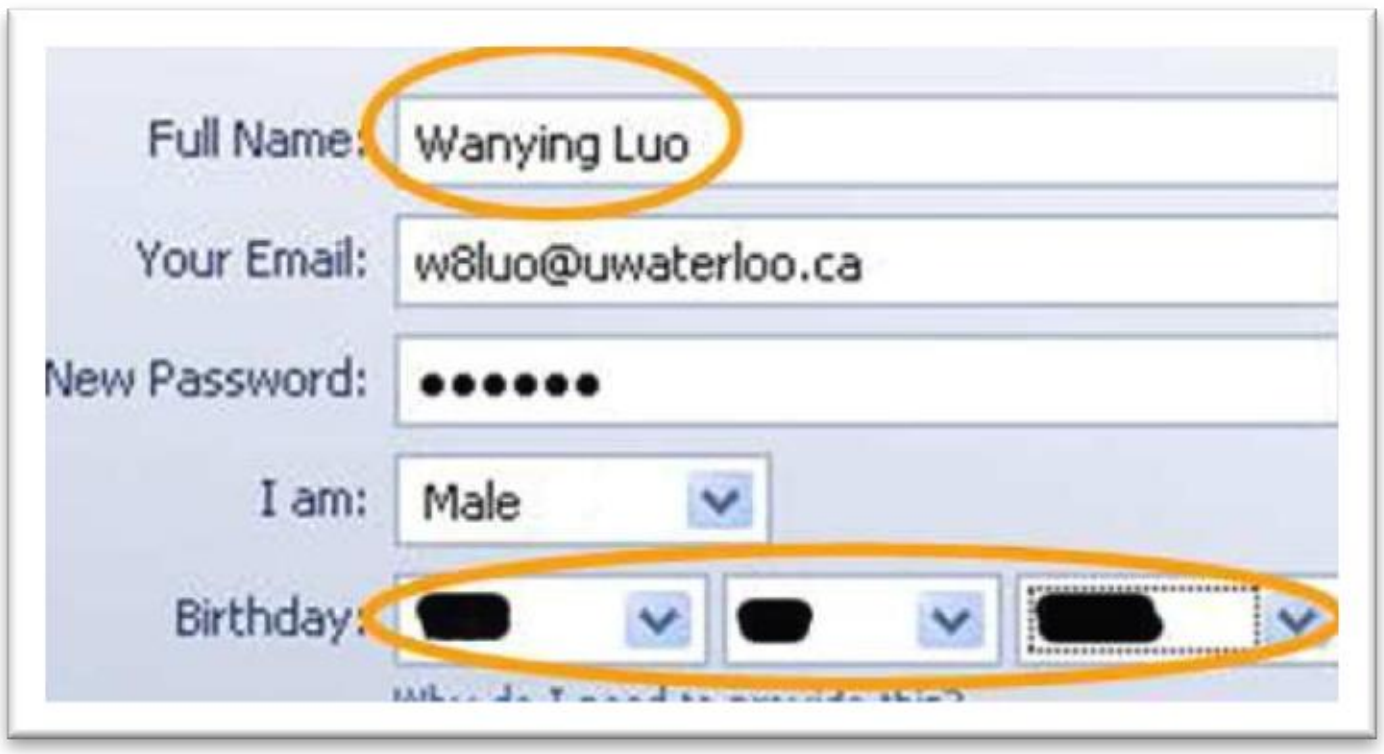

**Figure 23:** Les fausses informations après la confirmation [Luo, 2009]

**Locker [Tootoonchian, 2009]:** afin de dépasser la limite de la solution de FaceCloak**,** Locker Cache (crypte), mappe les données (par indexation) et les stocke dans une partie de stockage tierce tels que Picasa[14] ou Flickr[15].

**Renforcement des contrôles d'accès [Beato, 2009]** : Une autre approche complémentaire à NOYB. Dans cette solution, le plugin (extention Firefox) gère localement (coté client) les vrais accès aux informations. Le profil est divisé en deux types de classes, des classes de contenus et des classes de connexion. Le mappage entre les classes de contenus et les classes de connexions définit les droits d'accès (Figure 24). Chaque classe de connexion autorisée à voir une classe de contenu possède une clé de décryptage pour visualiser le contenu. Le point fort de cette solution est que le fournisseur du RS n'a aucune connaissance sur "qui a accès à quoi?". Malheureusement, les utilisateurs expérimentés et qui comprennent les mécanismes de cryptographie peuvent trouver une telle solution intéressante mais pour les utilisateurs novices, il est difficile de comprendre le principe de fonctionnement de la solution.

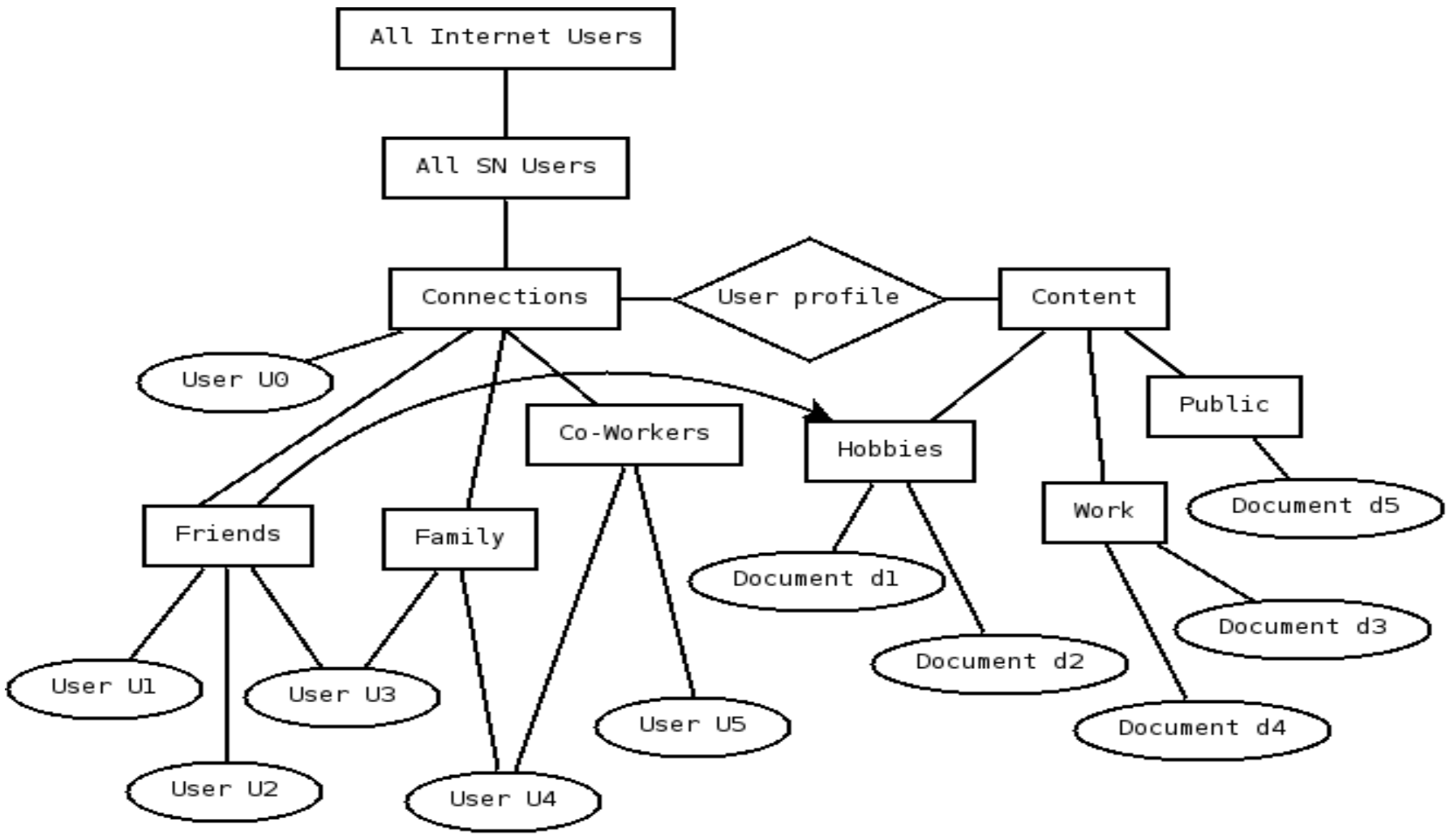

**Figure 24:** Mappage entre les classes **[Beato, 2009]**

[14] http://picasa.google.com/
[15] https://www.flickr.com/

Les plugins ont pour objectif d’assurer que les données des utilisateurs n’appartiennent qu'à eux et pas au fournisseur du RS. Ces solutions ne peuvent crypter que des données texte et ne peuvent pas encore être appliquées à d'autres médias tels que les photos. Ces plugins ne sont fonctionnels que sur des réseaux sociaux déjà existant et ils dépendent essentiellement de leur architecture.

### 5.2.2 Les solutions purement décentralisées

**Safebook [Cutillo, 2009]** : Safebook est un réseau social spécialement conçu pour protéger les utilisateurs contre les violations de la vie privée par les utilisateurs malveillants et les fournisseurs des RS. Safebook est caractérisé par une architecture décentralisée (peer-to-peer) qui se base sur la coopération entre les « peers » afin de protéger la vie privée des utilisateurs et d’éviter « the all-knowing » du fournisseur du RS. La sécurité des données et des profils des membres du Safebook est basée sur la confiance entre les membres dans la vie réelle car les systèmes P2P souffre du manque de confiance entre les pairs. En effet, le stockage et le routage des données sont réalisés par les pairs qui ont confiance l’un à l’autre dans le réseau social. Les nœuds dans Safebook forment deux types de réseaux de recouvrement :

- Un ensemble de Matryoshkas , qui sont des anneaux concentriques de nœuds construits autour du nœud de chaque utilisateur pour la fourniture d’un stockage de données de confiance, la récupération des données et la confidentialité des communications.
- Un substrat P2P qui fournit les services de recherche (par exemple par les tables de hashage distribuées -Distributed Hash Tables- (DHT)). En effet, Safebook est divisé en 3 couches :
    - ✓ La couche de réseau social centrée sur l'utilisateur implémente le niveau RS (ajout des amis, blogging…etc).
    - ✓ Le substrat de P2P implémente les services RS (Stockage, routage…etc.).
    - ✓ Internet agit comme une couche de communication.

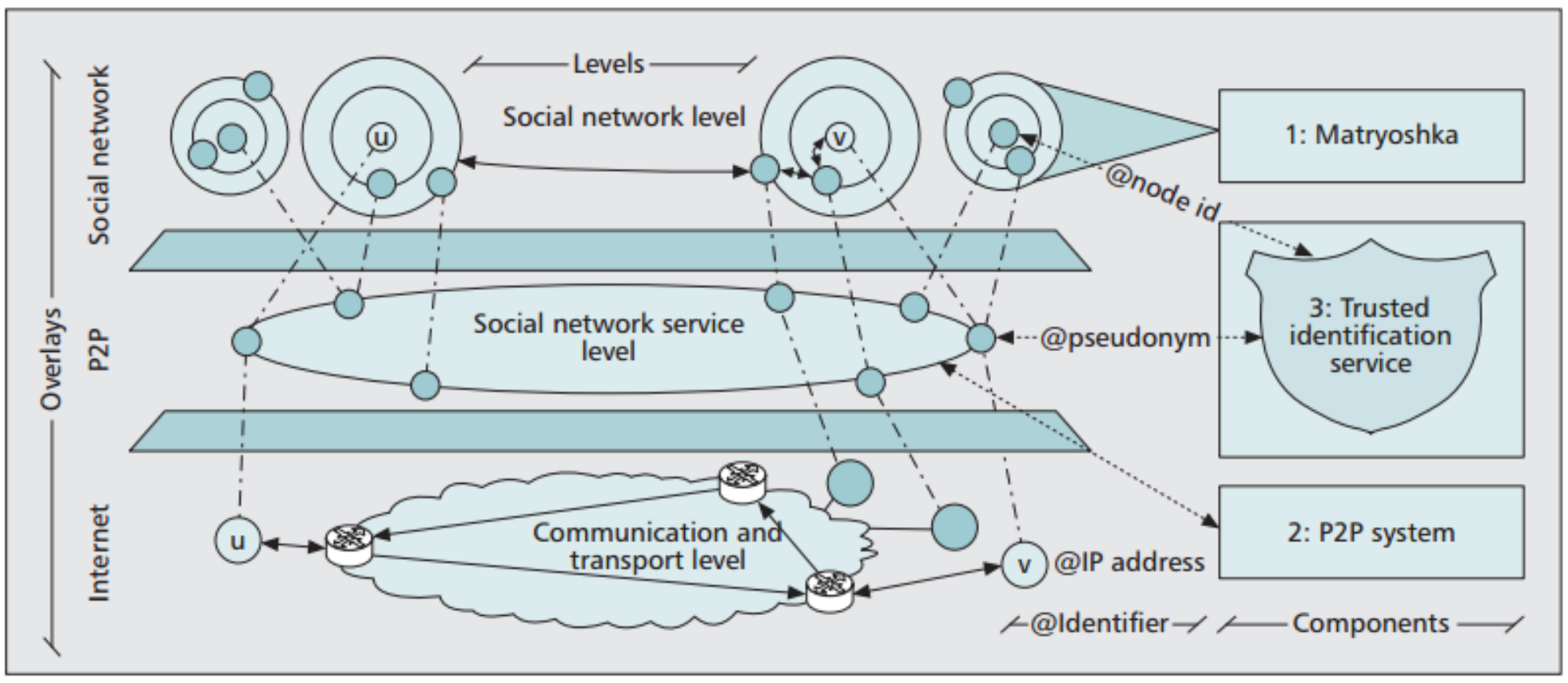

**Figure 25: les principaux composants de Safebook** [Cutillo, 2009]

Chaque partie en Safebook est représentée par un nœud qui est considéré comme un nœud d'hôte en Internet, un nœud de pair dans l'architecture P2P et un nœud membre dans la couche RS.

**eXO [Loupasakis, 2011]** est un système distribué offrant les services fondamentaux des réseaux sociaux tels que les mécanismes pour l'indexation de contenu et les métadonnées associées et les algorithmes efficaces pour les requêtes de recherche-pour-utilisateurs et recherche-pour-contenu. En eXO, chaque utilisateur se connecte à un nœud spécifique avec un identifiant de réseau unique. Le contenu partagé par un utilisateur peut être reproduit sur d'autres nœuds pour augmenter la disponibilité. Le système d'eXO est conçu de sorte que ses utilisateurs aient le contrôle total sur leur contenu et ressources. Les utilisateurs peuvent désigner leur contenu en tant que public ou privé pour définir lequel des utilisateurs peut accéder. Seuls les contenus publics sont indexés et peuvent être recherchés et consultés. En plus, les propriétaires peuvent rejeter les demandes d'accès faites par d'autres utilisateurs, même pour le contenu public.

### 5.2.3 Les solutions partiellement décentralisées

**SuperNova [Sharma, 2012]** est une architecture basée sur les super- pairs qui essaye de résoudre le problème de la disponibilité des données en réseau P2P en incitant les autres nœuds à stocker les données personnelles d'autres utilisateurs. Tout utilisateur du réseau P2P peut se porter volontaire pour devenir un super- pair.

**PeerSoN [Buchegger, 2009]** PeerSoN assure le contrôle d'accès par le cryptage. PeerSoN a une architecture composée d’un service consultation et un pair contenant les données des utilisateurs, tels que les profils d’utilisateurs. Le service de consultation stocke les métadonnées nécessaires pour trouver des utilisateurs particuliers, ainsi que les données particulières qu'ils stockent. Par exemple, cette métadonnée pourrait être leur adresse IP, informations sur les fichiers qu'ils ont stockés et les notifications pour les utilisateurs. Un pair qui veut communiquer avec un autre pair appelle le service de consultation pour récupérer les informations nécessaires pour localiser les pairs pertinents, puis se connecter directement à eux. Par conséquent, un utilisateur ne peut visiter le profil de ses amis que quand ils sont en ligne, offrant ainsi un niveau faible de disponibilité par rapport aux RS centralisés.

## II- Problématique de la vie privée dans les systèmes de e-learning

L'un des principaux avantages d'e-Learning est son adaptabilité aux besoins et aux préférences de l'apprenant. Pour ce faire, les systèmes E-Learning doivent collecter de grandes quantités d'informations sur l'apprenant. Ces informations peuvent être utilisées d'une manière inappropriée afin de générer des profits commerciaux, à des fins autres que la personnalisation, ou pour être partagées avec d'autres systèmes ou organisations d'e-Learning sans le consentement explicite de l'apprenant, ce qui viole le droit de l'utilisateur au respect de sa vie privée. Dans les systèmes d'e-Learning, les apprenants fournissent une grande quantité d'informations personnelles et font une confiance aveugle aux fournisseurs de ces systèmes. Les mécanismes de sécurité des systèmes E-Learning offrent certaines forme de protection de la vie privée par l'utilisation des paramètres de confidentialité qui permettent de contrôler la visibilité des données ou bien par la protection de la confidentialité des données contre les attaques externes; néanmoins elle ne protège pas les utilisateurs contre la vente et la divulgation de leurs informations par les fournisseurs des systèmes d'e-Learning qui possèdent tous les privilèges d'accès. L'utilisateur perd son droit à déterminer quels types d'informations peuvent être accessibles une fois que ses données sont divulguées à d'autres tiers. En effet, ni le fournisseur ni l'utilisateur ne peuvent contrôler la circulation des informations et la façon dont elles seront utilisées une fois divulguées.

Le respect de la vie privée est un droit fondamental. Cependant, la problématique de la vie privée est très peu abordée dans les systèmes d'e-Learning **[Hage, 2011][Yong, 11][May, 2011]**. Il n'existe pratiquement que la solution de Hage appelée « Privacy-Preserving E-learning » **[Hage, 2011]** qui représente une alternative des systèmes e-Learning standards.

## Le système "Privacy-Preserving E-learning" [Hage, 2011]

Pour satisfaire les divers besoins de la vie privée, Hage a défini quatre niveaux de vie privée et quatre niveaux de traçabilité. La solution permet aux apprenants de recevoir des relevés de notes et des diplômes anonymes. Les apprenants peuvent prouver qu'ils sont les propriétaires de ces derniers à des tiers (employeurs, d'autres systèmes d'e-Learning, etc.) sans divulguer leur identité réelle. En effet, pour que l'apprenant puisse prouver qu'il est le propriétaire du relevé ou du diplôme anonyme, Hage introduit le concept des « Blind Digital Certificates », un certificat numérique qui ne révèle pas l'identité de l'apprenant. Il propose d'utiliser des pseudonymes certifiés par des « CA ». La solution proposée est appelée « Anonymous Credentials for E-Learning Systems (ACES) », un ensemble de protocoles qui préserve la vie privée des apprenants. De plus, afin d'éviter l'utilisation abusive de l'anonymat, ACES empêche les tentatives de partager des relevés entre les apprenants.

### 1. La vie privée de l'apprenant

Selon Hage, l'apprenant peut avoir besoin de garder des parties de son profil privées pour deux types de raison: compétitives et personnels. Un exemple d'une raison compétitive : un personnage public connu (politicien par exemple) préfère garder son identité cachée pour se protéger des préjugés et de la connaissance de son public ou de ses adversaires. D'autre part, l'apprenant peut avoir besoin de sa vie privée pour des considérations personnelles. Par exemple, il peut avoir besoin de se protéger d'un tuteur biaisé ou d'éviter la pression et le stress dû à l'anxiété de performance pour ne pas être jugé par ses pairs selon ses notes. De plus, dans une recherche récente, les apprenants ont montré une préférence nette à la vie privée dans les systèmes d'apprentissage. **[Hage, 2011]**

## 2. L'effet de la vie privée sur l'apprentissage

Les recherches existantes ont démontré que les émotions ont un effet sur l'apprentissage **[Zins, 2007]** : contrairement aux émotions négatives, les émotions positives améliorent la performance de l'apprenant. La motivation de ces études est d'éviter les situations qui créent des émotions négatives, tout en motivant l'apparition de situations qui créent des émotions positives. Les tests de Hage avaient pour objectif de déterminer si la protection de la vie privée dans les systèmes d'e-learning a un effet positif ou négatif sur l'apprentissage. Pour ce faire, deux hypothèses ont été testées :

**Hypothèse 1:** Les apprenants se sentent timides et gênés car les autres peuvent voir leurs résultats, ce qui crée chez eux le stress. Les apprenants se stressent parce qu'ils veulent obtenir de meilleurs résultats de sorte qu'ils ne soient pas jugés par leur tuteur et leurs pairs comme des élèves faibles. « Privacy preserving e-Learning » permet d'éviter cette pression qui crée des émotions négatives et aide les apprenants à obtenir de meilleurs résultats.

**Hypothèse 2:** Dans cette hypothèse, il considère que la vie privée a un facteur négatif: Comme les résultats des tests vont rester privés, les apprenants auront tendance à devenir plus négligents et ils ne vont pas se soucier de leurs notes ils vont faire moins d'efforts pour juste passer.

Les deux hypothèses ont été testées dans deux environnements différents, le premier « sans protection de la vie privée » et le deuxième « avec protection de la vie privée ». Les apprenants ont passé deux tests de même niveau de difficulté dans les deux environnements. L'enregistrement des émotions était pris avant et après chaque test. Les résultats ont montré que les apprenants ont eu de meilleurs résultats dans l'environnement « avec protection de la vie privée » que dans l'environnement « sans protection de la vie privée ».

Les tests ont démontré que malgré l'apprenant est anonyme dans le système « Privacy preserving e-Learning » et les résultats des tests sont privés, l'obtention de bonnes notes reste un facteur important pour l'apprenant. Selon Hage, la protection la vie privée a un effet positif sur les apprenants. Ainsi les résultats d'un sondage effectué après le test des hypothèses ont montré que 97% des participants aux tests ont voté pour le respect de leur vie privée dans les systèmes d'e-Learning.

## 3. Le framework "privacy Preserving E-learning" [Hage, 2011]

Les apprenants ont des besoins et des exigences différents concernant leur vie privée. Les informations que l'apprenant veut protéger peuvent appartenir à ces cinq types :

### 3.1 Les types d'informations

- **L'identité (id):** représente l'ensemble d'informations qui permet de déterminer physiquement la vraie identité de l'apprenant. Cela comprend des données telles que son nom, son adresse et son numéro de carte étudiant.

- **Le profil démographique (dp: demographic profile) :** l'ensemble des caractéristiques démographiques de l'apprenant, comme l'âge, le sexe, la race, l'origine ethnique, la langue, etc.
- **Le profil de formation (lp : learning profile):** représente les informations liées à la formation tels que les qualifications, le style d'apprentissage, les intérêts et les objectifs de l'apprenant.
- **L'historique du cours (ch : course history):** la liste des cours que l'apprenant a suivis dans le passé, et leurs informations respectives telles que les activités de l'apprenant pendant le cours et sa note finale.
- **Les cours actuels (cc : current courses) :** liste des cours dans lesquels l'apprenant est actuellement inscrit et ceux qu'il fréquente, ainsi que les activités de l'apprenant dans chaque cours. Par exemple, une activité pourrait être la publication d'un message dans le forum, en utilisant l'un des objets d'apprentissage du cours, ou même faire un quiz ou un test.

### 3.2 Les niveaux de la vie privée

Les niveaux de la vie privée se basent sur les catégories des informations citées précédemment. Les niveaux sont comme suit :

1. **No Privacy:** l'apprenant ne souhaite pas, ou ne se soucie pas de garder ses informations privées.
2. **Soft Privacy:** L'apprenant veut garder son identité et son profil démographique privé, mais il permet l'accès de son tuteur à son profil de formation, d'historique des cours et des cours actuels.
3. **Hard Privacy:** L'apprenant veut garder son identité, son profil démographique, l'historique des cours et son profil de formation privés, mais il permet l'accès s à ses cours actuels.
4. **Full Privacy:** L'apprenant veut garder toutes les composantes de ses données personnelles secrètes.

### 3.3 Les niveaux de « tracking »

Un autre aspect à considérer, qui est indépendant des informations personnelles énumérées ci-dessus, est le tracking des apprenants dans un cours. Les niveaux de tracking que les différents apprenants pourraient accepter sont :

1. **Strong Tracking:** le système peut relier les activités réalisées dans tous les cours à l'apprenant, même si l'apprenant est anonyme. Dans ce cas, le système peut suivre l'apprenant « *u* » et tracer ses accès aux cours c1, c2 ... cn.

2. **Average Tracking:** le système peut faire le lien entre les activités dans un cours à l'apprenant « *u* », mais il ne peut pas les relier à d'autres activités dans d'autres cours. Par exemple, Alice crée deux pseudonymes, A1 et A2, un pour chaque cours, comme le système ne peut pas savoir qu'il y a une relation entre A1 et A2, ni que les deux pseudonymes appartiennent à Alice.

3. **Weak Tracking:** dans ce cas, malgré le système peut reconnaitre l'apprenant « *u* » comme visiteur régulier, il ne peut pas le lier à un cours ni tracer ses activités.

4. **No Tracking:** dans ce cas, le système n'est même pas capable de connaitre l'apprenant « *u* » comme un visiteur récurrent du système. (Similaire à l'utilisation d'un compte invité pour accéder à une démo du système E-Learning)

### 3.4 Les considérations à prendre en compte

Dans l'environnement « Privacy-Preserving E-Learning », l'apprenant doit fournir ses certificats durant le processus d'apprentissage, tout en maintenant sa vie privée. Par conséquent, l'apprenant doit être capable de présenter les preuves nécessaires pour pouvoir continuer sa formation et avoir son diplôme sans révéler ses informations personnelles. De plus, le système doit prendre en considération le fait que les apprenants peuvent profiter et abuser leur état d'anonymat et partager leurs certificats. Ainsi, ces certificats anonymes doivent être valables seulement à leur propriétaire légitime, et difficile, voire impossible à partager. La solution de Hage se limite au cas du « hard privacy » avec « average tracking ».

## 3.5 Anonymous Credentials for E-Learning Systems (ACES)

Le système ACES consiste en un ensemble de protocoles pour permettre à l'apprenant d'obtenir/ présenter des certificats anonymes de/ au système d'e-Learning. Pour les algorithmes suivants, ES signifie « E-learning System » dans lequel l'apprenant est inscrit et tous les calculs sont effectués mod(N).

- RSA blind signature (*u* : le message de l'apprenant)
- *N*, *e*, *d*: les paramètres RSA du ES
- Les sorties: *s*, la signature sur *u*.

**1.** l'apprenant choisit une valeur aléatoire *r non nulle*, tel que GCD(*r*, *N*) = 1 (premier entre eux). Il calcule $t = u{\cdot}r^e$, et envoie *t* à l'ES.

**2.** l'ES calcule $t' = t^d = (u \cdot r^e)^d = u^d \cdot r$, puis, il envoie *t'* à l'apprenant.

**3.** L'apprenant enlève le facteur aveuglant, *r*, en multipliant *t'* par $r^{-1}$, l'inverse de *r*. par conséquent, il obtient la signature valide du RSA sur *u*:

$s = S_d(u) = t' \cdot r^{-1} = u^d$, de telle façon, n'importe qui peut vérifier la validité *s* en utilisant la clé publique *e* du ES.

**Algorithme 1: Algorithme d'obtention des signatures aveugles [Hage, 2011]**

### 3.5.1 L'obtention d'une « Blind Signature » sur un Pseudonyme

Soit un apprenant *L* qui veut s'inscrire dans un nouveau ES. A cet effet, il a besoin d'avoir des certificats de l'ES, *E0*, où il a effectué sa formation dans le passé. Avant qu'il soit inscrit dans le nouveau ES, *E1*, il choisit un pseudonyme u, pour être utilisé dans E1. Mais, tout certificat obtenu de *E0*, et qui sera présenté à E1, doit être liée à *u*, sans dévoiler *u* à *E0* ni *L* à *E1*. Donc, *E0* signe aveuglément le pseudonyme *u* avec les certificats appropriés, en utilisant la signature aveugle RSA.

### 3.5.2 L’obtention des certificats anonymes

Les certificats anonymes sont définis aux niveaux suivants : L’apprenant, le cours, le relevé de notes et le diplôme. L’ES génère les certificats anonymes et EE signifie l’entité externe.

#### 3.5.2.1 « Anonymous Learner Credential » (ALC)

Les certificats anonymes de l’apprenant sont utilisés pour prouver à une entité externe (EE) que l'apprenant a été, ou est encore inscrit dans un ES donné, même si son identité est inconnue. Ceci est similaire à l'obtention d'un certificat d'inscription d’une université. En d'autres termes, l'ES fournit une lettre disant que l'apprenant « L » est, ou a été, un étudiant dans son institution.

Le certificat anonyme de l’apprenant se compose de deux types de données: le message, m, décrivant le certificat d'inscription, et un pseudonyme que l'apprenant a l'intention d'utiliser dans le EE, ce qui pourrait être un autre ES ou une autre organisation. Le certificat anonyme de l’apprenant est généré comme suit (le symbole || désigne la concaténation).

1. L’apprenant, identifié par *L* dans l’ES, crée un pseudonyme, *u*, dans l’EE.

**2.** L’apprenant L, demande de l’ES de signer *m*, qui contient l’attestation d’inscription.

**3.** l’ES signe *m* avec sa clé secrète telque: $S_{SK}(m)$, et envoie de résultat à l’apprenant.

**4.** l’apprenant L forme $M = u||m||S_{SK}(m)$, et demande à l’ES d’appliquer une signature aveugle sur *M*.

**5.** L'ES signe aveuglément *M*, il obtient *s* = RSA_blind_signature(*M*), et puis il envoie *s*, qui est effectivement le certificat anonyme de l’apprenant L.

**Algorithme 2: L'obtention du certificat anonyme par l’apprenant [Hage, 2011]**

Il est important d'utiliser un canal de communication sécurisé entre les ES et l'apprenant de sorte à empêcher un intrus, d'intercepter la $S_{SK}$ (M) à l'étape 3 de ce protocole, et d’effectuer les étapes 4 et 5 à la place de l'apprenant réel. Le reste des certificats anonymes, tel que le certificat anonyme du cours (pour prouver que l'apprenant a terminé avec succès un cours donné), les relevés de notes anonymes et les diplômes anonymes s’obtiennent tous de la même manière expliquée ci-dessus pour l’obtention du certificat d’inscription.

#### 3.5.2.2 Prévention contre partage des certificats anonymes entre les apprenants

Chaque diplôme anonyme est identifié par un pseudonyme u secrètement choisi par l'apprenant. Après avoir validé les diplômes anonymes, l'EE insère le pseudonyme dans une liste appelée « Revocation of Anonymous Credentials List » (RACL). Elle contient tous les certificats anonymes qui ont été présentés à différents EE, de sorte q’aucun certificat anonyme ne peut être utilisé deux fois. En effet, la validation d'un diplôme anonyme donné comprend deux étapes:

- ✓ La vérification de la signature de ES.
- ✓ La recherche dans RACL d'un double de u.

Cette approche empêche deux apprenants de partager le même certificat délivré par l'ES.

#### 3.5.2.3 Présentation du certificat anonyme à un tiers

Après avoir obtenu un diplôme anonyme depuis l'ES, l'apprenant le donne à une EE, qui vérifie si le diplôme anonyme est valide en utilisant la clé publique de l'ES pour vérifier qu'il a été correctement signé par l'ES en question. Si le certificat anonyme est valide, l'EE vérifie qu'il ne fait pas partie de la RACL.

#### 3.5.2.4 Légitimité des certificats anonymes

Pour prouver que l'apprenant est le propriétaire légitime du certificat, il a besoin de créer un certificat numérique aveugle (BDC : Blind Digital Certificate) pour chaque entité avec laquelle il travaille. Le BDC est un certificat numérique qui ne révèle pas les informations personnelles de l'apprenant. Soit P l'ensemble des informations personnelles de l'apprenant. L'apprenant génère une paire de clés publique / privée (PK, SK), puis il crypte P tel que $y = E_{PK}(P)$ et demande d'avoir un certificat numérique sur $z = [y, PK]$ de la part d'une CA. z est un certificat numérique aveugle, et *y* devient l'identité numérique que l'apprenant pourra utiliser dans l'ES.

Si l'apprenant a besoin de prouver à son manager, par exemple, qu'il a terminé la formation / le cours de l'ES avec succès, le certificat numérique aveugle peut être «dévoilé» à son manager, pour qu'il puisse vérifier l'authenticité de l'ES et déchiffrer *y* en P pour s'assurer que l'apprenant est le propriétaire du certificat issu de l'ES.

## Conclusion

Comme les réseaux sociaux ont beaucoup avantages dans le domaine de communication, ils ont également caché derrière ce grand succès, des objectifs latents mais très dangereux sur la vie privée de leurs utilisateurs.

Les systèmes centralisés tels que Twitter, Linkedin ou Facebook offrent plusieurs avantages en termes de qualité de service. Ces systèmes se caractérisent par une haute disponibilité de données et de services, ainsi que le développement et le déploiement de mécanismes de sécurité très avancés pour la protection contre les pirates. Les fournisseurs de ces réseaux en ligne donnent la possibilité à leurs utilisateurs de contrôler la visibilité de leurs données par des paramètres ajustables, ce qui offre une certaine protection de la vie privée. Cependant, cela ne protège pas l'utilisateur contre la violation réelle de sa vie privée par la vente de ses données. La vente des données représente aujourd'hui un grand business pour les fournisseurs de services en ligne. A cause de l'architecture centralisée qui permet aux fournisseurs d'avoir le control total sur les données et leurs termes d'utilisation qui ne respecte pas le principe de la souveraineté des données, les utilisateurs ont perdu le control sur la circulation de leurs données. Les plugins de protection de la vie privée ont été proposés afin de donner la possibilité aux utilisateurs de garder la souveraineté sur leurs données. Cependant, l'inconvénient principal de ces solutions est qu'elles nécessitent que l'utilisateur et ses amis aient une certaine expérience en cryptographie. En effet, la plupart des utilisateurs en ligne sont des utilisateurs novices et, en général, ils ne sont pas conscients de ces problèmes. De plus, si le fournisseur détecte des activités de cryptographie sur son système, cela cause la suspension directe du compte.

Les RS distribués sont apparus pour résoudre le problème de la souveraineté de données inhérente aux RS centralisées et pour donner aux utilisateurs plus de liberté en leur permettant de déployer leur propre réseau social. Ces systèmes ont participé à une tentative de vaincre le monopole des RS centralisés représentés par Facebook, mais jusqu'à présent, aucun des systèmes proposés n'a été adopté à très grande échelle. Ces solutions offrent l'avantage aux utilisateurs de garder le control sur leurs données et se protéger contre leur vente. Cependant, ces solutions héritent les inconvénients des architectures décentralisées, tel que le problème de disponibilité des données. De plus, avec les mécanismes de duplication utilisés dans le réseau (surtout pour les données publiques), il est assez difficile de suivre la façon dont ces données sont diffusées et à exécuter le droit à l'oubli quand l'utilisateur décide de quitter le réseau ou supprimer un contenu.

La problématique de la vie privée dans les systèmes e-learning est particulièrement différente et il semble que les chercheurs évitent cette problématique dans les systèmes d'e-Learning vu la difficulté qu'elle présente pour trouver un compromis entre la collecte d'informations sur l'apprenant pour l'adaptation du processus d'apprentissage aux besoins spécifiques de l'apprenant et son droit à la vie privée, qui ne veut plus dire assurer la confidentialité des données mais donner le droit à l'utilisateur de contrôler l'accessibilité et l'utilisation de ces données. En effet, la problématique de la vie privée dans les systèmes d'e-learning reste un débat entre les chercheurs afin de prouver si les systèmes d'e-learning sont infectés ou non par les menaces à la vie privée. De plus, Le respect de la vie privée a son impact sur la personnalisation de l'apprentissage qui se base essentiellement sur les informations collectées sur les apprenants (Tracking). Ces informations sont utilisées afin de personnaliser l'expérience d'apprentissage, de capitaliser les forces de l'apprenant, de cibler ses faiblesses et d'adapter l'expérience d'apprentissage en fonction de ses besoins et son style d'apprentissage. Par conséquent, le peu d'informations disponible sur l'apprenant (en raison de ses préférences de protection) aura ses retombées sur la personnalisation du processus d'apprentissage. C'est un grand défi de trouver un équilibre de manière à satisfaire à la fois les besoins de la protection de la vie privée et de la personnalisation. Cependant, les chercheurs essayent par ces « Anonymous credentials » de faire bénéficier les apprenants de tous les services d'e-Learning et d'avoir leurs bulletins et diplômes tout en restant anonymes. Cela évite beaucoup de problèmes d'absence d'informations sur les activités des apprenants. La difficulté se présente au niveau de la démonstration d'éligibilité de l'apprenant au moment de la présentation du diplôme et au moment de l'empêchement des abus de l'anonymat.

D'autre part, l'avènement du E-Learning 2.0 qui utilise essentiellement les outils du web2.0 introduit une nouvelle série de défis qui doivent être abordés pour protéger la vie privée des apprenants.

# Chapitre 3.
# L'utilisation des agents pour l'apprentissage social

# Introduction

Les systèmes multi-agents prennent aujourd'hui une place de plus en plus importante en informatique, particulièrement dans le domaine de l'intelligence artificielle et les systèmes distribués. C'est une discipline qui s'intéresse aux comportements collectifs produits par les interactions entre agents. Ces agents, qui représentent des entités autonomes et flexibles, sont très utiles pour la modélisation, la conception et l'implémentation de systèmes intelligents, complexes, ouverts et dynamiques tels que les systèmes d'apprentissage en ligne.

Ce chapitre introduit la notion d'agents et celle des systèmes multi-agents. Nous présenterons ensuite quelques plateformes d'e-Learning qui utilisent la technologie « agents ».

## 5. Qu'est-ce qu'un agent

Pour Ferber **[Ferber, 1989]**: « Un agent est une entité autonome, réelle ou abstraite (physique ou virtuelle), qui est capable d'agir sur elle-même et sur son environnement, qui, dans un univers multi-agent, peut communiquer avec d'autres agents, et dont le comportement est la conséquence de ses observations, de ses connaissances et des interactions avec les autres agents ».

## 2. Caractéristiques des agents : [Jennings, 1998 ] [Chaib-Draa, 2001]

Le paradigme agent possède des caractéristiques qui le distinguent des autres paradigmes logiciels tels que les systèmes orientés objets. Un agent peut posséder les caractéristiques citées ci-dessous :

- ✓ **l'autonomie :** l'agent est capable de prendre des décisions et d'agir sans l'intervention d'un tiers (humain ou agent), comme il peut contrôler ses propres actions ainsi que son propre état interne ;
- ✓ **la capacité représentationnelle ou la perception:** l'agent peut avoir une vision très locale ou plus large de son environnement et des agents qui l'entourent ;
- ✓ **la réactivité :** quand l'agent reçoit des informations de son environnement, il doit être capable de réagir en conséquence et de répondre à temps.
- ✓ **l'intentionnalité ou pro-activité :** l'agent n'est pas seulement capable de réagir aux changements de son environnement mais il peut même exhiber un comportement proactif orienté vers ses objectifs et prendre des initiatives au moment approprié.
- ✓ **la communication** : l'agent peut communiquer avec les autres agents.
- ✓ **l'anticipation :** l'agent possède plus ou moins les capacités d'anticiper les événements futurs ;
- ✓ **la rationalité :** les agents rationnels peuvent évaluer leurs actions selon des critères et choisir les meilleures actions pour atteindre l'objectif.
- ✓ **l'adaptabilité**: l'agent peut apprendre de ses expériences passées et sait résoudre les nouveaux problèmes à partir de ces dernières.
- ✓ **la sociabilité** : un agent interagit avec les autres agents (logiciels et humains) quand la situation l'exige afin d'accomplir ses tâches ou d'aider les autres agents dans leurs activités et atteindre leurs buts.

## 3. Classification des agents

Nwana **[Nwana, 1996]** propose sept catégories d'agents :

- ✓ **Les agents réactifs:** cette catégorie d'agents n'a pas de mémoire des expériences passées et n'a pas une représentation explicite de son environnement, elle réagit quand un stimulus se présente (stimulus|action).
- ✓ **Les agents collaboratifs :** ce sont des agents autonomes, qui coopèrent et négocient avec les autres agents afin d'atteindre des ententes quand il s'agit de la résolution distribuée de problèmes et éviter les conflits d'intérêt entre les agents;
- ✓ **Les agents d'interface :** ces agents fournissent une assistance à l'utilisateur. Ils sont capables de s'adapter aux préférences de l'utilisateur;
- ✓ **Les agents mobiles :** ceux sont des agents qui ont la capacité de se déplacer d'eux-mêmes d'un site a un autre pour se rapprocher des données ou des ressources;
- ✓ **Les agents d'information ou d'Internet :** ils sont chargés d'administrer, manipuler ou collecter les informations de plusieurs sources distribués (sites internet…etc.)
- ✓ **Les agents hybrides :** il combine plusieurs caractéristiques des autres agents.
- ✓ **les agents intelligents :** Selon Nwana, les agents logiciels intelligents n'existent réellement pas encore mais d'après Jennings et Wooldridge **[Jennings, 1998]**, un agent intelligent se caractérise par son autonomie, sa réactivité, sa capacité à agir ainsi que sa sociabilité. Selon **[Rouane, 2006]** ils sont dotés aussi de la capacité d'apprentissage.

## 4. Les systèmes multi-agents (SMA)

Un système est dit à base d'agents si l'abstraction clé de sa modélisation est celle d'un agent. Le système peut être mono-agent ; comme les systèmes d'intelligence artificielle classique(IA) qui modélise le comportement d'un seul agent intelligent et essaie de simuler dans une certaine mesure les capacités du raisonnement humain. Le système peut également être à base de plusieurs agents interactifs, dans ce cas on parle de systèmes multi-agents. Ces systèmes qui ont vu le jour avec l'avènement de l'intelligence artificielle distribuée (IAD) s'intéresse aux comportements intelligents qui sont produits par la coopération des agents dans un système distribué[16]. De plus, les agents permettent de réduire la complexité et le temps de la résolution d'un problème par sa division en sous-problèmes, chaque sous-problème est affecté à un agent intelligent indépendant dit « résolveur ». Pour y parvenir ; la résolution du but commun est établie par l'organisation et la coordination des activités des agents. **[Chaib-Draa, 2001]**

### 4.1 Caractéristiques des SMA

Selon **[Chaib-Draa, 2001] :**

1. chaque agent a des informations ou des capacités de résolution de problèmes limitées, les agents ont chacun un point de vue partiel.
2. il n'y a aucun contrôle global du système.

[16] Un ensemble d'ordinateurs indépendants qui apparaît à un utilisateur comme un système unique et cohérent **[Labri, --]**

3. les traitements et les données sont distribués, ainsi que les compétences, les rôles et les buts des agents.
4. Les agents forment une organisation sociale : la manière dont le groupe est constitué, a un instant donné, pour pouvoir fonctionner. **[MAZYAD, 2013]**
5. L'interaction : c'est la caractéristique principale qui distingue un SMA d'une collection d'agents indépendants. La motivation des agents pour interagir dépend de leurs buts, leurs capacités à réaliser certaines tâches et les ressources dont ils disposent. C'est à travers l'interaction que les agents peuvent combiner les efforts et s'entraider afin de réaliser une tache ou d'atteindre conjointement un but particulier ou bien entrer en compétition et négocier. **[MAZYAD, 2013]**

### 4.1.1 Les types d'interaction entre agents [Chaib-Draa, 2001]

**- La coopération** : travailler ensemble à la résolution d'un but commun.
**- La coordination** : organiser la résolution d'un problème de telle sorte que les interactions nuisibles soient évitées ou que les interactions bénéfiques soient exploitées.
**- La négociation** : parvenir à un accord acceptable pour toutes les parties concernées.
**- La communication :** La communication dans les systèmes multi-agents est à la base des interactions et de l'organisation des agents. Il existe deux modes de communication dans la littérature : la communication indirecte qui se fait par transmission de signaux via l'environnement et la communication directe qui correspond aux échanges des messages entre les agents. **[MAZYAD, 2013]**

### 4.2 Les avantages des systèmes multi-agents [MAZYAD, 2013]

L'utilisation des systèmes multi-agents présente une série d'avantages car ces systèmes ont hérité des bénéfices traditionnels de la résolution distribuée et de l'intelligence artificielle **[Chaib-Draa, 2001]** :

- **L'ouverture** : c'est la capacité d'ajouter ou de retirer dynamiquement dans le système des agents, des fonctionnalités ou des services.
- **La souplesse de l'outil informatique** : ils permettent de rendre la programmation plus simple, de modifier le comportement des agents, d'ajouter ou de supprimer des actions possibles. **[Le Bars, 2003]**
- **La décentralisation** : la décentralisation des agents permet de supporter l'échec individuel de l'un des éléments ou l'ajout d'un nouveau sans dégrader le système dans son ensemble.
- **La résolution distribuée de problèmes** : un problème peut être décomposé en sous parties, chacune pouvant être résolue de façon indépendante pour aboutir à une solution stable.
- **La vitesse d'exécution**: grâce au parallélisme, plusieurs agents peuvent travailler en même temps pour la résolution d'un problème.

## 5. Normalisation des agents FIPA

Les efforts de normalisations visent à rendre compatibles les différents systèmes d'agents. Tous les efforts actuels ont un but commun qui consiste à définir une norme capable de s'imposer sur les différents systèmes d'agents afin de permettre leur interopérabilité. L'un des standards disponibles concernant les systèmes multi-agents, on trouve la norme FIPA (Foundation for Intelligent Physical Agents).

### 5.1 FIPA (Foundation for Intelligent Physical Agents)

FIPA[17] est une organisation de standardisation fondée en 1996 à Genève (Suisse) dont l'objectif est d'assurer l'interopérabilité entre les agents et les applications à base d'agents hétérogènes. FIPA définit des spécifications concernant la communication entre les agents, le transport des agents, la gestion des agents, l'architecture abstraite et les applications. La communication des agents représente le noyau du système FIPA. Le but des modèles FIPA n'est pas de décrire une structure XML qui ne standardise que la structure syntaxique des documents ou de décrire des configurations physiques, ce n'est pas le champ d'application des normes FIPA de fournir des lignes directrices strictes pour l'implémentation des plateformes multi-agents. La norme FIPA présente un modèle de référence pour la gestion des agents (Agent Management Reference Model), étant une base pour les développeurs qui veulent concevoir et implémenter leur propre plate-forme multi-agents. **[ORZ, 2010]**

Les spécifications de FIPA sont reparties en cinq catégories : **[Poslad, 2001]**

- **Les applications** : spécifient les domaines d'application sur lesquels peuvent être déployés des agents FIPA.
- **Les architectures abstraites** : ensemble de spécifications qui concernent les entités abstraites nécessaires pour le développement d'un environnement d'agents.
- **La gestion des agents** : ces spécifications traitent le contrôle et la gestion des agents dans/entre les plateformes d'agents.
- **Le transport des messages d'agents** : ces spécifications traitent la représentation et le transport des messages à travers les différents protocoles réseaux.
- **La communication des agents** : La communication des agents est basée sur l'envoi de messages. Le langage « FIPA ACL (Agent Communication Language – ACL)» est le langage standard des messages et impose le codage, la sémantique et la pragmatique des messages.

L'architecture d'un système respectant les spécifications de la norme FIPA contient trois composantes obligatoires:

- ✓ ACC – Agent Communication Channel : gère la communication entre les agents.
- ✓ AMS – Agent Management System : supervise l'enregistrement des agents (attribution des identifiants AID), leur authentification, leur accès et l'utilisation du système.
- ✓ DF – Directory Facilitator: fournit la possibilité d'intégrer des services décrits et référencés dans un annuaire.

Ces trois éléments sont automatiquement activés au démarrage de la plateforme d'agent.

[17] Foundation for Intelligent Physical Agents – http://www.fipa.org

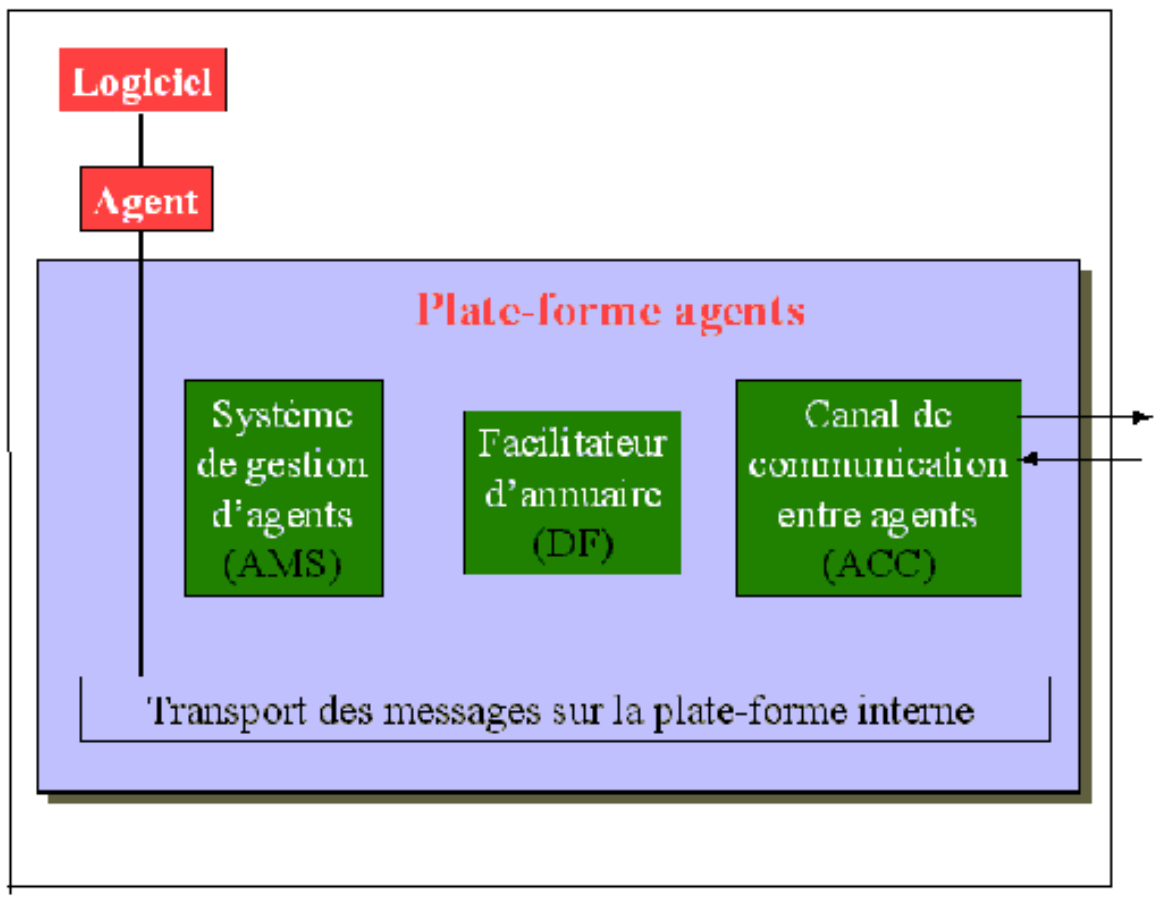

**Figure 26:** Le modèle de référence pour une plate-forme multi-agents FIPA **[Politechnica,2002]**

## 6. La plateforme de développement des systèmes multi-agents, JADE

Il existe plusieurs plateformes de développement des systèmes des SMA. Parmi celles les plus connues, on trouve JADE (Java Agent DEvelopment Framework) **[Bellefemine, 2004]** développée par CSELT (Groupe de recherche de Gruppo Telecom, Italie)**.** JADE est une plateforme implémentée en Java qui permet l'implémentation des systèmes multi-agents et des applications conformes aux standards FIPA pour assurer l'interopérabilité entre agents. La plateforme Jade se compose principalement de deux composants: Une plateforme agents compatible FIPA qui est fournie avec une Licence open source et des packages qui ont pour but de simplifier le développement des systèmes multi-agents. Les agents communiquent par l'échange asynchrone de messages écrits en langage FIPA ACL. Ces échanges se basent sur des protocoles d'interaction fournis par FIPA. Des mécanismes sont fournis pour authentifier et vérifier les droits associés aux agents lors de l'échange de messages, ces derniers sont transférés via des canaux SSL[18]. **[MAZYAD, 2013]** La plate-forme JADE est en développement continu. Elle a été utilisée avec succès par de nombreux organismes de recherche et commerciaux. **[ORZ, 2010]**

## 7. Les plateformes e-learning

Les plateformes e-Learning Open Source continuent à évoluer etprogressivement elles commencent à intégrer des éléments du Web 2.0 : fonctionnalités, ergonomie, design, « serious games » et pour certaines un réseau social propre à la plateforme, comme Chamilo **[Chamilo, 2012]**. En 2012, le répertoire de THOT[19] compte plus de 180 plateformes dont 150 plateformes commerciales et 25 plateformes open-source[20] alors que le répertoire CMS SPIP[21] cite 63 plateformes open source. Les initiatives open source comprennent parmi les leaders Moodle, Sakai, ATutor et WhiteBoard. Nous présentons dans ce qui suit les plateformes qui supportent le travail collaboratif et qui utilisent le paradigme agent mais ceux sont des plateformes peu connues. Les plateformes connues comme Sakai, Moodle et Claroline ne seront pas présentées dans ce travail car elles utilisent seulement le paradigme objet. Cependant, les systèmes d'apprentissage social et de travail collaboratif, les plus connus en littérature, utilisent des agents de type « intelligents ». Pour cette raison, nous présentons dans ce qui suit que les plateformes à base d'agents intelligents.

[18] SSL : (Secure Sockets Layer) C'est une norme de sécurité de transfert qui fournit le chiffrage des données, l'authentification du serveur et l'intégrité du message.
[19] http://cursus.edu/institutions-formations-ressources/formation/13486/plates-formes-learning-formation-2013/#.U_C_QMV5Np4
[20] http://cursus.edu/institutions-formations-ressources/formation/13486 consulté le 26/09/2013
[21] http://icp.ge.ch/sem/cms-spip/spip.php?article1783

### 7.1 Plateformes d'e-Learning à base d'agents intelligents

Durant les années 1990, Chan[22] a présenté la stratégie du compagnon d'apprentissage afin de faciliter le processus d'apprentissage. Cette stratégie simule le comportement d'un deuxième élève (pair ou un ami) qui apprend avec l'élève. Depuis, de nombreuses autres approches ont été développées. Par exemple, la stratégie du compagnon perturbateur qui sème le doute et qui sert à tester la confiance des élèves en soi face à l'apprenant perturbateur (troublemaking learning companion). Depuis cette initiative, la technologie des agents a été appliquée dans divers types d'applications pour l'éducation : à la conception d'environnements pairs-assistance (peer-help), les agents de recherche d'information, les agents de traitement des informations de l'élève, les agents de recueil ou de génération des Feedbacks, les agents pédagogiques, les agents de tutorat…etc. Une étude empirique effectuée pour évaluer l'efficacité des agents intelligents dans l'enseignement en ligne a montré que les agents peuvent améliorer le taux d'achèvement, la satisfaction des apprenants et la motivation. **[Nedev, 2008]**

**DALE (Distributed Adaptive Learning Environment) : [Pu, 2004]** DALE intègre des agents et des objets d'apprentissage dans un système distribué. Dale utilise des agents intelligents pour adapter dynamiquement le contenu aux besoins particuliers des apprenants.

**F-SMILE Multi-Agent System: [Virvou, 2002]** Le système fonctionne dans un environnement multi-agents d'apprentissage intelligent qui peut fournir un tutorat adaptatif basé sur la modélisation de l'apprenant à travers les services Web. F-SMILE se compose d'un agent « learner modelling (LM)», d'un agent Conseiller « Advising Agent », un « Tutoring Agent » et un Agent dirigé par la parole « Speech-driven Agent ». L'agent LM observe l'apprenant de façon permanente et au cas où il soupçonne que l'utilisateur rencontre des difficultés, cherche à savoir quel est la cause du problème. L'agent cherche les actions prises par l'apprenant dans des situations similaires dans le passé. Ces actions alternatives sont envoyées à l'agent « conseiller», qui est responsable sur la sélection de l'action la plus appropriée pour être proposée à l'apprenant concerné. La sélection de la meilleure action alternative est basée sur les informations collectées par l'agent. Au cas où le problème de l'utilisateur vient d'un manque de connaissances, « Tutoring Agent » est activé afin de générer d'une manière adaptative une leçon appropriée aux besoins de l'apprenant. Lorsque le conseil et la leçon sont préparés, ils sont envoyés au « Speech-driven Agent », qui est chargé de rendre l'interaction avec l'utilisateur semblable aux interactions humaines.

[22] Chan, T.-W., 1990. Learning companion systems. In: In C. Frasson. Ablex Publishing Corporation, pp. 6‑33.

**aLFanet Multi-Agent System: [Nedev,2008]** AlFanet offre des services éducatifs adaptés aux besoins individuels et collaboratifs de l'apprenant. Les besoins individuels sont basés sur les informations personnelles de l'utilisateur (par exemple, les styles d'apprentissage, les objectifs d'apprentissage, les préférences d'apprentissage), alors que les besoins de collaboration se concentrent sur les relations entre les utilisateurs (par exemple, contact avec les élèves ayant des caractéristiques similaires, l'accès aux commentaires pertinents ou les documents envoyés par d'autres apprenants, de collaborer avec un groupe de travail particulier). Les modèles pédagogiques avancés, basés sur le concept de l'apprentissage actif et adaptatif, peuvent être spécifiés par les auteurs du cours au moment de la conception du cours. L'objectif principal des systèmes adaptatifs est de fournir à l'utilisateur un accès efficace au site en présentant d'abord les liens et les matériaux qui pourraient l'intéresser. Pour résoudre ce problème, différentes formes du modèle utilisateur sont appliquées. Il existe deux approches pour le développement de ces modèles. La première est d'appliquer des règles prédéfinies pour les situations qui surviennent dans le cours pendant l'interaction des étudiants, ce qui exige de connaître les situations à l'avance. La deuxième est d'apprendre les modèles d'apprenants à partir des données collectées de leurs interactions avec les autres utilisateurs. Dans ce cas, pendant leur interaction avec le système, les utilisateurs fournissent beaucoup d'expériences qui sont imprévisibles et qui dépendent beaucoup de leur profil (les intérêts, les besoins, les connaissances, les préférences, le style d'apprentissage, etc.). A partir de là, les profils et les expériences similaires apparaissent et peuvent être utilisés pour faciliter les tâches d'adaptation pour les apprenants et les tuteurs. Ce type d'adaptation, appelé adaptation de l'interaction, s'occupe de fournir l'aide et le support nécessaires pendant l'interaction dans le cours. De cette façon, les tuteurs et les apprenants obtiennent un accès adaptatif aux services qu'ils utilisent, aux contenus et aux activités de collaboration avec lesquels ils travaillent, aux utilisateurs qu'ils ont besoin de contacter…etc. Le modèle d'adaptation est basé sur une architecture multi-agent qui fonctionne d'une manière autonome pour résoudre la tâche d'adaptation demandée sur la base des connaissances que ces agents peuvent obtenir.

**Help&Learn**: C'est un système multi-agent dont le but est d'améliorer la gestion des connaissances dans des communautés d'apprentissage collaboratif (knowledge management). Dans Help&Learn, un utilisateur peut demander de l'aide (joue le rôle de demandeur d'aide (Helpee)) ou bien fournir l'aide demandée (joue le rôle de fournisseur d'aide (Helper)). Les connaissances sont échangées sous forme d'objets d'aide (HelpItems). Ainsi, qu'ils soient apprenants ou enseignants, tous les utilisateurs sont des pairs qui collaborent pour partager leurs connaissances. Dans ce système, des assistants personnels sont utilisés pour maintenir le profil de l'utilisateur ainsi que sa base de connaissances tandis que d'autres agents sont utilisés pour trouver le meilleur pair capable de fournir de l'aide et fournir ainsi une meilleure gestion des ressources et des connaissances.

**MASCE [Mahdi, 2010]** : C'est un système multi-agents pour l'apprentissage collaboratif à distance. Ce système complète l'apprentissage traditionnel en présentiel. En effet, MASCE permet à l'apprenant de revoir les documents de cours, de demander de l'aide et d'évaluer l'aide fournie afin de permettre au système d'avoir une liste des meilleurs aidants, d'interagir avec leurs tuteurs ou les autres apprenants à l'aide des outils de travail collaboratif (chat, email). L'objectif de ce projet est de pouvoir continuer l'apprentissage entamé en classe et de permettre au tuteur d'interagir avec les apprenants afin de les aider dans leur apprentissage, de les évaluer et de suivre leurs activités.

**I-Help** : **[Brooks, 2006] [Vassileva, 2009]** Le système I-Help est un système de mise en relation entre pairs conçu pour aider les apprenants engagés dans des activités de résolution des problèmes. Le système permet aux étudiants de donner et de recevoir de l'aide de façon synchrone ou asynchrone. Il se base essentiellement sur deux outils : un outil de chat « **I-Help 1 - on -1** » Peer2Peer (Private discussion : entre « a helpee » (demandeur d'aide) et « a helper » (aidant qui peut être un élève ou un expert)) et un forum « **I-Help Pub** » pour publier les questions des apprenants (Public discussion). « Greer » et son équipe **[Brooks, 2006]** s'intéressent au développement des algorithmes de localisation des ressources (numériques et humaines) qui sont spécifiques aux demandes d'aide de l'apprenant et au développement des mécanismes de déduction pour le traitement du « learner modele » afin de localiser les meilleurs pairs capables d'aider l'apprenant.

## 7.2 Comparaison des systèmes étudiés

Les systèmes d'apprentissage que nous avons présentés dans cette section permettent deux types d'interaction :

- Interaction « humain-agent » : comme « Speech-driven Agent » qui interagit avec l'utilisateur d'une manière semblable aux interactions humaines.
- Interaction « humain-humain » : Comme le système I-help qui facilite les interactions sociales entre les apprenants.

Nous présentons aussi les architectures utilisées pour construire ces systèmes d'apprentissage dans le tableau (Tableau 1).

| Système | Interaction humain-agent | Interaction humain-humain | Architecture |
|---|---|---|---|
| DALE | Oui | Non | Hybride |
| F-SMILE | Oui | Non | Client-serveur |
| aLFanet | Oui | Oui | Client-serveur |
| Help&Learn | Oui | Oui | Peer2peer |
| MASCE | Oui | Oui | Client-Serveur |
| I-Help | Oui | Oui | Peer2peer |

Tableau 1: Comparatif de l'existant

Nous remarquons que les propositions des chercheurs concernant l’utilisation des agents intelligents dans le domaine d’e-Learning peuvent être réparties en deux types d’objectifs :

1- Pour améliorer les processus d’adaptation des expériences d’apprentissage en profitant des avantages des SMA qui facilitent la modélisation, la décomposition des traitements et des interactions (DALE, F-SMILE et AlFanet).
2- Pour améliorer les requêtes de recherches des ressources numériques ou les pairs (humains) ayant les caractéristiques nécessaires pour pouvoir aider l’apprenant. (Help&Learn, I-Help et MASCE).

## Conclusion

Les technologies d’Internet s’intègrent de plus en plus dans les systèmes d’e-Learning et offrent beaucoup d’avantages dont notamment celui de faciliter l’enseignement et l’accessibilité aux ressources pédagogiques. L’isolement des apprenants dans ces systèmes d’apprentissage crée chez les apprenants une certaine autonomie qui les rend plus indépendants de leurs enseignants. D’autre part, les outils du Web ont rendu les systèmes d’e-Learning plus interactifs où l’apprenant joue un rôle actif dans l’enrichissement des connaissances partagées dans le système, surtout celles qui sont le produit des interactions sociales entre les utilisateurs. Ce changement a modifié le rôle de l’enseignant, qui était auparavant le seul détenteur des connaissances. Son rôle actuellement dans ces système est d’être un facilitateur d’apprentissage qui guide les apprenants à bien utiliser les connaissances disponibles en dehors de lui. Cependant, les systèmes d’e-Learning doivent développer des outils permettant de gérer d’une manière efficace et bénéfique les ressources produites par les interactions des pairs d’apprentissages et les cours créés par les enseignants. Vu l’originalité des solutions que proposent les multi-agents, les systèmes d’e-Learning font souvent appel aux SMA pour résoudre les problèmes de la recherche d'objets pédagogiques ou pour la récolte et traitement des données des apprenants afin de suivre leur progression, de les motiver, les guider et éviter leur abandon. D’autres recherches plus avancées proposent des solutions plus intelligentes permettant d’adapter le processus d’apprentissage aux besoins de l’apprenant. Ces solutions permettent d’évaluer le niveau de l’apprenant dans chaque domaine et de sélectionner les ressources d’apprentissage les plus appropriées à son niveau et ses caractéristiques. D’autres solutions essaient de simuler l’intelligence humaine pour créer des agents capables de prendre le rôle du tuteur ou les camarades de classe dans les plateformes qui favorisent l’isolement de l’apprenant. D’autres solutions encore utilisent les agents pour renforcer la dimension sociale de l’apprentissage en aidant l’apprenant à retrouver les meilleurs personnes qui peuvent répondre à ses questions et l’aider à résoudre son problème et à le garder en contact avec les autres pour éviter son isolement.

# Partie 2.

# Conception

# Chapitre 4.

# Description de la proposition :

# Système« ApprAide »

# Introduction

« ApprAide » représente un système d'apprentissage social qui a pour objectif de créer un environnement de partage de connaissances et un milieu d'entraide entre les apprenants et les enseignants. Comme il représente aussi un moyen pour faciliter la recherche des réponses aux questions, une base de tests en ligne que l'apprenant peut passer pour valider ses connaissances et un système pour faire des cours de soutien ponctuels en ligne avec les enseignants. Nous présentons dans cette partie une description générale de notre système « ApprAide » qui prend en compte le problème de protection de la vie privée des apprenants.

# 1. Description du système « ApprAide »

« **ApprAide** » (Acronyme : **Appr**entissage et **Aide**) est un système qui permet de mettre en relation les apprenants et les enseignants dans le but de s'entre-aider, d'échanger des informations et de partager des connaissances. C'est un système d'apprentissage qui complète l'apprentissage en classe, en aidant l'apprenant à trouver des personnes avec lesquelles il peut interagir pour poser ses questions, réviser avec eux ou avoir des cours de soutien en ligne avec les enseignants connectés. Nous distinguons deux types de demande d'aide :

1) Demande d'aide synchrone : dans ce cas l'apprenant se connecte à la plateforme pour chercher un enseignant connecté afin d'avoir un cours de soutien ponctuel ou pour chercher un apprenant qui peut se porter volontaire pour l'aider à résoudre son exercice ou lui expliquer ce qu'il ne comprend pas.
2) Demande d'aide asynchrone : dans ce cas l'apprenant pose sa question et il la laisse en attente d'une réponse. Ce type de demande est ouvert au public, à la différence du premier, il a la chance de recevoir plusieurs réponses de personnes différentes (enseignants et apprenants) pouvant ouvrir une discussion riche entre les membres du système.

Le système vise à offrir aux apprenants des outils facilitant les interactions et la recherche des personnes qui peuvent les aider. Le système est composé de deux applications : la première est une application web hébergée sur le serveur. Elle se charge de l'inscription et l'authentification des utilisateurs ainsi que le stockage des données publiques partagées par les utilisateurs. La deuxième est une application cliente qui gère les échanges privées et le stockage des données privées de l'utilisateur.

## 1.1 L'application web

L'application web permet aux utilisateurs de s'inscrire. Il existe deux types d'inscription :

a. l'inscription des apprenants qui ne requière que le remplissage d'un formulaire simple qui ne demande que des informations nécessaires à son authentification.
b. l'inscription des enseignants qui nécessite le téléchargement de quelques documents (tel que le diplômes) attestant le niveau de l'enseignant et justifiant son recrutement.

Les inscriptions des enseignants doivent être validées par l'administrateur qui vérifie le dossier du candidat-enseignant avant son enregistrement comme « enseignant-Freelancer[23] » sur le système. Après l'étude du dossier, l'enseignant recevra un message sur le site contenant la décision. Si le candidat est accepté, il sera réorienté vers une page de téléchargement de l'application cliente.

Nous résumons les cas d'utilisation de l'application web dans le diagramme de la figure 27 :

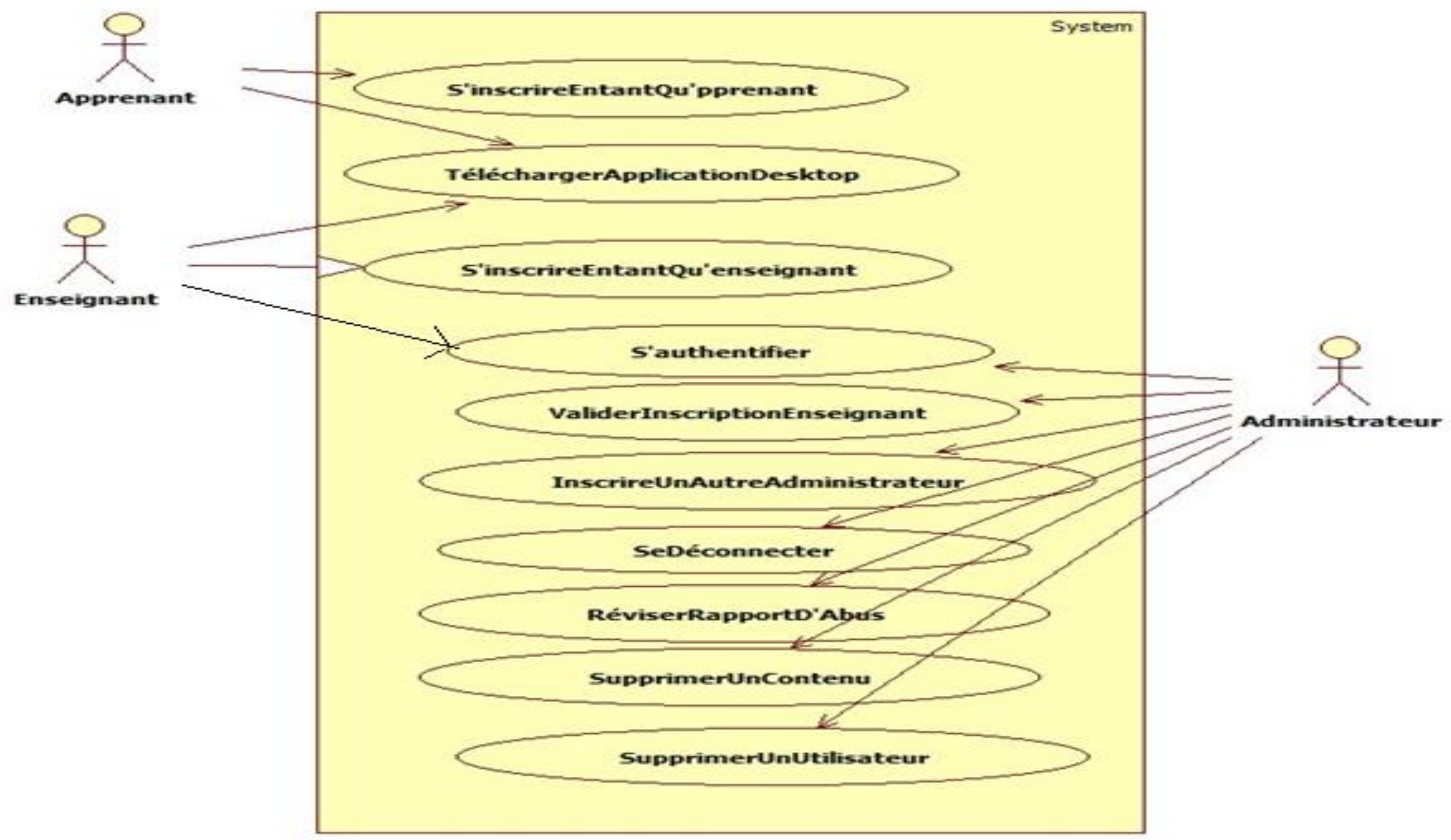

**Figure 27: Les cas d'utilisation de l'application web**

Le tableau 2 explique quelques cas d'utilisation :

| Cas d'utilisation | Détail |
|---|---|
| S'inscrireEntantQu'Apprenant | L'inscription de l'apprenant consiste en le remplissage d'un formulaire contenant que les informations nécessaires pour son authentification (un pseudonyme, un mot de passe et l'adresse mail). Si le formulaire est validé, l'apprenant est invité à télécharger l'application « ApprAide». |
| S'inscrireEntantQu'Enseignant | L'inscription de l'enseignant consiste en le remplissage d'un formulaire contenant les informations nécessaires pour son recrutement. Si l'enseignant est accepté par l'administrateur, il recevra un lien pour le téléchargement de l'application cliente « ApprAide ». |

[23] Qui exerce sa profession de façon indépendante.

| | |
|---|---|
| ValiderInscriptionEnseignant | L’administrateur peut accepter ou refuser la candidature de l’enseignant. |
| RéviserRapportD’Abus | L’administrateur peut réviser les rapports d’abus concernant les contenus ou les utilisateurs signalés. |

Tableau 2: Les cas d'utilisation de l'application web

Le schéma ci-dessous (Figure 28) représente les classes et les fonctionnalités associées à chaque type d’acteur de l’application web.

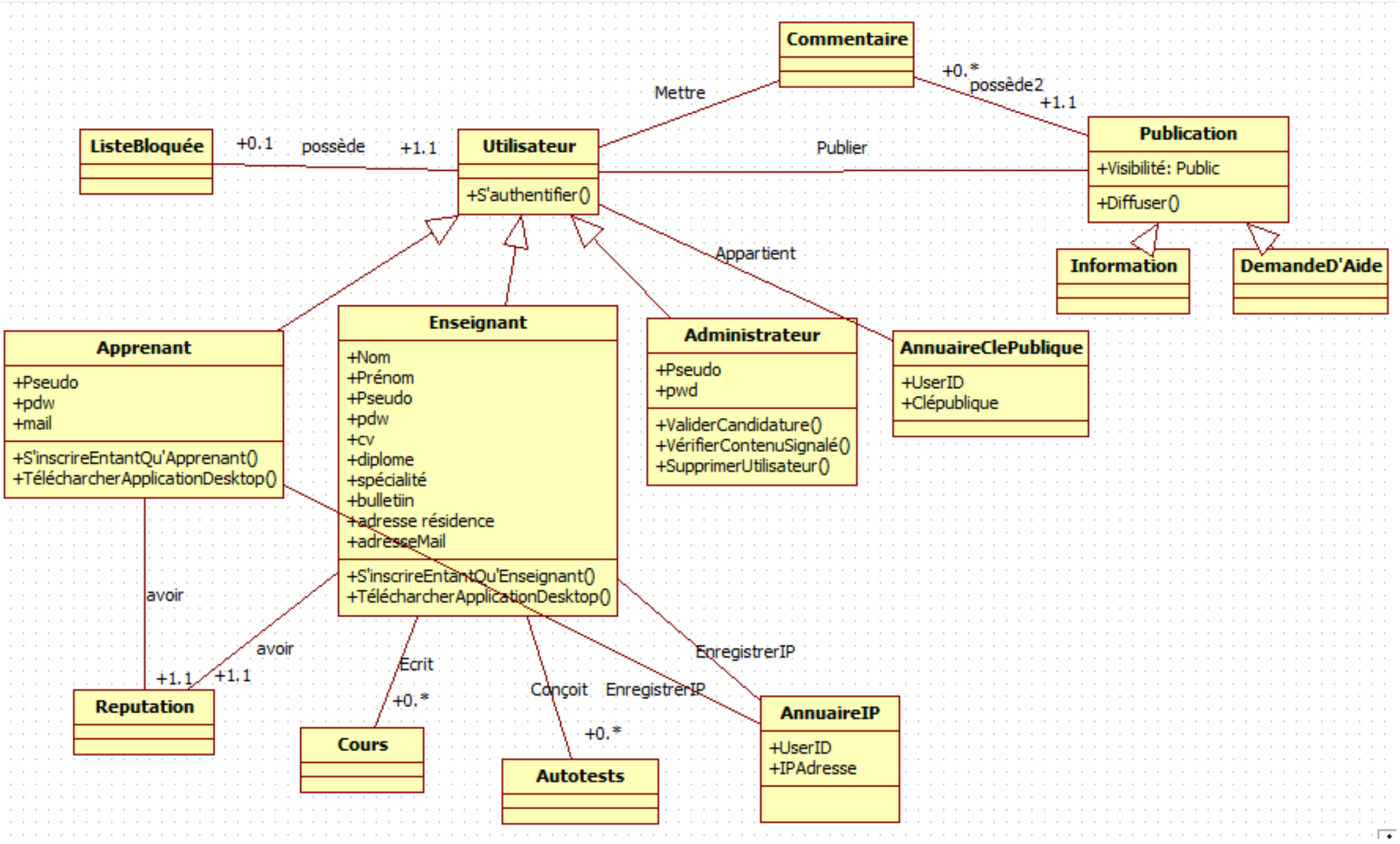

**Figure 28: Diagramme de classe de l'application web.**

### 1.2 L’application cliente « ApprAide» :

Cette application permet l’accès à tous les services de notre système. C’est à partir de cette application que l’utilisateur peut s’authentifier, interagir avec les utilisateurs de notre système et télécharger les cours et les autotests. Les outils intégrés à cette application sont les suivants :

### 1.2.1 Les outils

#### 1.2.1.1 Un outil de réseautage social

Cet outil est utilisé pour les demandes d’aide asynchrones. C’est un système où les utilisateurs peuvent poser des questions, partager des informations et des connaissances, publier des statuts ou commenter et évaluer les publications. Les publications sont diffusées sur un fil d’actualité qui ressemble à celui de Facebook et Linkedin. Les questions et les informations publiées sont diffusées sous forme de «news feeds». Avant le partage d’une question ou d’une connaissance sur le réseau social, l’utilisateur doit préciser la matière et le niveau d’étude afin de permettre aux aidants d’apporter les réponses adéquates et se limiter juste au niveau d'étude de l'apprenant. Les réponses aux questions sont publiées sous forme de commentaires. Les commentaires et les publications peuvent être évalués par les utilisateurs du système. L'utilisateur peut filtrer les publications qu'il veut voir sur son fil d'actualités. Il peut aussi chercher des contacts, ajouter des amis, créer des listes de contacts, éditer son profil, ajuster ses paramètres de confidentialité et bloquer, signaler ou supprimer des utilisateurs (les personnes avec lesquelles il ne veut pas interagir). Le diagramme (Figure 29) suivant montre les cas d'utilisation de cet outil :

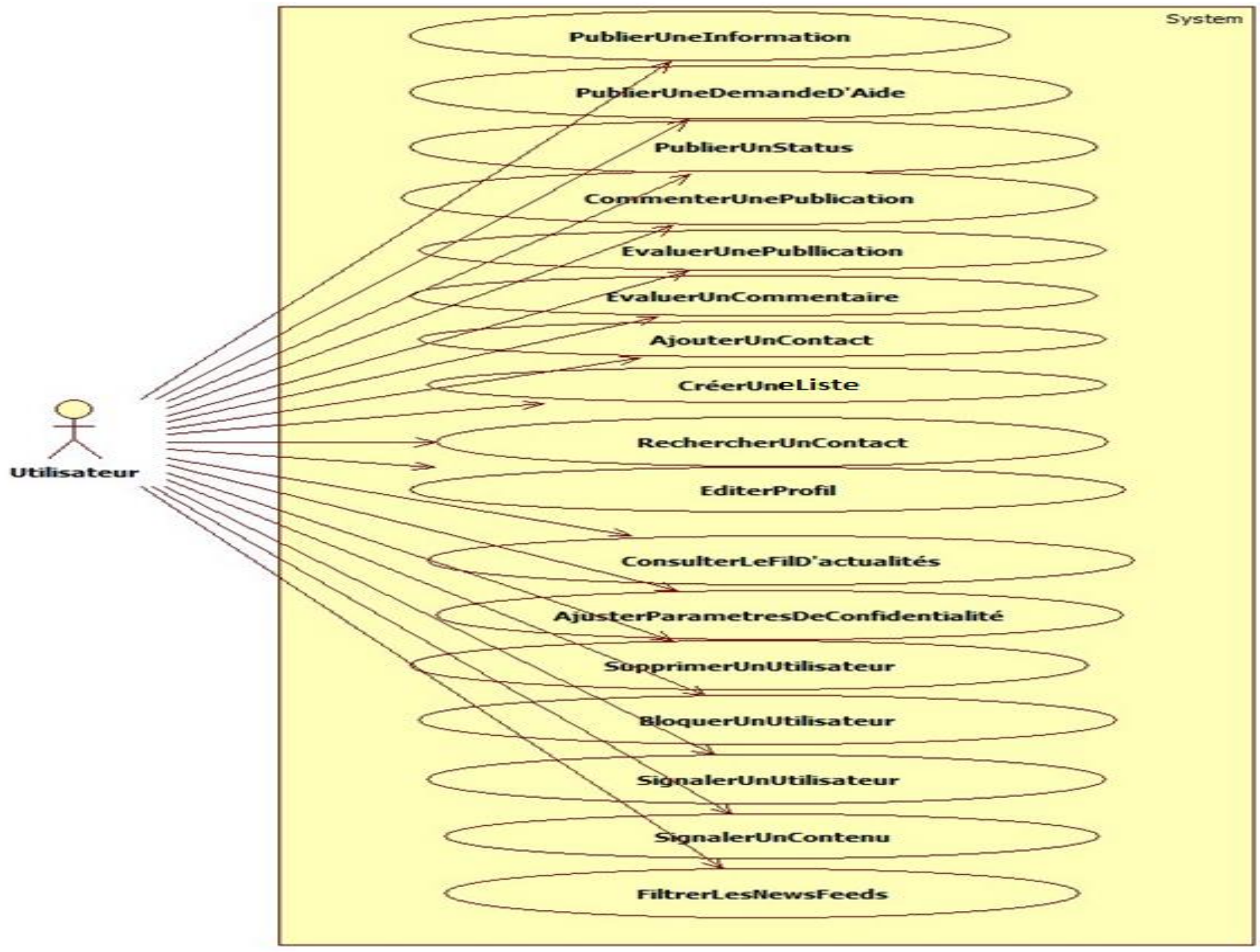

**Figure 29: Les cas d'utilisation du réseau social**

Le tableau 3 présente une description de quelques cas d'utilisation :

| Cas d'utilisation | Détails |
|---|---|
| PublierUneDemandeD'Aide | L'utilisateur peut publier une publication de type « Demande d'aide ». L'utilisateur doit spécifier la matière, le niveau d'étude ainsi que l'année d'étude auxquels la publication appartient. |
| PublierUnStatut | Dans notre système l'utilisateur peut exprimer ses émotions ou son avis concernant un projet ou un sujet particulier. |
| AjusterParamètresDeConfidentialité | Chaque publication est accompagnée d'un petit menu pour aider l'apprenant à spécifier l'audience de la publication. |
| SignalerUnContenu | L'utilisateur peut signaler une publication ou un message. Le contenu, ainsi que les pseudonymes du propriétaire du contenu et la victime sont envoyés à l'administrateur pour prendre une décision. |
| FiltrerLesNewsFeeds | L'utilisateur peut filtrer les flux d'actualité par matière ou par type de publication (demande d'aide, Information, Statut) |
| Tableau 3: Les cas d'utilisation du réseau social | |

Selon Greer **[Greer, 2002],** Cet outil a plusieurs avantages:

- Mieux stimuler les interactions entre les apprenants, ce qui développe de meilleures compétences de communication chez les apprenants, les motiver à continuer à apprendre avec les autres, avoir le réflexe d'aller chercher des solutions chez les autres quand ils trouvent des problèmes au lieu de se décourager.
- La rédaction des réponses développe l'esprit de synthèse chez l'apprenant.
- Le fait de travailler et interagir avec les autres, développe l'esprit critique chez les apprenants.
- Poser et publier une question, motive l'apprenant à y trouver la réponse lui-même.
- Celui qui veut répondre à la question posée, même s'il a la réponse, il va chercher et vérifier avant sa publication, et cela améliorera son apprentissage.
- Les commentaires des autres aident l'apprenant à savoir s'il est sur le bon chemin, si les autres apprenants ont des problèmes similaires et à comparer son état d'avancement par rapport aux autres apprenants (s'ils sont de la même classe ou école).
- Les enseignants, quant à eux, peuvent aider les apprenants, en les orientant et les encourageant. Ils peuvent même détecter les difficultés rencontrés pendant la résolution des problèmes, les notions mal assimilées pendant le cours et les lacunes des apprenants (les fautes d'orthographes, de calcul, de logique...etc.).

#### 1.2.1.2 Un tableau virtuel

Cet outil est utilisé pour les demandes d'aide synchrones. Le tableau virtuel peut être utilisé pour faire des cours de soutien en ligne (enseignant-élève). De même, Il est accessible aux élèves pour qu'ils puissent l'utiliser avec leurs camarades pour la résolution des exercices. Les cours synchrones donnés sont des cours de soutien ponctuels, i.e.: quand l'apprenant a besoin de trouver un enseignant pour avoir des réponses à ses questions. Cet outil a pour objectif d'aider l'apprenant à avoir l'assistance au moment où il en a besoin. Le tableau intègre aussi un petit espace de « Chat » qui peut être utilisé pour les échanges synchrones des messages privés.

#### 1.2.1.3 Un outil de test de connaissance

Un outil de création et de conception des tests ouvert aux enseignants. Après la validation du contenu du test par l'enseignant, le test est mis en ligne afin de permettre aux apprenants l'accès à ce dernier. Les élèves peuvent passer les tests d'une manière libre afin de valider leurs connaissances. Les résultats des évaluations sont privés. Cet outil sert essentiellement à détecter les points faibles et les points forts des apprenants.

### 1.2.1.4 Les annonces des cours synchrones en ligne

Cela permet aux apprenants de savoir le programme des cours de soutien en groupe qui seront organisés en ligne par les enseignants, contrairement aux cours individuels qui se font selon la demande de l'apprenant. Les cours de soutien en groupe sont affichés sous forme d'annonces.

Le schéma ci-dessous (Figure 30) représente les classes et les fonctionnalités associées aux enseignants et aux apprenants de l'application cliente « ApprAide ».

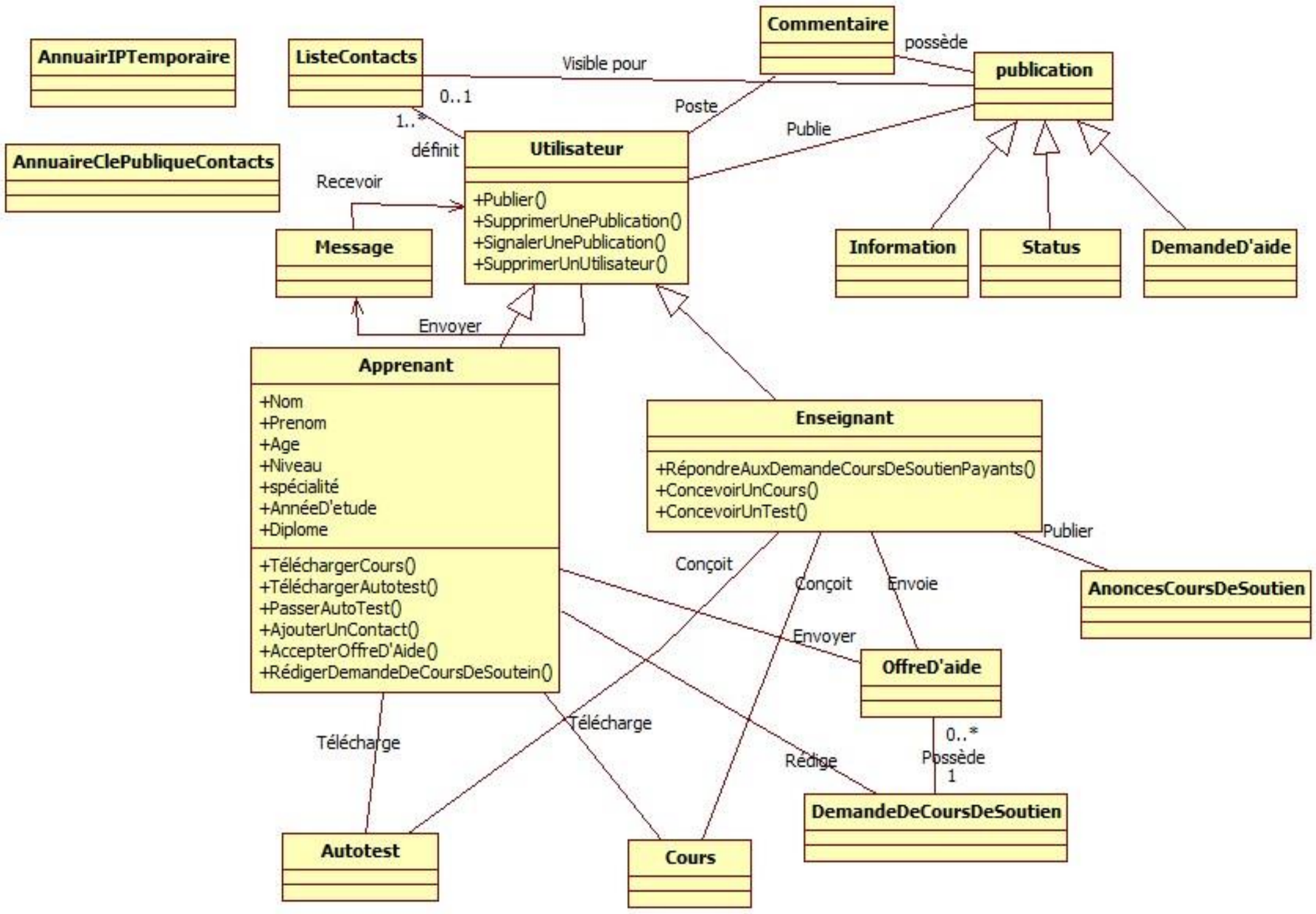

**Figure 30: Diagramme de classe de l'application "ApprAide"**

## 2. Description des rôles des utilisateurs

Dans notre système, il existe trois catégories d'utilisateurs : le tuteur, l'apprenant et l'administrateur.

### 2.1 Le tuteur

L'enseignant joue le rôle d'un tuteur ou un facilitateur d'apprentissage qui est en interaction active avec les apprenants. Il a pour rôle d'aider les apprenants, de répondre à leurs questions, de les guider et les orienter vers les bonnes solutions. De plus, son rôle est de concevoir des tests pour évaluer les connaissances des apprenants, de donner des cours de soutien en ligne et de publier des contenus pédagogiques.

### 2.2 L’apprenant

L’apprenant a un rôle actif dans la création des contenus pédagogiques. L’apprenant peut être un demandeur d’aide « aidé » ou un fournisseur d’aide « aidant ». Un apprenant « aidé » peut demander un cours de soutien en ligne d’un enseignant, comme il peut demander des informations ou des réponses à ses questions sur le réseau social. Il peut même chercher les personnes capables de l’aider (spécifier le public visé de sa publication ou les compétences du fournisseur d’aide). Le demandeur d’aide peut utiliser la discussion privée (discussion instantanée) pour poser ses questions et discuter avec les fournisseurs d’aide. L’apprenant « fournisseur d’aide » peut répondre aux questions posées par d’autres apprenants, participer aux discussions instantanées, utiliser le tableau virtuel pour expliquer et faire des exercices avec ses camardes

### 2.3 L’administrateur

L’administrateur a pour rôle de veiller sur le bon fonctionnement du système et d’agir en cas de problème. L’administrateur peut:

- Valider l’inscription des enseignants après vérification de leurs dossiers de candidature.
- Réviser les rapports d’abus sur les profils, publications et commentaires signalés.
- Supprimer les contenus inappropriés.
- Supprimer les utilisateurs non souhaités.

## 3. Conception des profils :

Nous avons basé la conception des profils des apprenants sur les spécifications établies par IMS LIP et PAPI [**Oubahssi, 2005]**. Nous avons gardé les éléments de profil qui sont nécessaires pour le bon fonctionnement de notre système. Nous avons ainsi ajouté de nouvelles classes pour prendre en compte les nouveaux types de données créées par l’ajout des activités de réseautage social au système. Nous avons également supprimé quelques éléments de profil définis par IMS LIP car notre système ne gère pas des formations proprement dit. Le profil est représenté par un ensemble de classes. Deux types de classes sont à distinguer : des classes de « Contenus » et des classes de « Connexions ».

### I- Les classes de contenus

Ce sont des classes qui décrivent l’apprenant et ses activités dans le système. Nous énumérons les classes suivantes :

a. **Identité :** elle regroupe l’ensemble des informations qui décrivent l’apprenant d’une manière unique et qui ne sont pas liées à l’apprentissage. Cette classe contient les informations suivantes : son nom, son prénom, son e-mail, son numéro de téléphone et sa photo.
b. **Attributs démographiques :** C’est l’ensemble des informations suivantes : âge, date de naissance, sexe, nationalité et lieu de naissance.
c. **Activités :** deux classes d’activités se distinguent :

1- **Activités liées à l'apprentissage :** demande d'un cours de soutien en ligne, accès à un cours ou un test…etc. Cette classe contient ainsi la liste des cours et des tests écrits et publiés par l'enseignant ou l'apprenant.
2- **Activités de réseautage social :** l'ajout des amis et éventuellement la suppression et le blocage des utilisateurs, l'écriture des commentaires sur les profils des autres utilisateurs et la recherche des aidants dans le réseau social.

**d. Critères de comparaison :** les sous-classes que regroupe cette classe servent non seulement à aider l'apprenant à connaitre son niveau et à aider le système à connaitre ses points forts et ses faiblesses afin de l'aider, mais servent aussi dans le processus de recherche des meilleurs aidants selon les critères définis par le demandeur d'aide. Par exemple, les agents du système peuvent lui conseiller de rejoindre un cours de soutien en ligne fait pour les apprenants faibles en mathématiques.

- **Auto-évaluations :** l'ensemble des résultats des autotests effectués par l'apprenant pour chaque matière.
- **Les évaluations :** l'ensemble des évaluations données par les enseignants et les apprenants, tels que son niveau dans une matière et sa bonne conduite (helpfulness). A chaque fin d'une aide synchrone, l'utilisateur est invité à évaluer le fournisseur d'aide ou le demandeur d'aide.
- **Niveau de connaissance dans un domaine** : l'utilisateur peut dire son niveau de connaissance dans un domaine ou une matière particulière. L'utilisateur peut choisir trois niveaux : élevé, intermédiaire et faible. Cela nous aidera à intégrer rapidement les nouveaux apprenants (qui n'ont pas encore effectué beaucoup de tests d'évaluation de connaissances) dans le processus de recherche des « aidants».
- **Préférences d'aide :** ensemble d'attributs qui définissent combien de fois l'utilisateur souhaite être contacté pour aider et pour combien de temps, les heures de ses disponibilités, le nombre de personnes il peut aider en même temps, les matières dans lesquelles il peut aider.

**e. Ses intérêts** : qui peuvent varier entre musique, sport, culture…etc.

**f. Certification et diplôme :** regroupe l'ensemble de certifications et de diplômes que possède l'enseignant ou l'apprenant.

**g. Les publications :** la liste des publications de l'utilisateur dans le réseau social.

**h. Sécurité :** contient le mot de passe de l'apprenant, les clés ainsi que l'ensemble des contrôles d'accès pour chaque type de données.

## II-les classes de connexion

Ces classes catégorisent les utilisateurs faisant partie de la liste d'amis de l'utilisateur. Une connexion peut être symétrique « amitié » (comme sur Facebook) ou asymétrique « abonnement » (comme sur Tweeter et Facebook). Par défaut, notre système définit les classes de connexion suivantes :

**a. Amis :** un « Ami » peut être un vrai ami, un membre de la famille, une connaissance, un ami d'un ami, ou même quelqu'un que l'utilisateur n'a jamais rencontré avant, sauf en ligne.

**b. Camarades :** Ses amis ou ses camarades de la même classe ou de la même école

c. **Aidants préférés :** Liste des enseignants et des apprenants que l'utilisateur souhaite contacter en priorité quand il a besoin d'aide. Cette liste aide les agents du système à choisir le meilleur « aidant » quand l'apprenant demande de l'aide.
d. **Enseignants :** Liste des enseignants que l'utilisateur connait dans la vie réelle.
e. **Famille :** Les membres de la famille, tels que ses parents, ses frères, ses cousins…etc.

De plus, le système gère automatiquement trois autres listes :

f. **Abonnés :** Un apprenant peut s'abonner sur le profil de son enseignant (si l'enseignant autorise les abonnements sur son profil) sans demander d'être son ami.
g. **Bloqués :** Cette liste contient l'ensemble des utilisateurs bloqués qui ne peuvent pas contacter l'apprenant ou être dans la liste de suggestion des aidants après la formalisation de la demande d'aide.
h. **Supprimés :** Ce sont les personnes que l'utilisateur a enlevées de sa liste des connexions. Ces personnes ne peuvent pas être suggérées au demandeur d'aide.

## Conclusion

« ApprAide » est un réseau de pairs (P2P de nœuds d'élèves et d'enseignants) conçu pour aider les apprenants quand ils s'engagent dans des activités de résolution de problème ou de révision. « ApprAide » facilite la recherche et la localisation des pairs capables d'aider l'apprenant. Deux types d'aide existent dans notre système : Les demandes d'aide asynchrones, qui permettent au demandeur d'aide de recevoir plusieurs réponses à sa question, de créer un échange riche entre les membres, d'affronter leurs idées et de développer l'esprit d'entraide chez eux. Par contre, les demandes d'aide synchrones permettent à l'apprenant de bénéficier de la présence, en temps réel, d'un enseignant ou d'un apprenant capable de répondre à ses questions. Le système peut être utilisé à des fins professionnelles, éducatives ou personnelles. C'est un système qui essaye de profiter de la popularité et l'acceptation des services des réseaux sociaux pour l'apprentissage. Cependant, avec l'ajout de l'extension social à l'apprentissage où les interactions sociales sont inévitables, la quantité des informations personnelles gérées par le système deviennent importantes, augmentant ainsi les risques de violation de la vie privée des apprenants.

# Chapitre 5.

# Description de la sécurité dans « ApprAide »

# Introduction

La sécurité est un critère très important pour le bon fonctionnement des systèmes d'e-learning. Les mécanismes de sécurité sont un moyen de protection des contenus pédagogiques et des données des utilisateurs contre toute tentative de destruction ou de modification malveillante des données. Les systèmes d'e-learning collectent de grandes quantités d'informations sur les utilisateurs afin de personnaliser l'expérience d'apprentissage. Les données collectées ne sont pas seulement liées à l'apprentissage mais elles sont aussi de nature personnelle, ce qui pose un problème de vie privée. Les informations saisies sur les profils des utilisateurs dans les systèmes d'e-learning reflètent, généralement, la vraie identité de la personne (vu la nature professionnelle des profils), facilitant ainsi leur identification sur le Web. Avec l'intégration croissante des outils d'apprentissage et du travail collaboratif dans les systèmes d'e-learning, où les interactions entre les membres de la communauté d'apprentissage sont inévitables, les déviations sont fréquentes dans l'utilisation de ces outils pour des buts hors apprentissage (surtout pour les communautés à accès public). Ces interactions sociales, qui peuvent déborder, augmentent la quantité d'informations personnelles gérées par ces systèmes qui sont, des fois, incapables de protéger la vie privée de leurs utilisateurs.

Actuellement, la vente des données privées des utilisateurs aux sociétés publicitaires génère beaucoup de profits pour les fournisseurs des services en ligne, violant ainsi la vie privée de leurs utilisateurs. La crédibilité des fournisseurs, en particulier les services d'apprentissage, et leurs vraies intentions envers la protection de la vie privée des utilisateurs sont aujourd'hui mises en question. La confiance aveugle en système d'e-learning et la négligence de la sécurité et la vie privée dans les recherches des approches de « tracking », ont participé aux réussites des violations de la vie privée. Jusqu'à présent, les chercheurs hésitent d'affirmer l'existence de ces menaces à la vie privée dans les systèmes d'e-learning. **[May, 2011][Hage, 2011][Yong, 2011]**

Les efforts fournis par les militants pour protéger la vie sont arrivées à imposer des lois protégeant ce droit mais ces lois ne sont pas unies et ne sont pas appliquées dans le monde entier. De plus, les solutions des chercheurs dans le domaine de la sécurité informatique sont encore nouvelles et ne satisfont pas les exigences de la vie privée. Les exigences de protection de la vie privée sont contre les intérêts des fournisseurs, c'est ce qui explique leur résistance à ces solutions. En effet, il existe quelques-unes en littérature, tel que « Privacy by design » permettant de respecter la vie privée des utilisateurs lors de la conception des systèmes, mais l'application de leurs spécifications pour des services déjà mis en ligne demande un changement significatif de code, et delà, toutes les solutions sont restées essentiellement théoriques.

Dans ce chapitre, nous présentons notre propre solution pour ce problème tout en gardant un compromis acceptable entre les intérêts opposés des systèmes d'e-learning et la protection de la vie privée. L'idée est de permettre à l'utilisateur de protéger lui-même sa vie privée sans dépendre de la bonne volonté du fournisseur du service.

## 1. Identification des menaces

Notre système doit protéger la vie privée des utilisateurs contre les menaces suivantes :

### 1) Les accès non autorisés et l'espionnage

L'utilisateur peut être victime des accès non autorisés aux informations générales du profil et aux contenus qu'il publie sur le système par d'autres utilisateurs malveillants ou par des organismes pour l'espionnage.

### 2) La vente des données

La vente des données personnelles de l'utilisateur ou leur partage avec d'autres organismes d'e-learning, parties tierces ou communautés scientifiques sans le consentement explicite de l'utilisateur sont parmi les actes de violation de la vie privée les plus dangereux, car une fois les données divulguées, ni l'utilisateur ni le fournisseur du service n'auront le contrôle sur la façon dont ces informations seront exploitées. Afin de maintenir les services, les fournisseurs doivent gagner de l'argent par la publicité.

### 3) Profilage

Une étude faite par **[May, 2011]** a montré que les utilisateurs n'apprécient pas le « tracking » même quand ils reçoivent des contenus personnalisés en retour. L'étude a même montré que les utilisateurs des systèmes d'e-learning considèrent le « profiling » comme un grand danger sur leur vie privée. A vrai dire, le profilage est une arme à double tranchons. Il peut améliorer l'expérience d'apprentissage, comme il peut être un moyen d'espionnage. Par exemple, le profilage à des fins publicitaires impliquera nécessairement l'utilisation des données personnelles, par conséquent, le profilage sans le consentement de l'utilisateur devient une menace.

### 4) La ruine de la réputation

La réputation est l'évaluation sociale du public envers une personne, un groupe de personnes ou d'une organisation. Il est un facteur important dans de nombreux domaines, tels que les entreprises, les communautés en ligne ou les statuts sociaux**.** En effet, si la réputation de l'utilisateur en ligne est endommagée, elle peut affecter sa crédibilité dans la vie réelle.

### 5) Prédateurs en ligne

Les adolescents (surtout ceux âgés de moins de 14 ans) sont le groupe d'âge le plus exposé aux attaques en ligne. Ils sont exposés à un risque très élevé d'être approchés par les prédateurs en ligne ou d'être victimes de la cyber-intimidation (cyber bullying). Les prédateurs en ligne peuvent facilement localiser et communiquer avec les victimes sur le réseau à travers les informations personnelles sur leurs pages de profil que les utilisateurs laissent à accès public.

### 6) Harcèlement

Un phénomène très connu dans les réseaux sociaux entre les jeunes adolescents. L'intimidateur a l'occasion de calomnier, d'abuser ou de révéler des informations facilement sur les RS (par exemple: il peut publier les résultats scolaires ou bien des photos embarrassants de sa victime). L'intimidateur peut continuer à harceler sa victime même si cette dernière le bloque ou le supprime. Beaucoup de victimes se suicident à cause de l'harcèlement continu. Les intimidateurs sont généralement de la même école de la victime.

## 2. Objectifs des utilisateurs pouvant exposer l'apprenant aux menaces à la vie privée

Comme notre système est ouvert au public où n'importe quelle personne peut rejoindre les membres de notre réseau d'apprentissage, nous pensons que les utilisateurs vont rejoindre le système pour différents objectifs (autres que l'apprentissage). Les catégories d'objectifs ainsi les comportements associés à chaque type d'objectifs viennent de nos constats des comportements des utilisateurs dans les réseaux sociaux habituels. En effet, les comportements des personnes dans ce réseau dépendent essentiellement de leurs personnalités. Notre classification des objectifs n'est qu'une analyse anticipée des comportements. Cependant, ces constats peuvent être prouvés ou améliorés par des recherches futures plus avancées en psychologie et sociologie.

### 1) Pour l'apprentissage et l'entraide

Ce type d'utilisateurs ne se connecte que pour avoir un cours de soutien, demander de l'aide ou aider les autres, travailler avec son groupe ou suivre les discussions auxquelles il a participé. Malgré la participation active de ce type d'utilisateurs dans le réseau social d'apprentissage, l'usage du réseau est approprié. Ces utilisateurs ne risquent pas de publier des contenus qui peuvent ruiner leur réputation. Cependant, cette participation active peut attirer les prédateurs. Comme ils sont des gens actifs et sociables, les prédateurs peuvent facilement débuter des communications avec leurs victimes de ce type.

### 2) Pour suivre les activités de leurs pairs

Les utilisateurs de cette catégorie d'objectif sont des personnes qui rejoignent le réseau pour suivre les activités de leurs pairs en ligne, récupérer les cours publiés par leurs enseignants ou consulter les informations publiées par leurs camarades. Ils sont la catégorie la moins active du réseau. Ils participent rarement par leurs commentaires dans le réseau social et ne communiquent pas avec des gens qu'ils ne connaissent pas. Pour cette raison, ils sont le groupe d'utilisateurs le moins exposé pour être approchés par les prédateurs. Leur usage du système est approprié, cependant, l'inactivité de ce groupe ne les aide pas à profiter des avantages de la participation dans l'apprentissage social.

### 3) Pour des raisons personnelles

Ce type d'utilisateurs rejoint le réseau social pour être en contact avec ses amis, sa famille et ses connaissances dans la vie réelle. Il échange beaucoup de messages privés et de publications contenant des informations personnelles. Généralement, son profil est riche de vraies informations sur son identité réelle. La vente de son profil causera la divulgation de ses données privées et ses massages. L'usage des outils de communication par ce groupe d'utilisateurs n'est pas à 100% inapproprié mais cela les expose aux menaces d'espionnage et les conséquences qu'elles peuvent engendrer.

#### 4) Pour connaitre de nouvelles personnes

D'après notre expérience sur Linkedin, il existe beaucoup d'utilisateurs qui rejoignent la communauté pour rencontrer de nouvelles personnes et de se divertir et ne pas pour des raisons professionnelles. Ils partagent beaucoup d'informations personnelles sur leurs profils et ils envoient souvent des messages contenant leurs numéros de téléphone, adresse mail…etc . Par conséquent, leur participation active dans le réseau est déviée de son objectif principal qui est l'apprentissage vers le partage des informations personnelles qui peut les exposer à tous types de menaces à la vie privée surtout leur réputation. En effet, un jeune adolescent peut ne pas être conscient des conséquences d'un tel type de comportement.

Nous avons identifié les objectifs des utilisateurs afin de déduire les types de menaces auxquelles l'utilisateur est exposé. Cette identification nous permettra de comprendre le comportement de l'utilisateur, de l'assister pendant l'ajustement de ses paramètres d'accès et veiller sur sa sécurité sur le réseau.

## 3. Modèles de l'attaquant

L'attaquant dans notre système peut être :

- Un utilisateur malveillant qui essaie de violer la vie privée d'un autre utilisateur du réseau social.
- Un attaquant externe.
- nous même! (en tant que fournisseur du service).

Dans ce qui suit, nous présenterons notre modèle de protection de la vie privée. C'est un système qui protégerait les données privées contre les accès non autorisés venant des utilisateurs malveillants et le fournisseur du service. En effet, le système doit être conçu afin de ne pas permettre même aux fournisseurs de violer la vie privée de ses utilisateurs et de donner à l'utilisateur le contrôle total de ses données.

## 4. Architecture du système

La Figure 31 présente une vue d'ensemble de l'architecture du système « ApprAide». Le système est basé sur une architecture hybride (Centralisé-P2P) dans laquelle un serveur central est responsable sur le stockage des données publiques, tels que les publications à accès public partagées par les utilisateurs ou les cours et les autotests conçus par les enseignants. L'application cliente est installée sur l'ordinateur de l'utilisateur. Elle est responsable sur la gestion des échanges privés, P2P, entre les utilisateurs et le stockage des données privées.

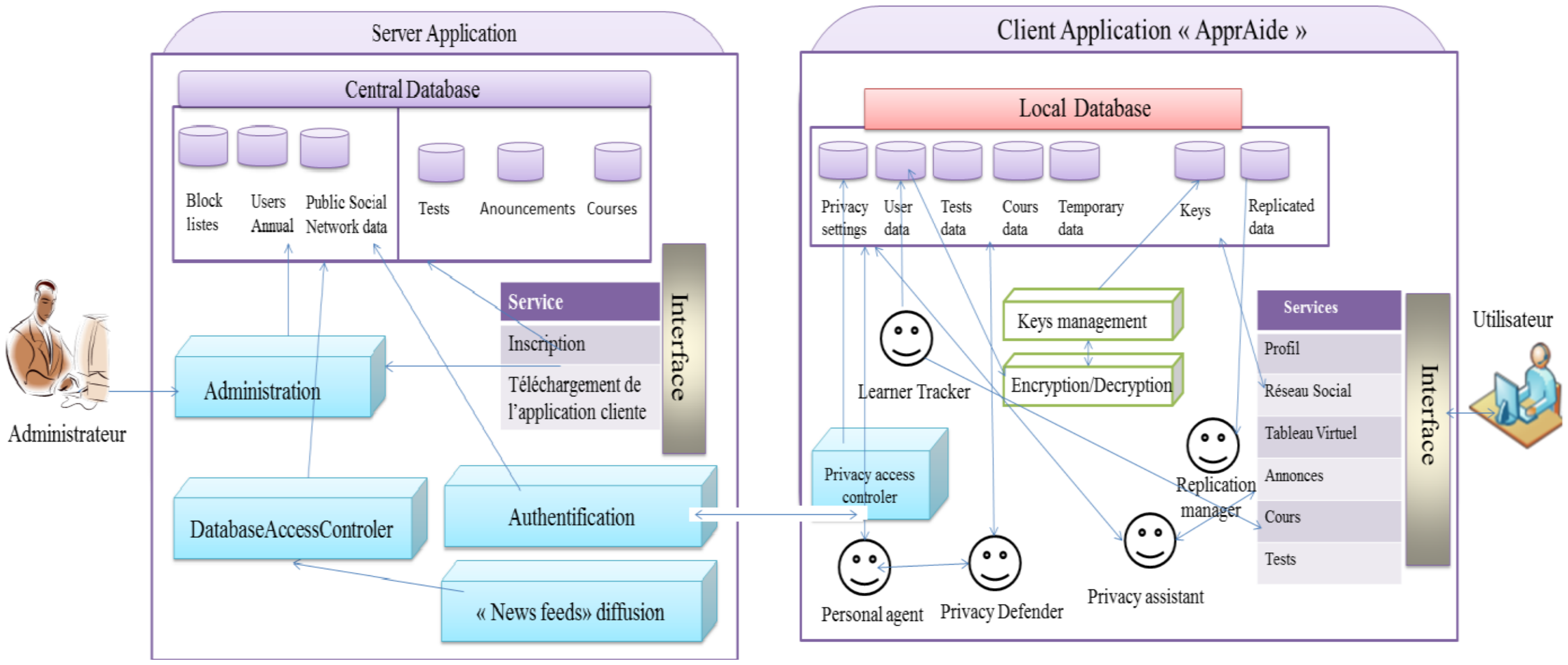

**Figure 31: Architecture du système « ApprAide »**

### 4.1 Application Serveur

L'application serveur est composée des modules suivants :

1- **Central Database** : la base de données stocke les informations d'inscription des utilisateurs ainsi que les contenus à accès public tels que les cours et les autotests conçus par les enseignants. Les informations et les demandes d'aide à accès public sont aussi sauvegardées sur le serveur afin de faciliter leur diffusion à tous les clients du système.

2- **Services** : ce module offre deux services ; un service d'inscription et un service de téléchargement de l'application cliente « ApprAide ». Pour pouvoir s'inscrire et utiliser l'application cliente, l'utilisateur doit remplir un formulaire d'inscription en ligne. Contrairement à l'inscription des apprenants, l'inscription des enseignants doit être validée par l'administrateur.

3- **Administration** : Le module « Administration » permet à l'administrateur de :
   - Valider ou refuser les dossiers des enseignants inscrits.
   - Supprimer des comptes.
   - Réviser les rapports de signalement d'abus.

4- **Authentification** : Ce module permet de gérer l'authentification des utilisateurs qui se connecte à partir de l'application cliente.

5- **« News feeds » diffusion** : ce module gère la diffusion des actualités aux applications clientes connectées.

6- **DatabaseAccessControler** : Ce module gère les demandes d'accès à la base de données.

## 4.2 Application cliente «ApprAide»

L'application « ApprAide» est chargée de fournir les services nécessaires pour l'apprentissage et pour faciliter les interactions sociales. L'application cliente est composée des modules « **Services** », « **Local Database** », « **Keys management** », « **Encryption/ Decryption»,** ainsi qu'un système multi-agents.

### 4.2.1 Les modules

1- **Services** : Ce module s'occupe de tous les services d'apprentissage.
2- **Local Database** : Ce module stocke toutes les informations privées liées à l'apprentissage ou les interactions de l'apprenant avec les autres membres du réseau.
3- **Keys management** : la génération et le stockage des clés de cryptage et de décryptage sont gérées par ce module.
4- **Encryption/Decryption** : Ce module gère le cryptage et le décryptage des données échangés.

### 4.2.2 Le système multi-agents dans l'application « ApprAide »

Chaque agent dans notre système a un rôle particulier **:**

**1- Personal agent :** L'AP prend la place de l'humain dans le système. Il exécute toutes les requêtes de l'utilisateur qui doivent passer par lui d'abord et qu'il retransmet ensuite aux autres agents et retourne la réponse à l'utilisateur. Il est responsable aussi des négociations entre agents pour le choix du meilleurs aideurs. Il est responsable sur la préparation de la liste des demandeurs d'aides qui lui semble correspondre aux critères définis par l'utilisateur.
**2- Learner tracker :** il trace toutes les activités de l'utilisateur : les outils qu'il a utilisé, les demandes d'aide, les demandes d'amitié, les accès aux tests. Ainsi, il sauvegarde tous ses résultats des tests et les évaluations.
**3- Privacy Assistant :** Il assiste l'utilisateur pendant le paramétrage de ces droits d'accès. Il peut apprendre ces préférences de protection de la vie privée à partir des paramètres définis dans le passé ou bien à partir de ses réponses aux questionnaires. Son rôle est de sensibiliser les apprenants de la plateforme aux risques sur leur vie privée et à la visibilité de leur information.
**4- Privacy Defender :** Son rôle est de détecter les violations des droits de distribution des contenus (quand l'utilisateur essaye de partager avec ses amis un contenu protégé par un droit interdisant sa distribution) ou bien à la réception des flux RSS, l'agent peut détecter les contenus que l'utilisateur n'a pas le droit de consulter. A chaque détection d'un contenu que l'utilisateur n'a pas le droit de le voir, il le rejette. Le fil d'actualité du réseau social est géré par cet agent. Aussi, il a pour rôle la suppression automatique (sans l'autorisation de l'utilisateur) des contenus dupliqués sur sa base appartenant à un autre utilisateur (si le propriétaire demande de les supprimer).
**5- Replication Manager :** L'agent responsable sur la sauvegarde de toutes les activités de réplication interne et externe.

# 5. Description de notre solution pour la protection de la vie privée des apprenants

## 5.1 La sécurité des publications partagées sur le réseau social

### 5.1.1 Les publications et les métadonnées

Dans notre système, chaque publication est composée d'un contenu et un ensemble de métadonnées. Pour chaque publication l'utilisateur doit définir les métadonnées associées à la publication. Les métadonnées que nous avons définies sont les suivantes :

1- **Types de la publication :** les publications peuvent être de 4 types : Demande d'aide, partage d'une information (connaissance ou une nouvelle), partage d'un document ou un statut.
2- **Matière :** s'il s'agit d'une demande d'aide (question) ou une connaissance, l'utilisateur doit préciser la matière à laquelle appartient la connaissance ou la question.
3- **Niveau :** l'utilisateur doit préciser le niveau d'étude (CEM, Lycée...etc.).
4- **Paramètres de confidentialité :**

L'utilisateur définit dans ce paramètre la liste des personnes avec lesquelles il souhaite partager sa publication. Les paramètres de partage définissent les droits d'accès et ils sont définis comme suit:

- **Partager avec moi-seulement :** seulement le propriétaire peut voir sa publication.
- **Partager avec une liste particulière :** Seuls les utilisateurs qui appartiennent à la liste précisée par l'utilisateur peuvent accéder et voir la publication.
- **Partager avec une liste de personnes** : L'utilisateur dans ce cas définit une liste de personnes (choisies individu par individu), qui peuvent appartenir à des listes différents, et qui peuvent voir la publication.
- **Partager avec tous mes amis** : Tous les utilisateurs qui font partie de la liste d'amis peuvent voir la publication. (à l'exception des abonnées qui ne peuvent suivre et voir que les publications à accès public de l'utilisateur)
- **Partager avec tout le monde « Public»**: Ce type de publication est destinée à être consultée par tous les utilisateurs du système.

5- **Droit de distribution** :

A cause de l'architecture décentralisée que nous avons choisie, l'utilisateur peut être victime d'un partage non autorisé de ses données par les utilisateurs qui peuvent visualiser ses données. L'utilisateur peut décider les droits de distribution de ses contenus. L'utilisateur peut définir deux droits de distribution :

- **Pas de distribution** : L'utilisateur interdit les contacts qui peuvent voir le contenu de le partager. L'option de partage est automatiquement désactivée.
- **Distribution**: l'utilisateur autorise que son contenu soit partagé avec seulement quelques utilisateurs (par exemple : l'enseignant veut que son contenu soit partagé qu'avec les membres d'un groupe particulier ou l'utilisateur veut que son contenu soit repartagé qu'avec la liste des personnes déjà définie dans les paramètres de « partager avec » (les amis en commun).

Par défaut, les publications à accès public sont autorisées à être redistribué avec n'importe quel groupe d'utilisateurs.

Les métadonnées sont exprimés en code XML et associés au contenu avant que le contenu soit dupliqué sur les machines des pairs. La figure 32 montre un exemple de la génération de ces métadonnées:

```
<OWNER  ID_Apprenant="5484" pseudonym= "Marwa"/>

< ID_TypeContenu="demande d'aide"/>

<Content_ID= 17/>

<Science> Mathématique <Science>

<Level>Lycée</Level>

<AUDIANCE name="Mes enseignants" type="Classe de Connexion"
Audiance_ID="CC2"/>

<RIGHTS>

<ACCESS/>

<RELPLICATION_PROTECTION> Crypté </ RELPLICATION_PROTECTION >

< distribution> Non < distribution/>

<DUPLICATION_AUTORISATION/>

</RIGHTS>
```

**Figure 32:Exemple d'attribution des droits**

En effet, les métadonnées que nous utilisons définissent des droits. Ces droits sont de trois types : droits d'accès, droits de distribution et droits de duplication. Les métadonnées sont associées à chaque publication et elles représentent un moyen lisible par les agents qui leur permet de protéger le propriétaire du contenu quand le contenu est dupliqué sur la machine d'un autre utilisateur. Le but est de permettre à l'utilisateur de définir la liste d'audience pour chaque type de publication. Les listes d'audience sont définies comme suit (Figure 33) :

```
<AUDIANCE name="Mes enseignants" type="Classe de Connexion"
Audiance_ID="CC2"/>

<LISTE>

<USER_ID> Enseignant_80</USER_ ID>

<USER_ID> Enseignant_1088</USER_ID>
```

| |
|---|
| <USER_ID> Enseignant_4852</USER_ID><br></LISTE><br></AUDIANCE> |
| **Figure 33: Exemple d'une liste d'audience** |

Lors de la duplication, le nom de la liste est supprimé, seulement les identifiants des utilisateurs de la liste sont gardés.

### 5.1.2 Sécurité des données dupliquées

L'inconvénient principal de l'approche P2P est la disponibilité faible des machines. Les machines sont sujettes à des défaillances, les redémarrages, les power-offs et les déconnexions du réseau. La disponibilité raisonnable nécessite la réplication des données. Toutefois, en raison de la nature personnelle des données, la réplication doit être faite avec prudence. La réplication des données peut être faite avec des données chiffrées sur les hôtes auxquels l'utilisateur ne fait pas confiance et il peut distribuer les clés de déchiffrement aux utilisateurs autorisés. La difficulté est que tous les utilisateurs autorisés doivent avoir la clé de décryptage. La deuxième solution est de répliquer le texte en clair sur des pairs de confiance chargé de protéger les accès à ses données et de ne pas les dévoiler (amis proches ou membres de la famille). Cela simplifie énormément la gestion des clés. Notre solution consiste en le chiffrement des données partagées avec des classes définies de connexions avant leur réplication (à l'exception du partage du contenu avec une liste de personnes appartenant à plusieurs classes de connexion où le contenu est dupliqué seulement en clair sur leurs bases). D'autre part, les données partagées à accès public sont répliquées en clair sur le serveur. Les contenus privés sont stockés en clair dans la base locale de l'apprenant et le module « Privacy Access Controler » se charge de protéger les données contre les accès non autorisés. L'algorithme suivant (Algorithme 3) résume ce que nous avons avancé:

| |
|---|
| Si contenu partager avec :<br>Moi-seulement-> Stocker le contenu en clair, sur la machine du propriétaire seulement et pas de duplication.<br>Une classe de connexions-> Dupliquer le contenu chiffré sur les machines des utilisateurs.<br>Liste de personnes appartenant à plusieurs classes de connexion-> Dupliquer en clair sur les machines des personnes précisées dans la liste seulement.<br>Public-> Dupliquer en clair sur le serveur et pas de duplication sur les machines des pairs. |
| **Algorithme 3: Les conditions de chiffrement et de duplication des contenus** |

Nous avons divisé le profil en deux types de classes « Contenu » et « Connexion » afin de bien gérer les accès et le chiffrement des contenus. L'algorithme de cryptage utilisé est le

« RSA ». La figure 34 illustre le renforcement des droits d'accès par le cryptage avant la réplication.

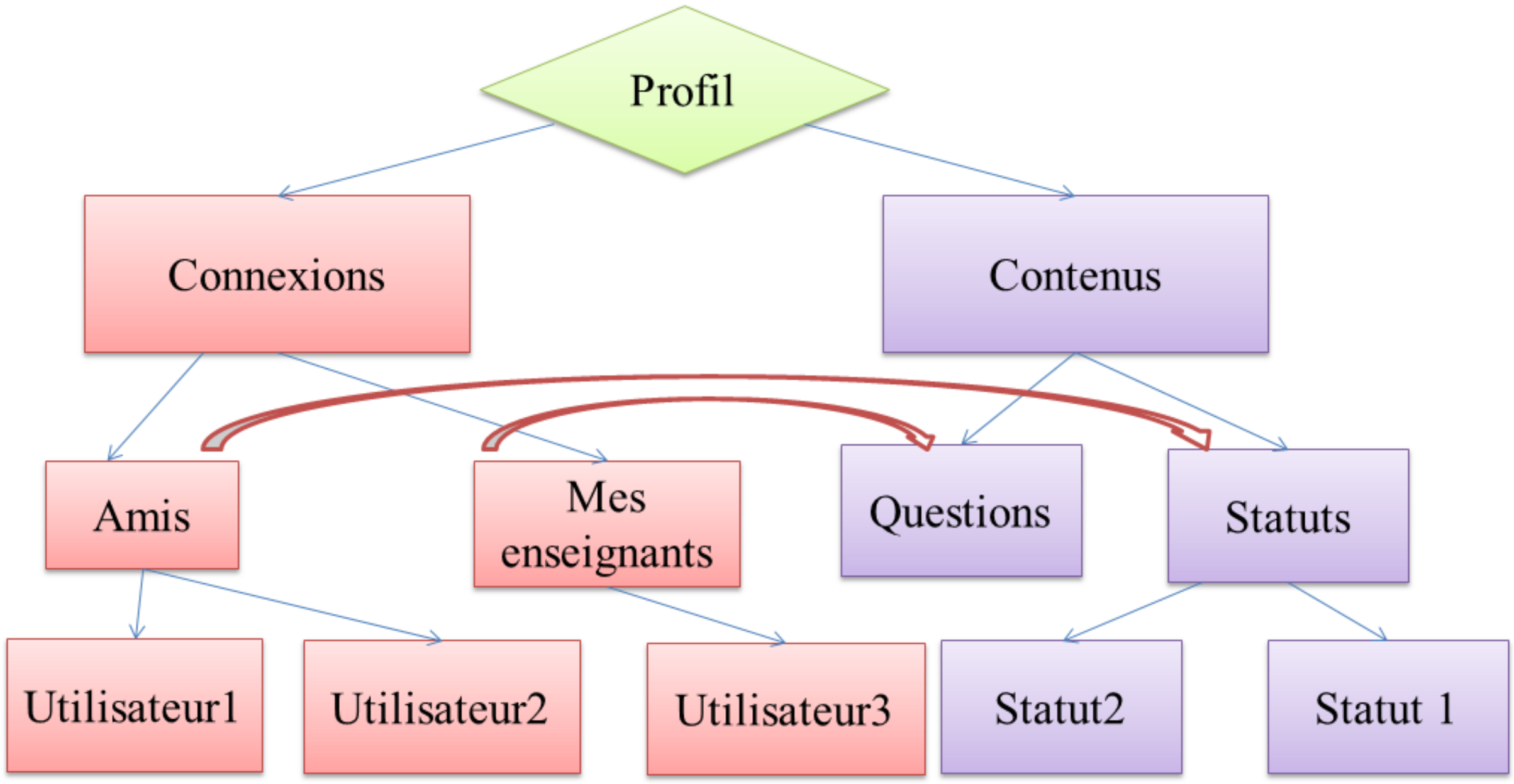

**Figure 34 : Renforcement des droits d'accès**

L'algorithme 4 décrit les étapes de duplication par le propriétaire de contenu et les membres de son réseau d'amis :

Accéder à un contenu ()

Id_type= Type (contenu)

Si accès à partir de la machine du propriétaire

- Récupérer « id_utilisateur » du demandeur d'accès.

- Vérifier la classe de connexion ou la liste de personnes associée au contenu.

Si « une classe de connexion »

Si (utilisateur_autorisé)

- Permettre l'accès.
- Dupliquer le contenu chiffré avec les métadonnées des droits

Sinon

Si « une liste de personnes »

Si (utilisateur_autorisé)

- Permettre l'accès.
- Dupliquer en clair sur sa machine avec les métadonnées des droits

Sinon -refuser l'accès Fsinon

Fsinon
FSI

Si accès à partir de la machine d'un pair (pas un pair de la liste de personnes)

- Récupérer « id_utilisateur » du demandeur d'accès.
- Récupérer ID de la classe de connexion.
- Vérifier la classe de connexion associée au contenu.

Si (utilisateur_autorisé)

- Permettre l'accès.
- Dupliquer le contenu chiffré avec les métadonnées des droits

Sinon

- Refuser l'accès.

Si ID de l'utilisateur fait partie de la table des pairs de duplication

- Dupliquer sur sa machine le contenu chiffré avec les métadonnées des droits

Fsin

Fsi

Si accès à partir de la machine d'un pair de confiance à un contenu partagé avec « la liste de personnes »

- Récupérer « id_utilisateur » du demandeur d'accès.
- Récupérer « id_contenu » du contenu
- Vérifier « la liste des personnes autorisées ».

Si (id_utilisateur) appartient à « la liste de personne »
-Permettre l'accès.
- **Dupliquer en clair** sur sa machine avec les métadonnées des droits.
Sinon
-refuser l'accès

Fsinon

FSI

**Algorithme 4: Le control d'accès et de duplication des contenus**

Chaque classe de connexion possède deux clés, une clé publique et une clé privée. Ces clés sont envoyées à tous les utilisateurs appartenant à la classe concernée. L'agent « Replication Manager » s'active pour chaque demande d'accès afin de déterminer le droit à la duplication et sauvegarder l'historique des duplications. L'agent veille aussi sur l'organisation de la duplication afin de ne pas envoyer des données déjà dupliquées sur la machine du pair. L'algorithme d'activités suivant (Figure 35) résume les actions prises après le paramétrage des droits d'accès et la publication du contenu par l'utilisateur.

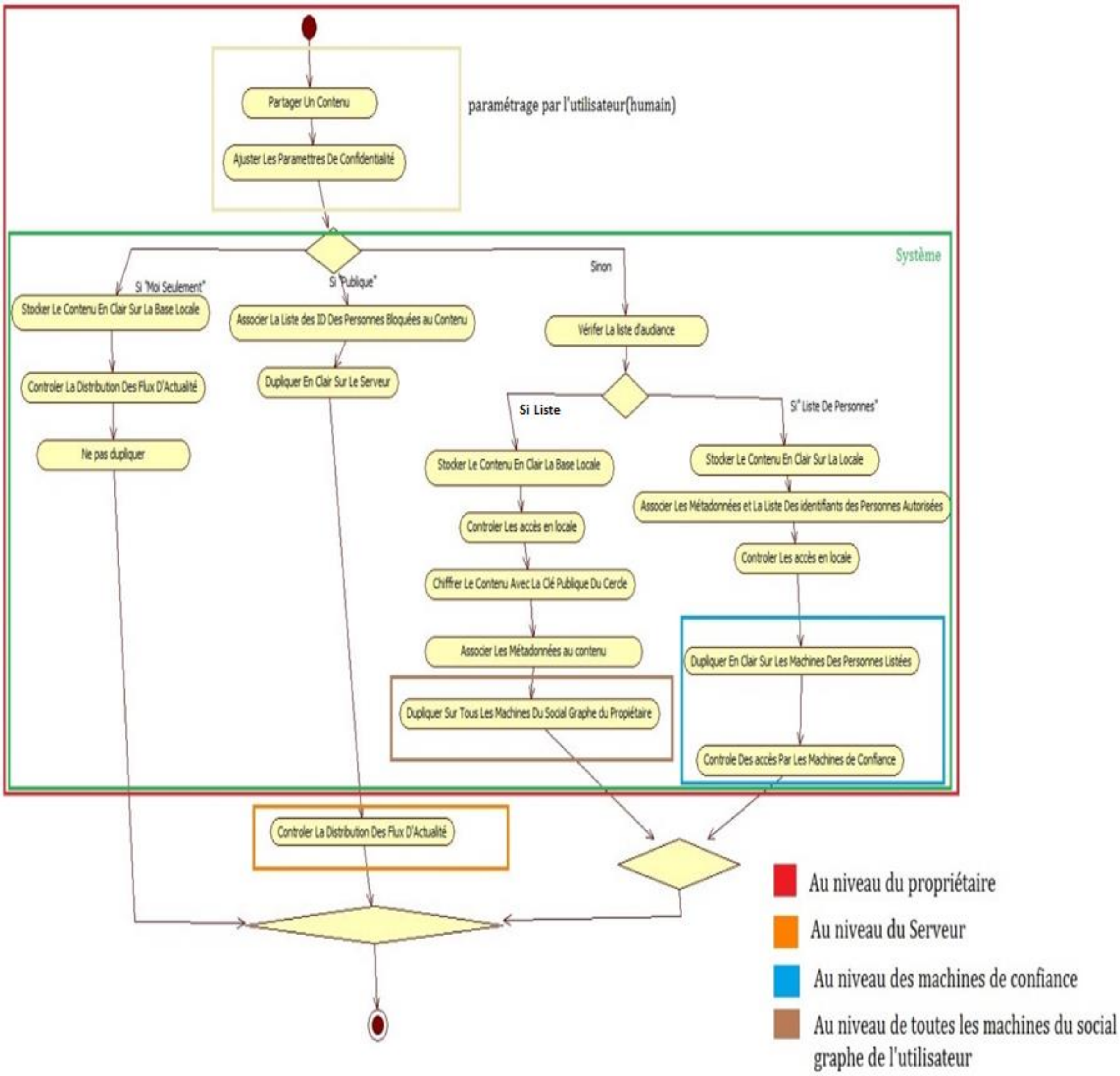

**Figure 35: Diagramme d'activité décrivant les actions prises lors de la duplication**

Le social graphe est le réseau d'amis de l'utilisateur. Les membres du réseau de l'utilisateur sont tous des pairs de duplication des données privées. Le serveur n'est sollicité que pour stocker les données publiques. Ce diagramme concerne seulement la réplication des publications partagées sur le réseau social. Les informations personnelles du profil sont stockées localement et elles ne sont pas dupliquées (à la manière des profils des utilisateurs du système d'exploitation Windows).

Le diagramme de séquence (Figure 36) montre les étapes de la duplication et la visualisation du contenu par l'utilisateur.

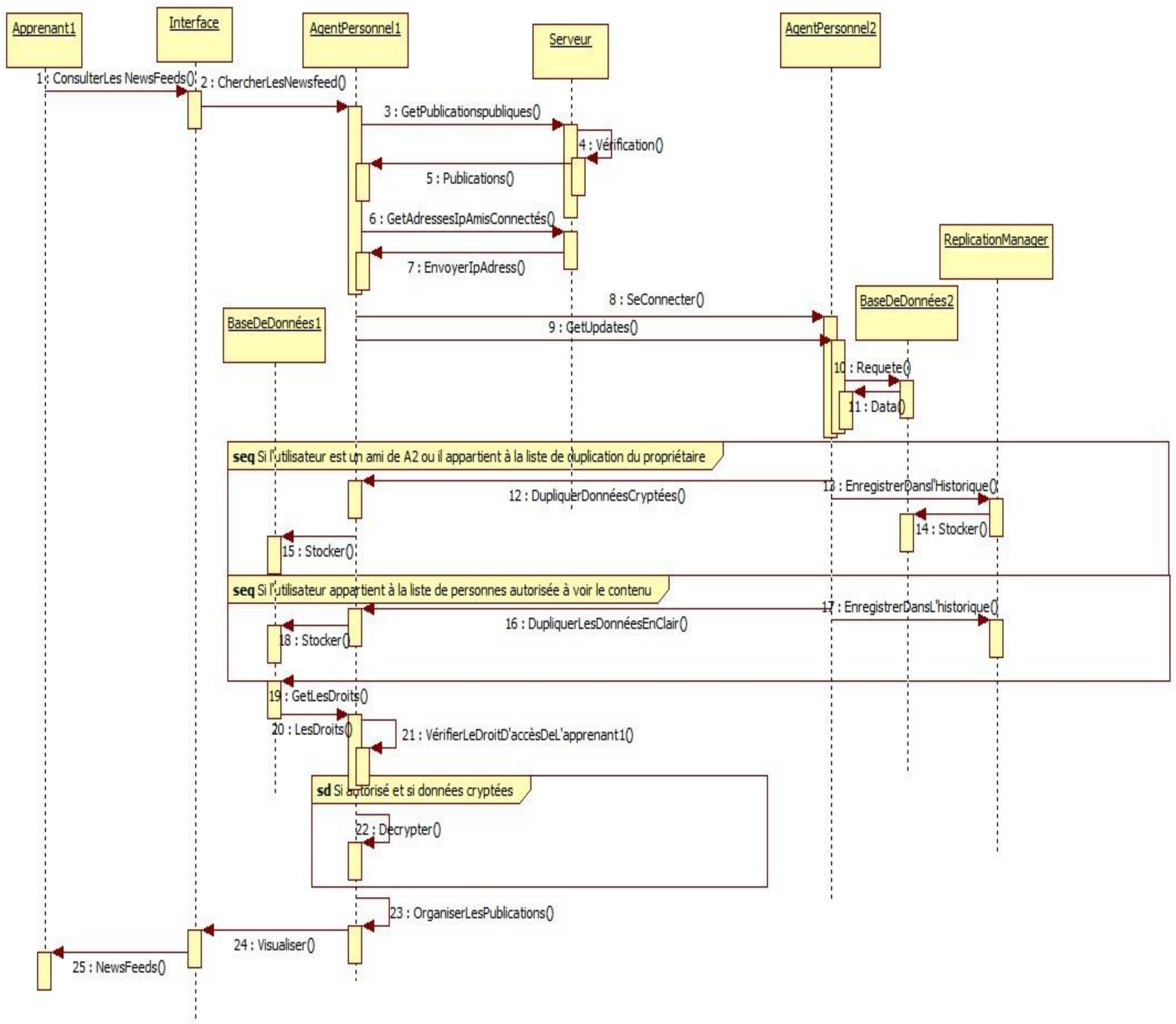

**Figure 36: Le diagramme de séquence de la duplication et la visualisation des données**

### 5.1.3 Evaluation de la proposition

Pour bien montrer la puissance de notre proposition, nous proposons les deux scénarios suivants :

**Scénario1 :**

Alice, à partir de son application « ApprAide », rédige et partage une publication qu'elle définit visible pour ses « Camarades » seulement. L'application associe les métadonnées correspondantes aux paramètres définis par Alice. Lors de la connexion de Bob, l'application duplique le contenu de la publication « chiffrée» par la clé publique de la classe « Mes camarades» dans la base locale de Bob. L'agent « Privacy Defender » de Bob vérifie les métadonnées, plus précisément, la liste d'audience définie par Alice (la propriétaire du contenu). Comme Bob fait partie des utilisateurs autorisés à voir le contenu, « Privacy Defender » décrypte le contenu par la clé secrète de la classe de connexions « Mes camarades » (la clé secrète est transmise à Bob par Alice lors de son affectation à la classe « Mes camarades ») et le visualise dans son fil d'actualité. Bob possède maintenant le contenu en clair et le partage de nouveau avec d'autres utilisateurs d'une classe quelconque de ses connexions. Le danger du partage du contenus par Bob (initialement publié par Alice) réside dans l'apparition du nom d'Alice aux amis de Bob (de la façon suivante : Bob **via Alice** : y-t-il quelqu'un qui a les cours de Mr. X, je n'ai rien appris des cours de Mme. Y). Bob veut aider Alice à trouver la personne qui peut l'aider et par bonne foi il rediffuse la publication d'Alice. Sans aucun moyen de protection des données d'Alice, des utilisateurs, qu'Alice n'a pas autorisé à voir son contenu, peuvent le visualiser. Et si Mme. Y est trop sensible et elle voit la publication d'Alice, cela pourrait créer des problèmes pour Alice. Afin d'éviter ce problème, quand Bob repartage le contenu d'Alice, l'application crypte le contenu d'Alice avec la clé publique de la classe de connexion que Bob a choisi. Cependant, l'agent « Privacy Defender » n'autorise pas tous les utilisateurs de la classe de connexions défini par Bob à accéder, pourtant ils possèdent la clé secrète de décryptage. En effet, seuls, les amis en commun entre la liste définie par Alice et la liste définie par Bob qui peuvent accéder, décrypter et visualiser le contenu.

**Scénario2 :**

Alice partage un contenu avec Bob et Alex seulement. Alex appartient à la classe de connexions « Famille » et Bob appartient à la classe « Mes camarades » d'Alice. A notre connaissance, en littérature, aucun travail ne propose une solution à ce cas car la génération des clés de chiffrement et de déchiffrement pour chaque cas rend la tâche de création et de gestion de nombreuses clés, devient extrêmement difficile. La plupart des travaux discutent l'attribution des niveaux de confiance à chaque classe de connexion afin de décider de crypter ou non le contenu, car ces solutions dépendent de la confiance existante entre les membres du réseau social pour protéger les données privées. En effet, nous n'utilisons pas des niveaux de confiance mais ce scénario est le seul qui utilise la notion de confiance. Nous pensons, que pour ce cas de scénario, Alice fait vraiment confiance en Alex et Bob, car elle les a choisis attentivement, un par un, parmi tous ses amis, pour voir le contenu. C'est le seul cas où la duplication des données se fait en clair. Cela facilite énormément la gestion des clés. Alex et Bob sont dans ce cas responsables de la protection de ce type de données. Cependant, nous mettons plus de restriction sur la duplication de ce type de données. En effet, seuls, Alex et Bob peuvent stocker ce contenu sur leurs machines.

### 5.1.4 Les paramètres d'accès par défaut

Pour bien protéger la vie privée des jeunes adolescents, l'agent « privacy assistant » règle les paramètres de confidentialité de l'utilisateur selon le tableau suivant :

| **Type de contenu** | **Moyen de protection** | **Paramètres** |
|---|---|---|
| **Identité** | Droits d'accès (Clair) | Famille |
| **Attributs démographiques** | Droits d'accès (Clair) | Famille |
| **Activités de réseautage social** | Droits d'accès + Chiffrement | Amis |
| **Activités liées à l’apprentissage** | Droits d’accès (Clair) | Moi-seulement |
| **Critères de comparaison** | Droits d’accès (Clair) | Moi-seulement |
| **Ses intérêts** | Droits d’accès + Chiffrement | Amis |
| **Les publications** | Droits d’accès + Chiffrement | Amis |
| **Certification et diplôme** | Droits d’accès + Chiffrement | Camarades et Famille |
| **Les connexions** | Droits d’accès (Clair) | Moi-seulement |

**Tableau 4: les paramètres de confidentialité par défaut**

## 5.2 Sécurité des échanges

Tous les messages échangés entre les utilisateurs ou les utilisateurs et le serveur sont signés afin de garantir l’intégrité des messages (garantir que le contenu n'a pas été altéré entre l'instant où l'auteur l'a signé et le moment où le lecteur le consulte) et l’authentification de l’utilisateur (valider l'authenticité de l'utilisateur en question).

**1) L’envoi des messages** :

- Alice rédige le contenu.
- Elle prépare un condensat du contenu.
- Elle crypte le condensat avec sa clé privée.
- Elle concatène le contenu en clair et la signature, elle obtient le message signé.

Le message signé est ensuite crypté avec la clé publique du récepteur pour garantir la confidentialité du message à transmettre.

**2) A la réception du message** :

- Bob décrypte le message avec sa clé privée et obtient le message en clair avec la signature.
- Il calcule le condensat du message en clair.
- Il décrypte la signature avec la clé publique d’Alice.
- Il compare le condensat qu’il a obtenu avec celui d’Alice.

S’ils sont égaux, le message est alors authentifié est sauvegardé dans la base de l’utilisateur. Si non, le message est rejeté.

## 5.3 Droit à l’oubli

Afin d’assurer le droit à l’oubli ou bien le droit à la suppression des données, quand l’utilisateur le souhaite, nous proposons l’algorithme (Algorithme 5) suivant.

**Au niveau du propriétaire du contenu** :

Si (Demande de suppression d’un seul contenu) :

- Récupérer identifiant du contenu id_contenu.
- Envoyer « id_contenu + liste des connexions+ demande suppression » à l’agent personnel.
- Ajout de la signature et id de l’utilisateur au message de destruction du contenu.
  « Signature+id_utilisateur+id_contenu + demande suppression »

Pour chaque pair « connecté » :

- Envoie de la demande de suppression à tous les « privacy defender » des pairs.

FPour

**Au niveau du pair**

- Vérification de la signature

Si la signature est valide

- Suppression du contenu par l’agent « privacy defender ».
- Envoie d’un message de confirmation de suppression signé à l’agent personnel du propriétaire.

Fpour

**Au niveau du propriétaire du contenu** :

Pour chaque réception d’une confirmation de suppression

- Accuser la réception de la confirmation de suppression.

Fpour

- Construire une liste des pairs (déconnectés) qui n’ont pas reçu la demande.

Si liste des pairs « non connecté » != NULL

- Ajouter « liste des pairs non connectés » à la demande de destruction du contenu.
- Envoyer la demande au serveur.

**Au niveau du Serveur** :

Pour chaque connexion d’un utilisateur de la liste :

- Envoyer la demande de suppression à son « Privacy Defender».
- Enregistrer le message de confirmation de suppression signé par le récepteur.

Si Propriétaire Connecté :

- Envoyer les messages de confirmations reçues à l’agent personnel.

FSi

Sinon

Si (demande de suppression d'un compte utilisateur)

- Envoyer « id_user + liste des connexions+ demande suppression + signature de l'utilisateur » au serveur.
- Confirmer la suppression et déconnexion.

**Au niveau du Serveur** :

- Supprimer les données de l'utilisateur du serveur.

Pour chaque connexion d'un utilisateur de la liste :

- Envoyer la demande de suppression à son « Privacy Defender ».
- Enregistrer le message de confirmation de suppression signé par le récepteur.

Fpour

Fsinon

**Algorithme 5: Algorithme du droit à la suppression**

Du côté du serveur : le droit à l'oubli est un engagement du fournisseur envers ses utilisateurs. En effet, le droit à l'oubli n'est que de simples lignes de code de suppression des données sur le serveur. Un exemple d'un organisme qui applique ce droit est le site de l'université de **Nancy [Univ, 2014].** Cela est mentionné sur leur site car cette université participe avec ses équipes de recherche dans des projets de protection de la vie privée tel que le projet « CAPPRIS » (Collaborative Action on the Protection of Privacy Rights in the Information Society). **[CAPPRIS, 2014]**

## 5.4 Protection des résultats d'autotests

E-testing **[Hage, 2011]** permet d'évaluer les connaissances de l'apprenant après avoir effectué une activité d'apprentissage. E-testing offre de nombreux avantages comme la correction rapide et automatique des tests et les feedback. IMS QTI (Question and Test Interoperability) **[IMS, 2006]** fixe une liste de spécifications pour permettre l'échange des informations liées à l'examen, tels que les questions, les tests et les résultats. QTI nomme les tests « assesments », les questions et leurs informations respectives comme des éléments « items » et un groupe d'éléments dans un test en tant que « section ». Une section peut contenir plusieurs sections (Voir la figure 37):

| | |
|---|---|
| 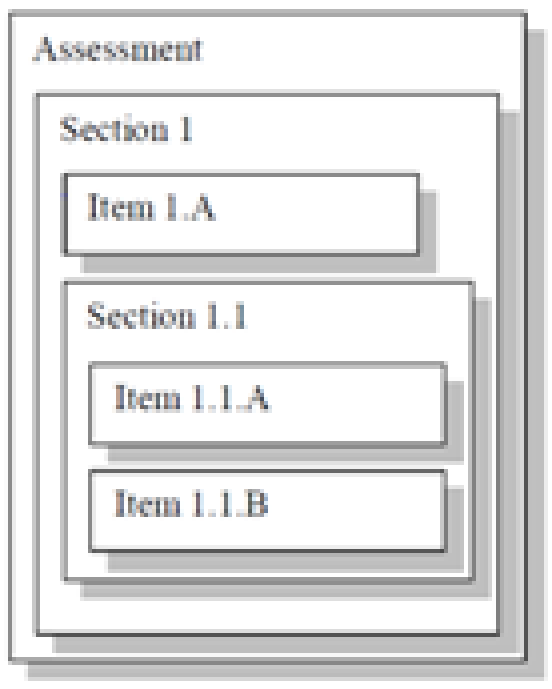  **Figure 37: Exemple d’un test [Hage, 2011]** | 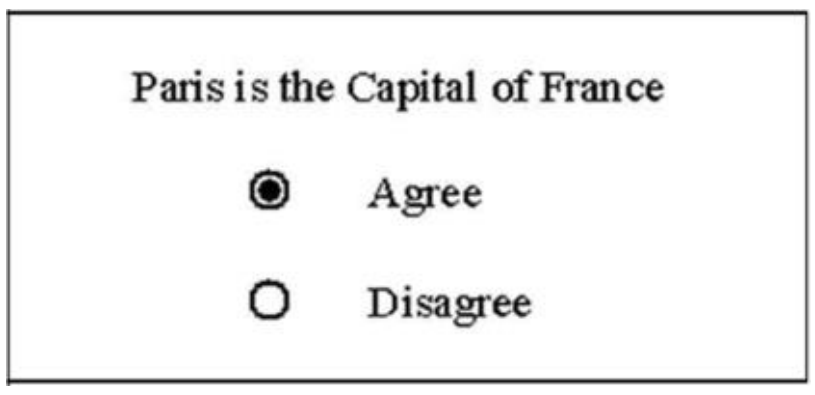  **Figure 38: Test de type vrai ou faux [Hage, 2011]** |

La figure 38 représente un exemple simple d’un test vrai ou faux. La figure 39 présente le code XML correspondant à la question de la figure 38 selon la norme IMS QTI.

```
<questestinterop>
<qticomment>
This is a simple True/False multiple-choice example using V1.2. The rendering is a
standard radio button style. Response processing is incorporated.
</qticomment>
<item ident="IMS_V01_I_BasicExample001">
        <presentation label="BasicExample001">
             <flow>
                   <material> <mattext>Paris is the Capital of France</mattext> </material>
                       <response_lid ident="TF01" rcardinality="Single" rtiming="No">
                           <render_choice>
                             <flow_label>
                                 <response_label ident="T">
                                      <material><mattext>Agree</mattext></material>
                                 </response_label>
                                 <response_label ident="F">
                                       <material><mattext>Disagree</mattext></material>
                                  </response_label>
                            </flow_label>
                           </render_choice>
                      </response_lid>
```

```
            </flow>
        </presentation>
</item>
</questestinterop>
```

**Figure 39: Code XML de la question de la figure 38 selon le standard IMS QTI [Hage, 2011]**

En effet, nous avons présenté cet outil pour deux raisons :

1) les résultats des autotests sont à caractère personnel et souvent les utilisateurs souhaitent les garder privés. L'analyse des comportements des apprenants pendant le test ainsi que les résultats obtenus aident à détecter les problèmes que l'apprenant rencontre pendant la résolution des exercices du test. La protection de la confidentialité des résultats de l'apprenant est obligatoire. En effet, quand l'utilisateur souhaite faire un test, le document XML contenant le test est téléchargé du serveur par l'application desktop. L'exécution du test ainsi que l'affichage et le stockage des résultats se fait localement, seul l'apprenant connait le résultat du test qu'il a effectué. L'agent personnel est capable d'analyser, comprendre et conseiller l'apprenant (du moment qu'il possède toutes les données sur l'utilisateur) sans divulguer aucune information sur l'utilisateur.
2) Dans les travaux de Greer **[Greer, 2002]**, les résultats des autotests effectués par les apprenants ont été utilisés comme un critère de recherche des meilleurs aidants dans le réseau.

## 5.5 Respect de la vie privée des utilisateurs lors de la recherche de meilleurs aidants

La recherche de meilleurs aidants concerne les demandes d'aide synchrones. Ce type de demandes utilise le tableau virtuel et l'outil de chat de notre système. Une demande d'aide synchrone peut être une demande de cours de soutien en ligne avec un enseignant du système ou une recherche d'un bénévole pour aider l'apprenant. Deux types de cours de soutien sont à distinguer : des cours de soutien individuels et des cours de soutien par groupe. Les cours de soutien en groupe sont annoncés dans la partie « annonces » du système afin de diffuser l'annonce à tous les utilisateurs du système. L'enseignant doit remplir un formulaire décrivant son cours comme suit (Figure 40) :

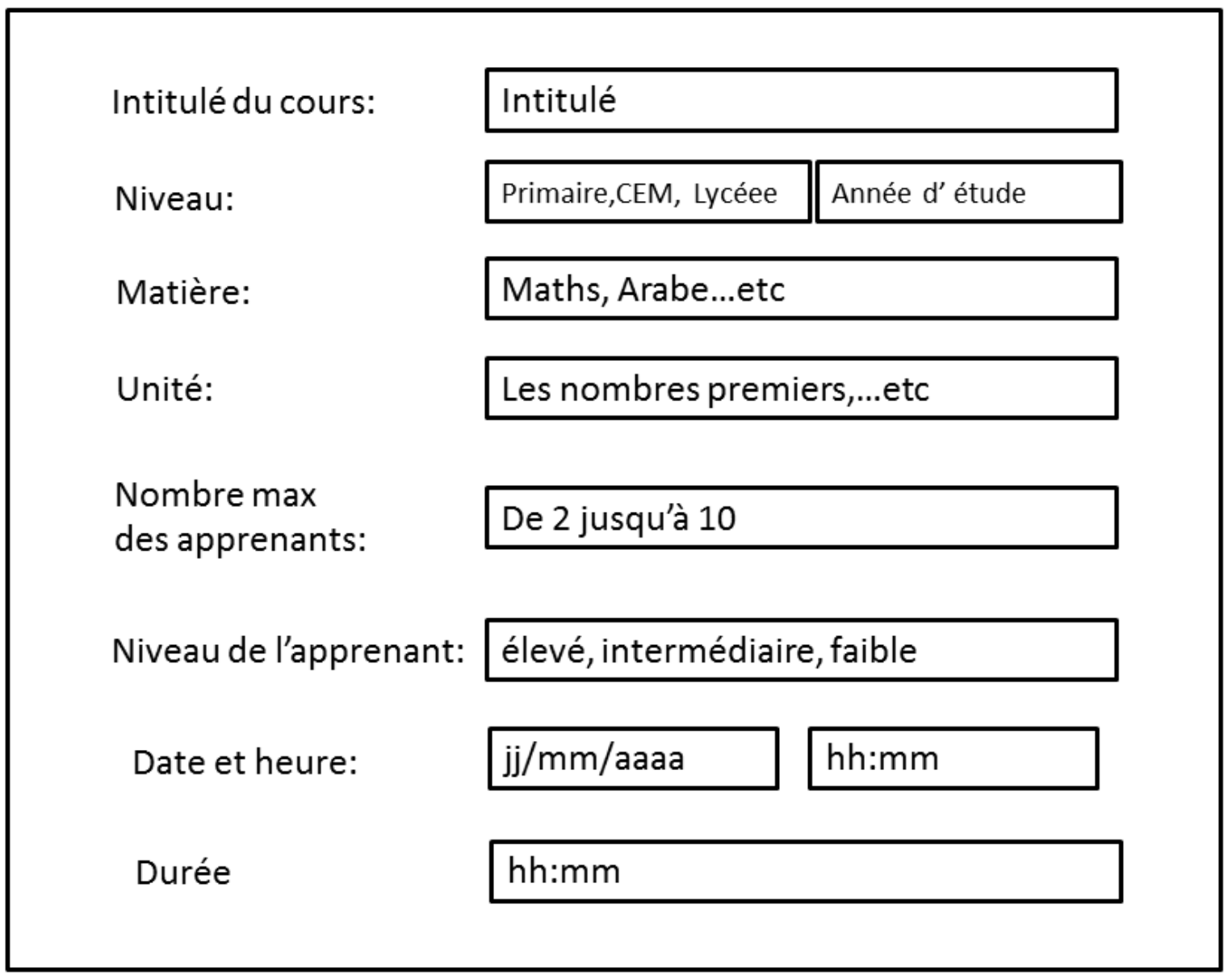

**Figure 40: Formulaire d'annonce du cours de soutien**

Ces annonces sont envoyées en format XML aux utilisateurs. A la réception des annonces, l'agent personnel trie les annonces selon les besoins de l'apprenant. Si l'apprenant est intéressé par une annonce particulière, il pourra rejoindre le cours en ligne. La recherche d'un cours de soutien individuel se fait de la même façon quand l'apprenant cherche « un aidant », qui peut être un enseignant ou un apprenant, pour une aide synchrone. Deux types d'aide existent, une aide gratuite (un apprenant ou un enseignant bénévole) et une aide payante (cours de soutien individuel ou en groupe avec un enseignant Freelancer). Cependant, le processus de recherche d'aidants est le même. La recherche d'un aidant se fait en 3 étapes :

### 1) Description de la demande par le demandeur d'aide :

L'aidé doit préciser dans sa demande

- Le niveau (Primaire, CEM, Lycée), l'année d'étude ($1^{ère}$, $2^{ème}$…etc), matière (maths…etc), et le chapitre.
- Le grade de l'aidant : enseignant ou apprenant.
- Si le grade choisi est celui d'un enseignant : bénévole ou Freelancer
- Niveau de l'aidant dans la matière précisée (élevé ou intermédiaire).
- La durée de l'aide.
- La description de l'aide si l'utilisateur le souhaite pour donner plus de détails.

L'agent personnel commence d'abord la recherche par l'envoi de la demande aux membres du réseau social de l'utilisateur. Au cas où aucune réponse n'est reçue, il envoie la demande au serveur pour la diffuser à tous les utilisateurs connectés au système.

2) **Réponses à la demande** :

A la réception de la demande d'aide par l'utilisateur, l'agent personnel vérifie les spécifications et leurs correspondances aux critères définis par le fournisseur d'aide. En effet, l'utilisateur définit la liste de spécifications suivante:

- Le niveau d'étude des personnes qu'il peut aider (du primaire…etc.).
- Les matières qu'il maitrise (Arabe, Français…etc.).
- La durée d'aide.
- Le nombre de personnes qu'il souhaite aider pendant sa connexion.
- Type d'aidant : bénévole (apprenant ou enseignant) ou Freelancer (Seulement les enseignants).

L'agent personnel vérifie et compare les spécifications du demandeur et du fournisseur d'aide. Si le profil du fournisseur d'aide correspond aux spécifications du demandeur d'aide, l'agent proposera à l'utilisateur une liste triée des demandes d'aide (S'il y'en a plusieurs, l'agent trie les demandes la plus correspondante aux profils du fournisseur jusqu'à la moins correspondante). Le choix final est laissé aux fournisseurs d'aide. La réponse du fournisseur d'aide est alors envoyée à l'agent personnel du demandeur d'aide.

3) **Réception des offres d'aide** :

A la réception de l'offre(s), l'AP vérifie la liste des utilisateurs bloqués ou supprimés par le demandeur d'aide. Après l'élimination de ces utilisateurs, l'AP vérifie pour chaque fournisseur d'aide, s'il y a eu des fréquentations dans le passé entre le fournisseur et le demandeur d'aide. Si c'est le cas, l'AP vérifie les évaluations du demandeur d'aide au fournisseur. Selon les évaluations, l'AP trie les offres et les affiche au demandeur d'aide. Ce dernier choisit une des offres affichées dans la liste. A la fin de l'aide synchrone, l'agent pose deux questions au demandeur d'aide (Figure 41)

Comment avez-vous trouvé son aide ?!

☐ Très utile ☐ Utile ☐ Pas du tout utile

Voulez-vous recevoir son aide une autre fois dans le futur?

☐ Oui ☐ Non

**Figure 41: Questionnaire du demandeur d'aide**

De même, à la fin de l'aide synchrone, le fournisseur d'aide est aussi invité à évaluer le demandeur d'aide :

Comment avez-vous trouvé son niveau ?!

❑ Excellent ❑ Bon ❑ Faible

Voulez-vous recevoir ses demandes d'aide une autre fois dans le futur ?

❑ Oui ❑ Non

**Figure 42: Questionnaire du fournisseur d'aide**

A la fin de ce processus de recherche, le demandeur choisi une des offres est la connexion « peer-to-peer » entre les deux utilisateurs s'établie. A la fin de la session, les évaluations sont échangées entre l'aidé et l'aidant. La figure 43 présente le diagramme des activités du demandeur d'aide et la figure 44 présentent le diagramme des activités du fournisseur d'aide :

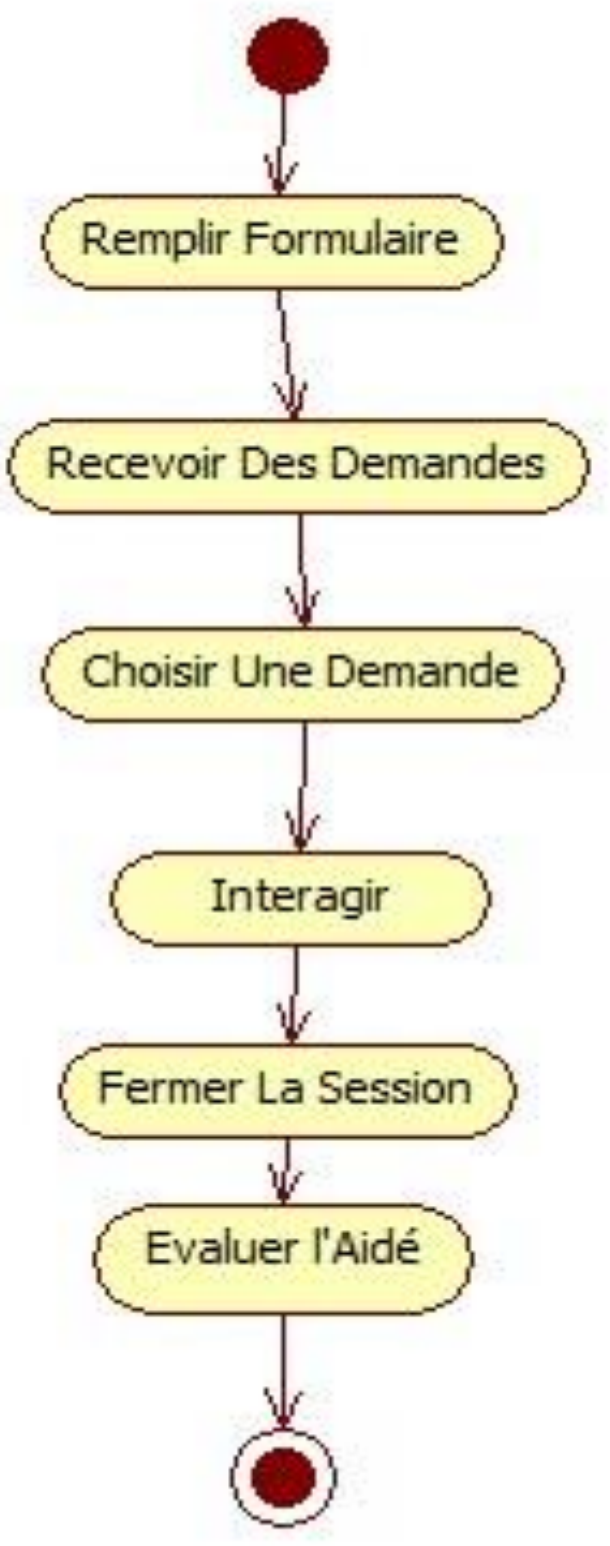

**Figure 43: Diagramme d'activité du demandeur d'aide.**

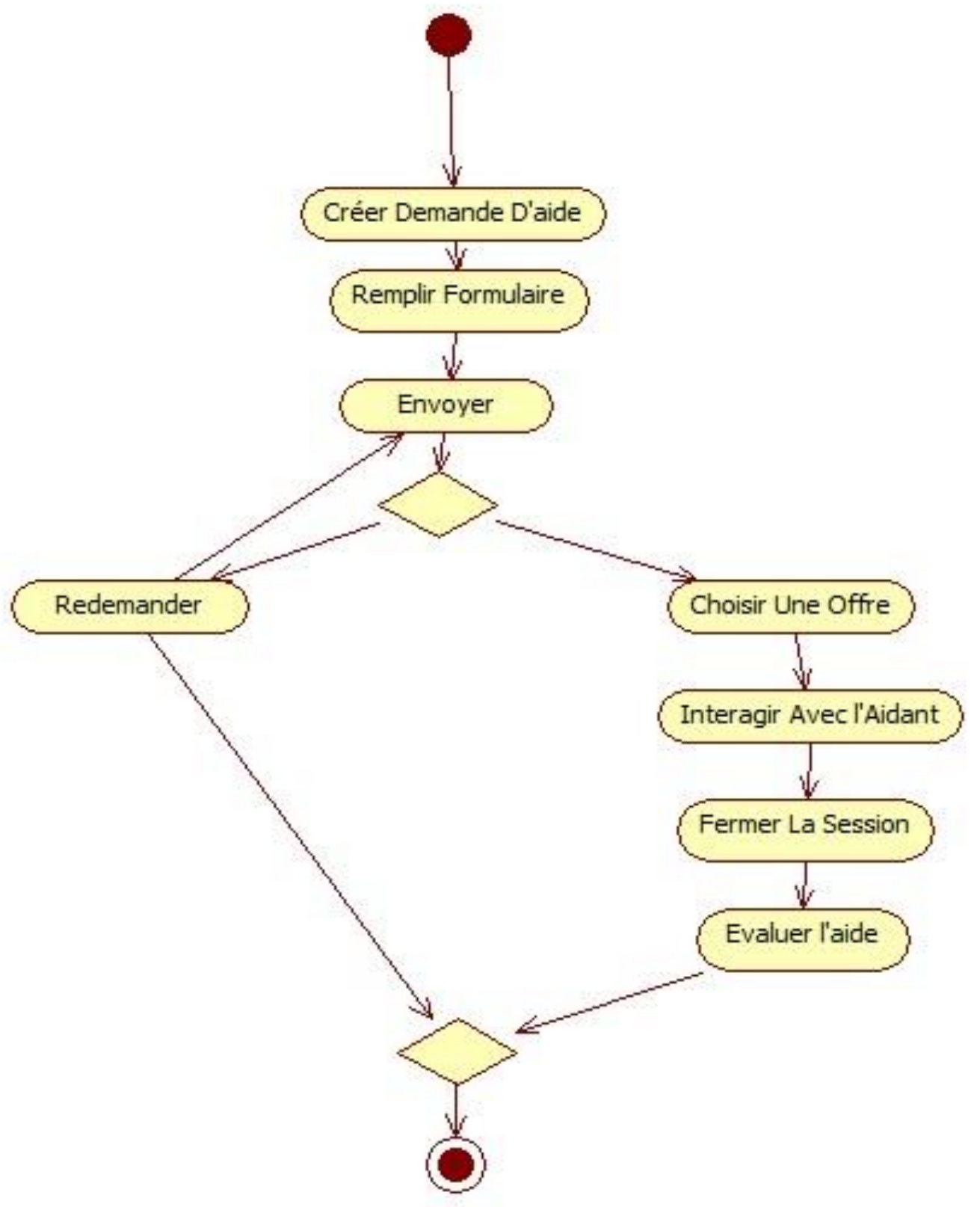

**Figure 44: Diagramme des activités du fournisseur d'aide.**

Le diagramme de séquence suivant (Figure 45) présente le processus de recherche de l'aidant :

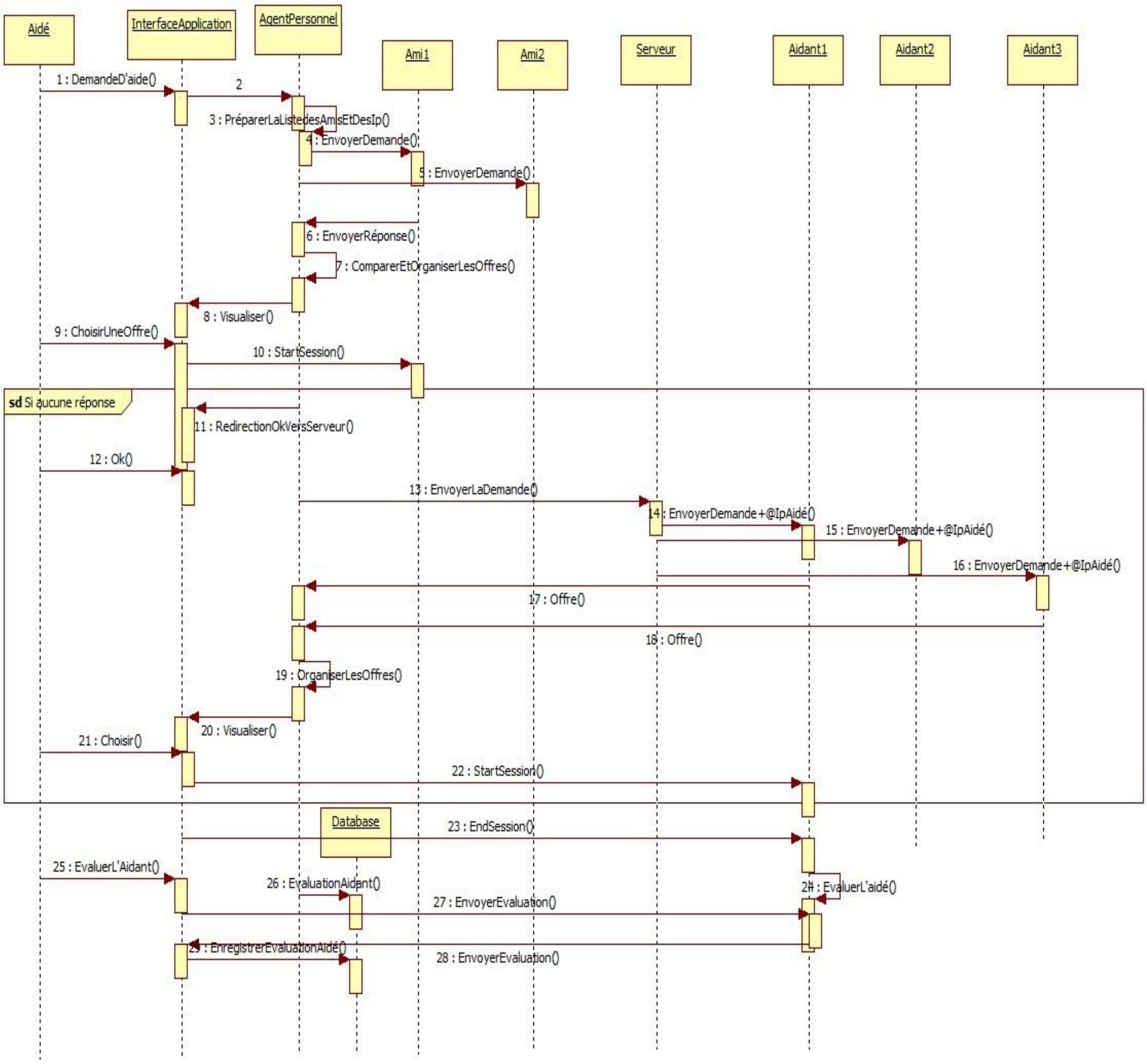

**Figure 45: Diagramme de séquence du processus de recherche d'aidant**

Cette solution ne viole pas la vie privée des utilisateurs car la réponse du fournisseur d'aide assure seulement que la personne possède un bon niveau dans la matière précisée. Cependant, en aucun cas, elle divulgue les résultats de l'utilisateur dans les autotests ou les évaluations qu'il a reçues par les autres pairs. La réponse confirme seulement qu'il possède les spécifications précisées. Si l'utilisateur ne répond pas sur une demande cela ne veut pas dire qu'on peut déduire qu'il ne possède pas les spécifications demandées ou qu'il a un niveau faible dans la matière car il se peut que l'utilisateur n'ait seulement pas choisi d'aider le demandeur d'aide.

### 5.6 Protection contre le harcèlement et gestion de la réputation des membres

En cas de signalement d'un abus par l'utilisateur, l'agent personnel de la victime filtre et cache immédiatement toutes les publications et les commentaires de l'utilisateur signalé. L'AP prépare un rapport d'abus pour l'envoyer à l'administrateur. Le rapport contient : le contenu signalé, l'identifiant de l'utilisateur signalé et une catégorisation de la menace. De même, quand l'utilisateur bloque un autre, il est demandé de donner la raison pour laquelle il a bloqué l'utilisateur. Les informations sur le blocage de l'utilisateur sont aussi envoyées aux serveurs afin de les analyser, catégoriser les utilisateurs et avoir une idée sur leur réputation. Quatre catégories de réputation des utilisateurs existent (Tableau 5):

| Id_utilisateur | prédateur | Intimidateur | Spam (Bloqué sans être signalé) | Total des signalements | Nombre des visites par « Privacy Assistant » | Révisé | Décision |
|---|---|---|---|---|---|---|---|
| Apprenant_898 | 0 | 4 | 0 | 4 | 15 | Vrai | Suspendu |
| Enseignant_204 | 3 | 0 | 6 | 9 | 58 | Vrai | Suspendu |
| Enseignant_258 | 0 | 0 | 0 | 0 | 0 | Faux | Aucune |
| Apprenant_269 | 0 | 3 | 0 | 1 | 6 | Vrai | 2 Fausses déclarations |

**Tableau 5: Gestion de la réputation des utilisateurs**

Comme les statistiques ne sont pas disponibles, nous utilisons des valeurs que nous définissons mais qui peuvent être améliorés dans le futur. Si l'utilisateur est signalé plus de trois fois par 3 utilisateurs différents. L'administrateur est sollicité pour réviser les rapports envoyés et prendre une décision. Si l'administrateur trouve que le contenu signalé n'est pas dangereux, il peut soustraire le nombre des fausses déclarations. Si le nombre de signalement dépasse 5 de 5 personnes différentes avant la révision de l'administrateur, le compte est suspendu temporairement jusqu'à la révision des rapports par l'administrateur.

Afin de sensibiliser les apprenants, spécialement les adolescents, aux dangers de parler avec des gens qui sèment le doute, l'agent « privacy assistant » est conçu pour ce but. « Privacy assistant » veille sur la sureté de l'utilisateur lors de l'utilisation du réseau social. Cet agent analyse les comportements de l'utilisateur sur le réseau social, ses messages, les différences d'âge et de sexe entre lui et ses amis…etc. Privacy assistant peut déduire l'objectif de l'utilisateur ainsi que les conséquences de ces activités à partir de l'analyse de ses activités.

Si le comportement de l'utilisateur peut lui causer des problèmes, l'agent lui conseille de régler ses paramètres et l'avertit des conséquences. Si « Privacy assistant » détecte un comportement dangereux de l'un des contacts de l'utilisateur, il le catégorise et puis il envoie une requête au serveur pour vérifier sa table de réputation. Le serveur enregistre le nombre des demandes de la réputation par les « privacy Assistants » pour chaque utilisateur afin de détecter les utilisateurs suspects qui n'ont pas été signalé. Afin de sensibiliser l'utilisateur au danger, « Privacy Assistant» affiche ce message (Figure 46) afin d'avertir l'utilisateur :

| |
|---|
| Il se peut que cet utilisateur est dangereux, voulez-vous le bloquer ?<br><br>☐ Oui ☐Non<br><br>//Si l'utilisateur choisit « non » et l'agent a trouvé que l'utilisateur suspect a été déjà signalé sur le serveur par d'autres utilisateurs, il affiche le message suivant :<br><br>Vous n'êtes pas la première victime de cet utilisateur, êtes-vous sûr de votre décision ?<br><br>☐ Oui ☐Non |
| **Figure 46: Message d'avertissement lors de la détection d'un comportement dangereux** |

Nous ne sommes pas contre l'analyse de comportement. Au contraire, les techniques d'analyse de comportement apportent beaucoup d'avantage à l'apprentissage des utilisateurs. L'analyse de comportement est dangereuse si les résultats de cette technique sont divulgués sans la connaissance ou le consentement explicite de la personne concernée. Dans notre cas, l'analyse de comportement est faite en local, sur la machine de l'utilisateur, et aucun autre pair n'a accès aux résultats de ces analyses.

## 6. Comparaison de notre proposition avec une proposition similaire de Ho Ai [Ho, 2012]

Notre travail s'inspire de la solution de Ho Ai **[Ho, 2012]**. Cependant, nous avons modifié et supprimé plusieurs étapes. Afin de bien montrer la différence entre notre proposition et celle de Ho Ai **[Ho, 2012]** nous présentons sa solution.

### 6.1 Privacy Watch [Ho, 2012]

C'est une proposition théorique qui a été implémentée sur le réseau social open source 'Elgg'[24]. La proposition encourage les fournisseurs d'accepter de stocker les données cryptées sur leurs serveurs (conformément au principe de la souveraineté des données). Le compromis que propose cette solution est de permettre aux fournisseurs de vendre seulement les données publiques qui sont stockées en clair sur le serveur. Les données privées sont cryptées avant d'être stockées sur le serveur du réseau social par une extension  Firefox « plugin » (Figure 47).

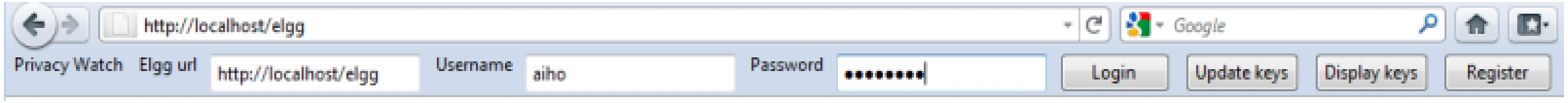

**Figure 47: l'extension Firefox de Privacy Watch [Ho, 2012]**

La solution propose aussi aux fournisseurs des RS d'utiliser dans leur architecture des politiques de confidentialités numériques appelées UPP (User Privacy Policy) **[Aïmeur, 2009]**. L'objectif est de permettre aux utilisateurs de communiquer leurs préférences de vie privée avant de permettre l'accès à leurs données, les utilisateurs peuvent exprimer leurs préférences en format de 'User Privacy Policy' (politique de confidentialité de l'utilisateur) (UPP). Le système avertira l'utilisateur quand la politique de confidentialité de l'application tierce rentre en conflit avec son UPP. Dans une telle situation, l'utilisateur a deux choix :

1. Accepter de changer son UPP suivant cette politique confidentialité.
2. Rejet de la demande d'accès de l'application.

#### 6.1.1 Les principaux éléments de l'UPP

Ho **[Ho, 2012]**, dans sa proposition change d'appellation  des différentes composantes du profil sur le réseau social ; les publications (les statuts, les photos, les vidéos…etc) sont appelées « Added content »  et les informations publiées sur les RS sont catégorisées en 4 niveaux de sureté. Ces niveaux sont définis comme suit :

1- **Healthy data:** cette catégorie regroupe les informations concernant l'utilisateur, comme son surnom, ses hobbies, les photos de la nature …etc. Ce type d'information ne peut pas être utilisé facilement pour connaitre l'identité de l'utilisateur.
2- **Harmless data:** Ceux sont les attributs démographiques tels que l'âge, le sexe, la religion et l'opinion politique. Cette catégorie d'informations est inoffensive car elle ne peut pas ruiner la réputation de l'utilisateur. Cependant, elle peut l'exposer aux menaces de profilage comme les sociétés publicitaires peuvent construire des profils détaillés par la collection de ces données.

[24] Un réseau social centralisé open source : http://www.elgg.org

3- **Harmful data:** ceux sont les publications et les photos inappropriées qui peuvent ruiner la réputation de l'utilisateur.
4- **Poisonous data:** C'est les informations personnelles qui peuvent l'exposer aux attaques d'usurpation d'identité (identity theft) tel que les informations financières, son vrai nom et son adresse.

Selon le niveau de confiance que voue chaque utilisateur à son ami, les listes d'amis sont classifiées en 4 groupes :

**1- Best Friends:** les personnes que l'utilisateur a en eux une confiance totale.
**2- Normal Friends:** ils peuvent être les membres de la famille, ou des amis dans la vie réelle.
**3- Casual Friends:** ceux sont les connaissances que l'utilisateur connait juste un peu.
**4- Visitors:** Ceux sont les visiteurs du profil de l'utilisateur.

De plus, Ho définit 4 niveaux de vie privée :

**1- No Privacy:** l'utilisateur ne s'inquiète pas sur sa vie privée et il veut que la plupart de ses informations soient publiques.
**2- Soft privacy**: l'utilisateur veut montrer ses « Poisonous data » à ses « Best Friends». « Casual » et « Normal Friends » sont autorisés à voir toutes les informations sauf « Poisonous data ». Les visiteurs sont autorisés à voir seulement « Harmless » et « Healthy » data.
**3- Hard privacy**: Comme "Soft privacy", "Normal Friends" ont toujours accès à « Harmful data » mais dans ce niveau, « Visitors » peuvent voir seulement « Healthy data » et «Casual Friends » ont accès seulement à « Harmless et Healthy data »
**4- Full privacy**: l'utilisateur ne permet pas les « Visitors » à accéder aux informations de son profil. De plus, l'accès à « Poisonous et Harmful data" est restreint à ses « Best Friends ». « Normal » et « Casual Friends » peuvent accéder seulement à « Harmless » et « Healthy data ».

L'UPP définit aussi 3 niveaux de « tracking » :

**1- Strong tracking**: L'utilisateur accepte d'être suivi sur le RS, quelques soient les moyens utilisés par ce but.
**2- Weak tracking**: L'utilisateur accepte que son nom apparaisse dans les listes d'amis des utilisateurs mais il ne permet pas d'associer des tags directs vers son profil.
**3- No tracking**: l'utilisateur ne souhaite pas que son profil soit mentionné sur le RS et il ne permet pas d'apparaitre son profil dans les résultats de recherche.

Les niveaux de traçabilité dans le contexte d'un RS signifient la visibilité des activités du profil aux utilisateurs du RS. Nous remarquons que les utilisateurs ne sont pas habitués à ce type de termes et c'est difficile de se rappeler des spécifications de chaque niveau.

### 6.1.2 Structure des UPP

La figure suivante (Figure 47) montre la structure de l'UPP :

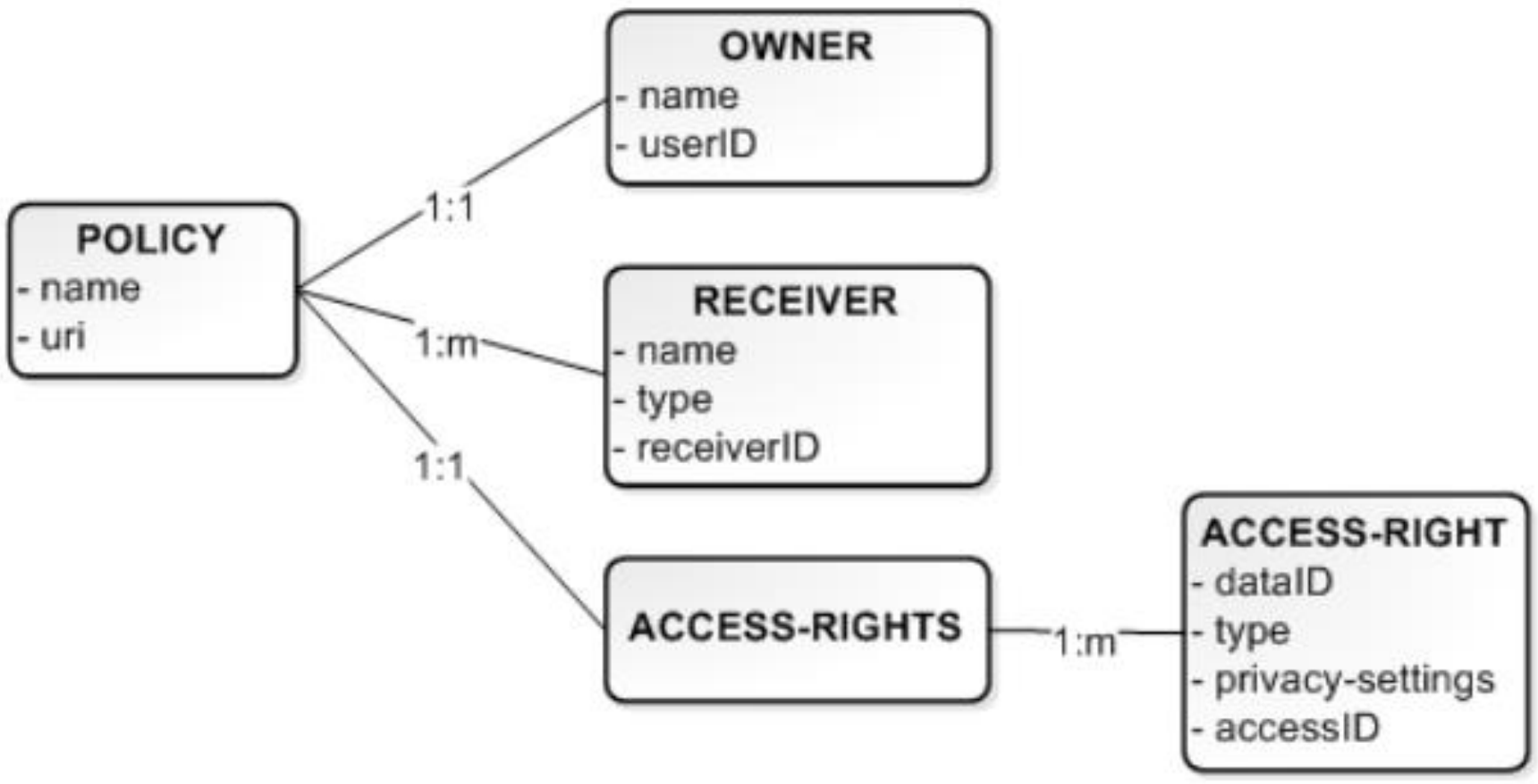

Chaque UPP est composé des éléments suivants :

**1- Elément « POLICY »**

Un UPP peut contenir un ou plusieurs éléments « POLICY ». L'élément « POLICY » correspond à toutes les informations d'une politique et contient un propriétaire, à au moins un récepteur et un droit d'accès. Les attributs principaux de l'élément « POLICY » sont :

- **Nom (obligatoire)** : nom de la politique.
- **URI (optionnel)** : URI de la déclaration de confidentialité dans le langage naturel.

2- **Elément « OWNER »**

L'élément propriétaire donne des informations concernant l'émetteur de cette politique tel que : <OWNER name= "Cindy" userai="Cindy1234"/>

Cet élément a deux attributs obligatoires : Nom et user ID

3- **L'élément « RECEIVER »**

Cet élément contient une description précise de l'objet de cette politique comme : nom utilisateur, un groupe d'utilisateurs ou une application. Par exemple, cet élément peut apparaître comme suit :
<RECEIVER name="Friends" type="group" receiverID="CS1"/>

**4- L 'élément « ACCESS RIGHT »**

L 'élément « ACCESS RIGHT » indique comment le récepteur de l'information doit traiter cette information et il peut avoir des éléments **« ACCESS RIGHT»** multiples. Chaque élément **« ACCESS RIGHT»** contient les attributs suivant :

a. **DataID (optionnel):** Identifiant de d'information
b. **Access ID (obligatoire):** identifiant du droit d’accès.
c. **Data type (obligatoire):** Identité, attribut démographique, activité, réseau social, contenu publié...etc.

L’élément **« ACCESS RIGHTS »** peut avoir différentes valeurs, y compris « no_comment », « no_distribution » ou « no_ tracking ». La figure 49 présente un exemple de code d’un UPP :

```
<POLICY>
<OWNER name="Cindy" userID="Cindy1234"/>
<RECEIVER name="Bob" type="user" receiverID="Bob9990"/>
<ACCESS-RIGHTS>
   <ACCESS-RIGHT accessID="a11" type="added_content" privacy_concern="Harmless">
            <no_distribution/>
   </ACCESS-RIGHT>
   <ACCESS-RIGHT accessID="a21" type="tracking">
            <soft_tracking/>
   </ACCESS-RIGHT>
</ACCESS-RIGHTS>
</POLICY>
```

**Figure 48: Exemple d’un UPP [Aimeur, 2009]**

La figure 50 décrit un UPP de Cindy à Bob. Tous les contenus publiés de Cindy sont visible à Bob. Mais, Bob n’est pas autorisé à les diffuser ou les partager avec les autres.

From: Cindy
To: Bob

Dear Bob,
It is my pleasure to become your friend, however, I would like that you agree on these privacy conditions regarding my personal information:
- **Do not redistribute or share** my Added Content (including but not limiting to photos, music, videos, blogs, and comments)
- **Do not include** my name in your list of friends and **do not link** to my profile.
Thank you very much,

**Figure 49: Le message automatique que Bob reçoit de la part de Cindy [Aïmeur, 2009]**

### 6.1.3 Les inconvénients de la solution

L’inconvénient de cette solution est le nombre important des messages que l’utilisateur recevra à chaque violation ou demande d’accès. En effet, l’utilisateur est à chaque fois sollicité pour changer son UPP ou pour rejeter la demande d’accès. De plus, comme elle se base sur une architecture centralisée, elle dépend essentiellement de la bonne volonté du fournisseur du réseau social pour protéger l’utilisateur.

La solution proposée par Ho pour la limitation de l'audience d'un contenu partagé par un autre utilisateur (qui n'est pas le propriétaire du contenu) est gérée par le fournisseur du RS. En effet, actuellement sur Facebook et Google+ seulement les amis en commun, entre le propriétaire et celui qui repartage le contenu, peuvent voir le contenu partagé (si le contenu n'est partagé à accès public par le propriétaire) (Voir figure 51). L'utilisateur sur Facebook peut même désactiver l'option du partage de ses contenus.

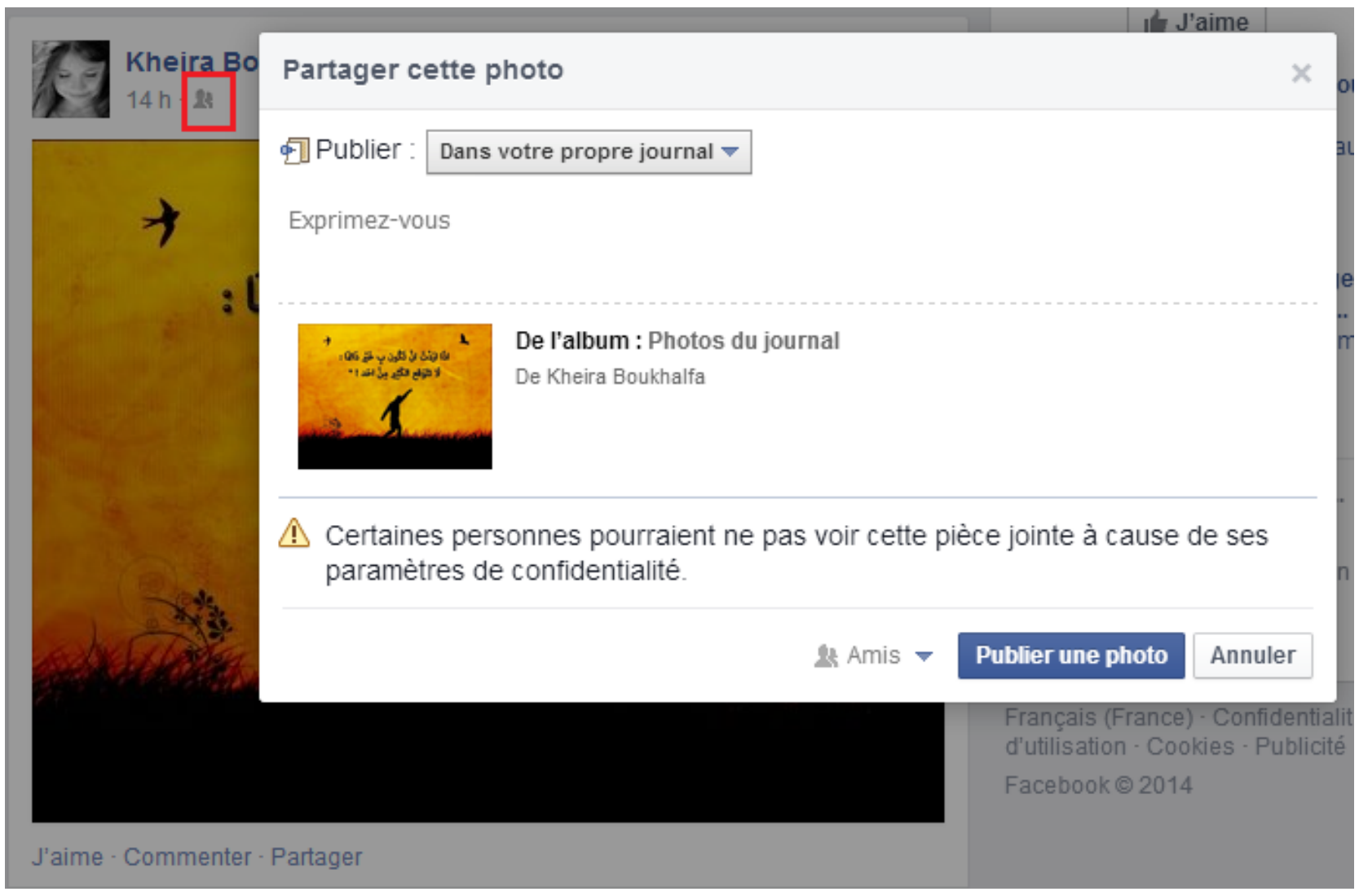

**Figure 50:** La restriction de l'audience du contenu repartagé par Facebook.

Un autre problème de la solution est qu'elle se base sur une architecture centralisée, ce qui permet au fournisseur de connaitre le social graphe de l'utilisateur et identifier l'identité réelle de l'utilisateur. Aussi, la solution propose des termes peu familiers aux utilisateurs tel que « Added Content », « Weak tracking » , « Harmfull » et « Poisonous data », « Soft Privacy », et les groupes d'amis tel que « Casual Friends » avec des accès différent selon les niveaux de confiance définit (par défaut) par Ho et que l'utilisateur ne peut pas les changer. Ces paramètres ne sont pas flexibles et sont définis d'une manière à être généralisés pour tous les contenus du même type. Ainsi, c'est difficile à l'utilisateur de se rappeler du sens de chaque paramètre avec la façon dont ils sont définis.

## 6.2 Discussion: la solution de Ho VS notre solution

La première différence entre notre solution et celle de Ho **[Ho, 2012]** est l'architecture du réseau social: nous avons opté pour une architecture décentralisée contrairement à « Privacy Watch » qui se base sur un réseau social centralisé. Deuxièmement, nous n'utilisons pas d'extension Firefox pour le cryptage et le décryptage des contenus. Les utilisateurs non expérimentés ne peuvent pas comprendre les principes de chiffrement et de déchiffrement, c'est pour cette raison que gérons ces mécanismes par l'application cliente « Appraide » où ces mécanismes de cryptage s'exécutent d'une manière invisible pour l'utilisateur. L'utilisateur n'aura aucune connaissance sur ce qu'il se passe dernière son ajustement de ses paramètres de confidentialité.

Afin de faciliter la compréhension et l'ajustement des paramètres, nous avons éliminé les niveaux de vie privée ainsi que les niveaux de « tracking » dans les RS car ce dernier est simple à gérer et ne nécessite pas la définition de plusieurs niveaux. En effet, il ne s'agit que d'une simple instruction de désactivation de suivi des activités. Par ailleurs, les tags, l'apparition du profil dans les résultats de recherche des utilisateurs, la visibilité du profil de l'utilisateur dans les listes d'amis ainsi que la diffusion de ses activités sur le réseau social sont des fonctionnalités différentes dans les réseaux sociaux et chacune d'elles peut être désactivée séparément des autres. Les niveaux de « tracking » limitent les choix de désactivation de ces fonctionnalités. Aussi, nous avons défini nos propres classes de connexions qui utilisent des noms de liste facile à comprendre et qui sont adaptées à notre contexte qui est l'apprentissage. Dans notre travail nous n'associons aucune relation entre les classes d'amis et les niveaux de confiance qui définissent les droits d'accès aux contenus. L'utilisateur dans notre solution est le seul à gérer et à définir ses droits d'accès pour chaque type de publication et pour chaque instance d'un type de publication. Cela donne plus de flexibilité des paramètres de confidentialité. De plus, nous ne permettons pas l'envoi des messages car cette solution peut déranger celui qui est en amitié avec l'utilisateur. En effet, à chaque changement de l'UPP, les amis de l'utilisateur seront informés sur changement de l'UPP. Malgré cette solution augmente la conscience des utilisateurs envers la vie privée de leurs amis, l'utilisateur avec un nombre important d'amis ne pourra pas se rappeler de chaque UPP de ses amis. Dans notre solution, rien n'est laissé à la bonne volonté de l'utilisateur. En effet, nous avons utilisé dans notre proposition quelques métadonnées définis par Ho et nous avons ajouté d'autres. Ces métadonnées aident les agents de l'application cliente « ApprAide » à comprendre et protéger les droits et les restrictions du propriétaire du contenu. Grace aux métadonnées, que nous avons défini pour la protection des données partagées et dupliquées sur notre réseau social décentralisé, les agents sont capables de détecter et de bloquer les violations localement avant que le contenu soit dévoilé à d'autres utilisateurs.

## Conclusion

Les efforts fournis par les militants dans ce domaine pour protéger la vie privée des utilisateurs notamment en Europe, Canada et USA ont quand même réussi à imposer certaines lois juridiques interdisant les fournisseurs des services en ligne de transgresser et de dévoiler la vie privée des citoyens. Les plaintes portées contre eux, leurs causent énormément de pertes financières. Plus les punitions sont sévères, plus ces violations diminueront et les fournisseurs se trouveront obligés de respecter la vie privée de leurs utilisateurs lors de la conception de leurs systèmes. Nous avons présenté dans ce chapitre une approche de conception qui respecte la vie privée des utilisateurs dans les systèmes d'apprentissage social. Une telle conception prouvera, en cas d'attaque en justice, que le fournisseur ne viole pas la vie privée de ses utilisateurs. L'approche, que nous avons présentée, a plusieurs avantages notamment en protection de la vie privée des propriétaires de contenus lors de la duplication des données et l'assurance du droit à l'oubli à nos utilisateurs. Cette solution a quelques inconvénients, en particulier le problème de disponibilité des données causé par l'architecture décentralisée du système ainsi que la difficulté qu'elle présente pour la gestion des mises à jours dans un système très dynamique comme le réseau social.

# Partie 3.

# Implémentation

# Chapitre 6.

# Réalisation du système « ApprAide »

## Introduction

La phase de réalisation consiste donne une forme concrète à la phase de conception. Cette partie décrit l'architecture physique du système, les outils utilisés pour l'implémenter, ainsi que des aperçus de l'application « ApprAide ».

## 1. Architecture du système

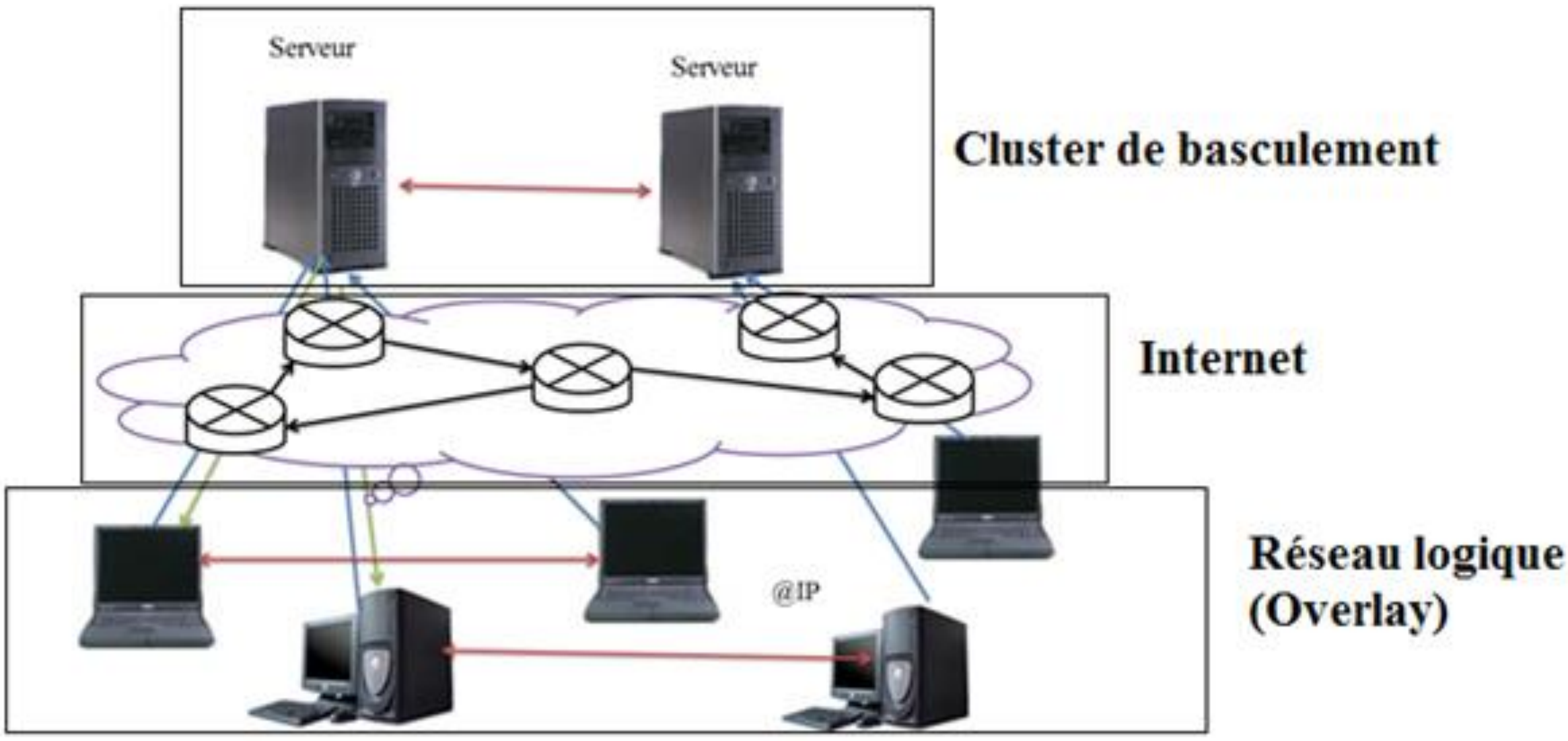

**Figure 51: Architecture Hybride**

Notre système suit une architecture hybride (Figure 52). En effet, nous avons opté pour une architecture décentralisée non seulement pour éviter le contrôle total sur les données des utilisateurs par l'entité centrale (dû aux systèmes centralisés) et permettre à l'apprenant d'être le seul propriétaire de ses données privées. Mais aussi, pour permettre à notre système d'effectuer les traitements de personnalisation sans violer la vie privée de l'apprenant. La décentralisation des données vers les bases locales de nos utilisateurs nous permet de tracer localement toutes les activités de l'apprenant, faire l'analyse de comportement et comprendre ses besoins afin d'adapter le processus d'apprentissage. Cette décentralisation permet à l'utilisateur de garder le contrôle sur ses données privées et de bénéficier des avantages des techniques de « Learner Modelling ». De cette manière, l'utilisateur est protégé contre la vente de son profil à d'autres tiers sans son consentement.

Nous avons ajouté un autre serveur à l'architecture pour éviter l'indisponibilité du serveur, supporter la charge des requêtes des clients et améliorer le temps de réponse. La redondance des données sur des serveurs secondaires est utilisée pour faire face aux pannes du serveur et aux échecs des services. Dans cette architecture, le serveur sert à authentifier les utilisateurs et agir comme un annuaire facilitant la recherche des utilisateurs. Cet annuaire contient les informations relatives aux utilisateurs (identifiant, adresse IP, etc.). Pour des raisons de disponibilité, nous avons centralisé l'accès aux cours, aux autotests et aux contenus à accès public. Ces derniers n'ont aucun impact sur la vie privée des utilisateurs.

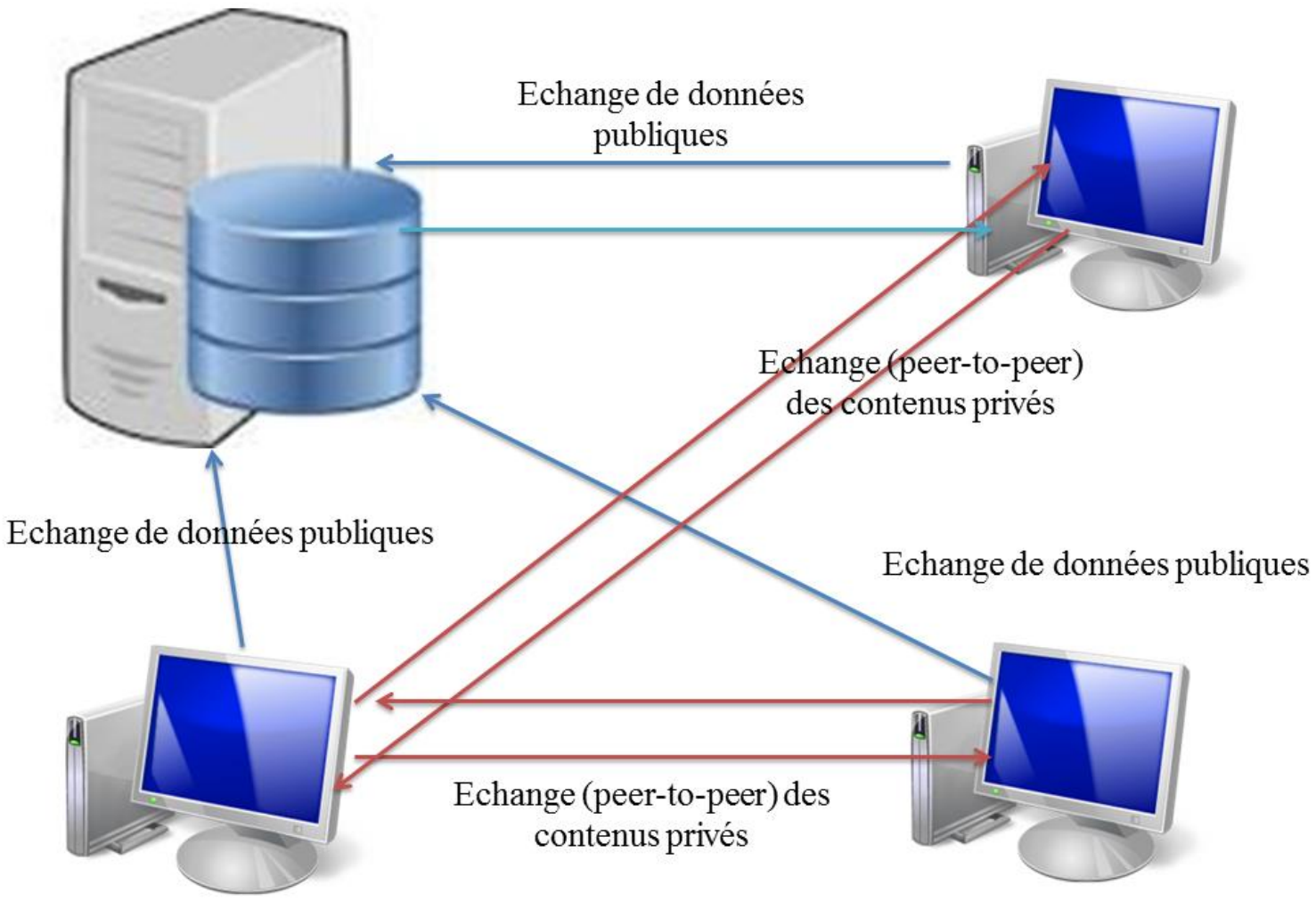

**Figure 52: Echange des données privées et publiques**

La base centrale (du serveur) ne contient pas les données privées des utilisateurs. L'échange et la duplication des données privées sont faits selon le modèle P2P (Figure 53) où les machines des utilisateurs sont à la fois client et serveur. La duplication des données locales de l'utilisateur sur les autres nœuds du réseau, permet de surmonter les limitations concernant la faible disponibilité des données dans le système P2P.

## 2. Les outils et les langages utilisés pour le développement de la solution

Le système est composé de deux parties :

**Un site Web:**

Pour implémenter le site web et les fonctionnalités du serveur, nous avons utilisé WAMPSERVER version 5.5.12. Le site web a été implémenté en utilisant :

- Le langage de script PHP pour implémenter la logique des traitements.
- Le langage JavaScript pour dynamiser le site Web.
- Le langage CSS pour faire le design.
- Le SGBD MySQL pour réaliser les bases de données.

**Et une application desktop :**

L'application desktop permet à l'utilisateur d'accéder à tous les services de notre système. L'application a été implémentée en utilisant :

- Le langage java pour l'implémentation des interfaces et des traitements.
- La plateforme JADE pour l'implémentation des agents.
- Le SGBD SQLite pour le développement de la base de données locale.

## 3. La réalisation de la solution :

### 3.1 Le site Web

Le site Web sert à inscrire les utilisateurs et à permettre aux utilisateurs de télécharger l'application desktop « ApprAide ».

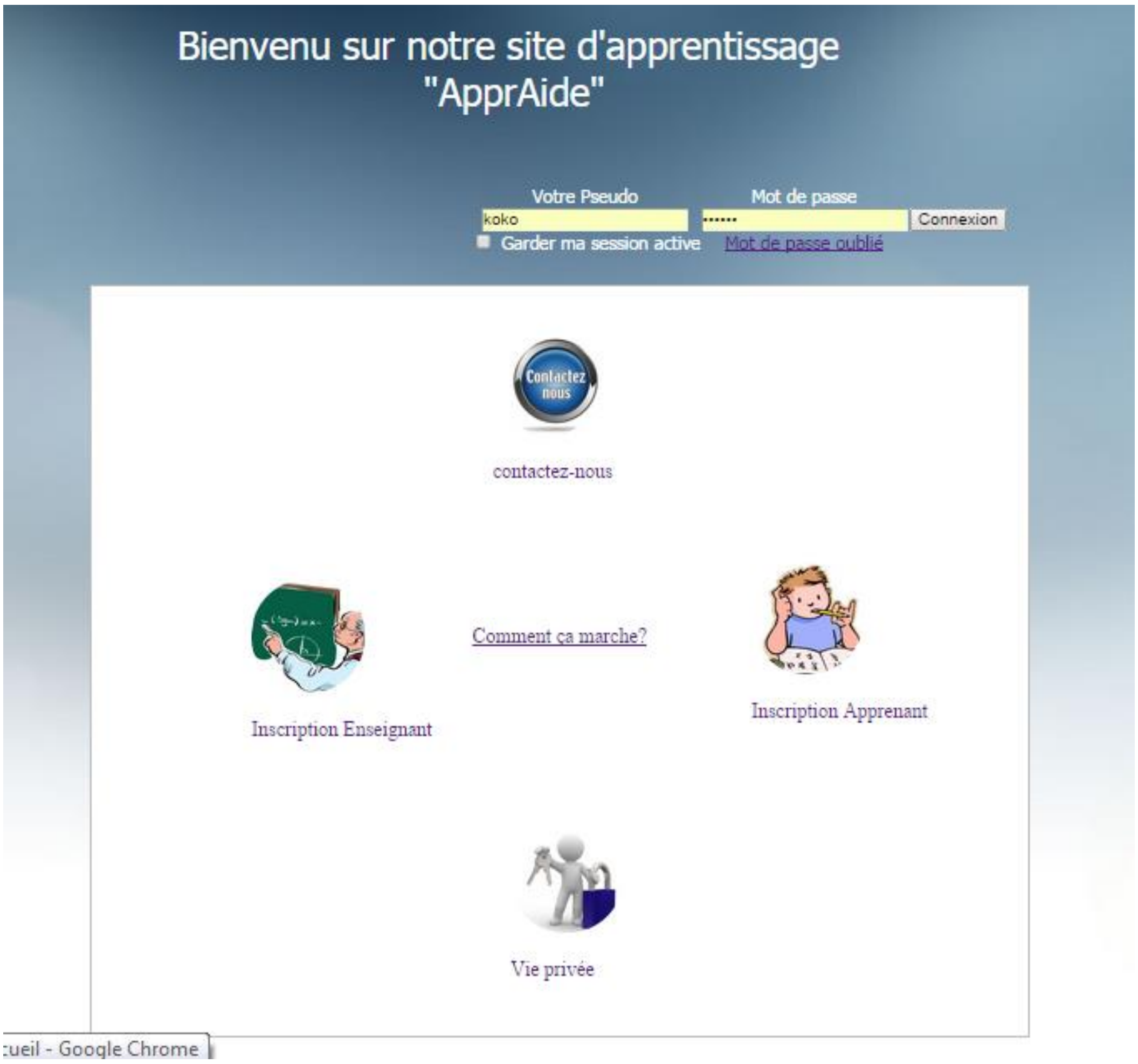

**Figure 53: Page d'accueil du site web "ApprAide"**

Deux types d'inscription existent sur le site « ApprAide »:

- **L'inscription des enseignants** :

Le formulaire d'inscription des enseignants contient les informations nécessaires pour le recrutement de l'enseignant. Après l'inscription de l'enseignant, son dossier est envoyé à l'administrateur. Si le dossier de l'enseignant est accepté par l'administrateur, lors de la prochaine connexion de l'enseignant, il recevra un message contenant le décisionet un lien vers une page de téléchargement de l'application desktop « ApprAide ». La figure suivante (Figure 55) montre le formulaire d'inscription des enseignants :

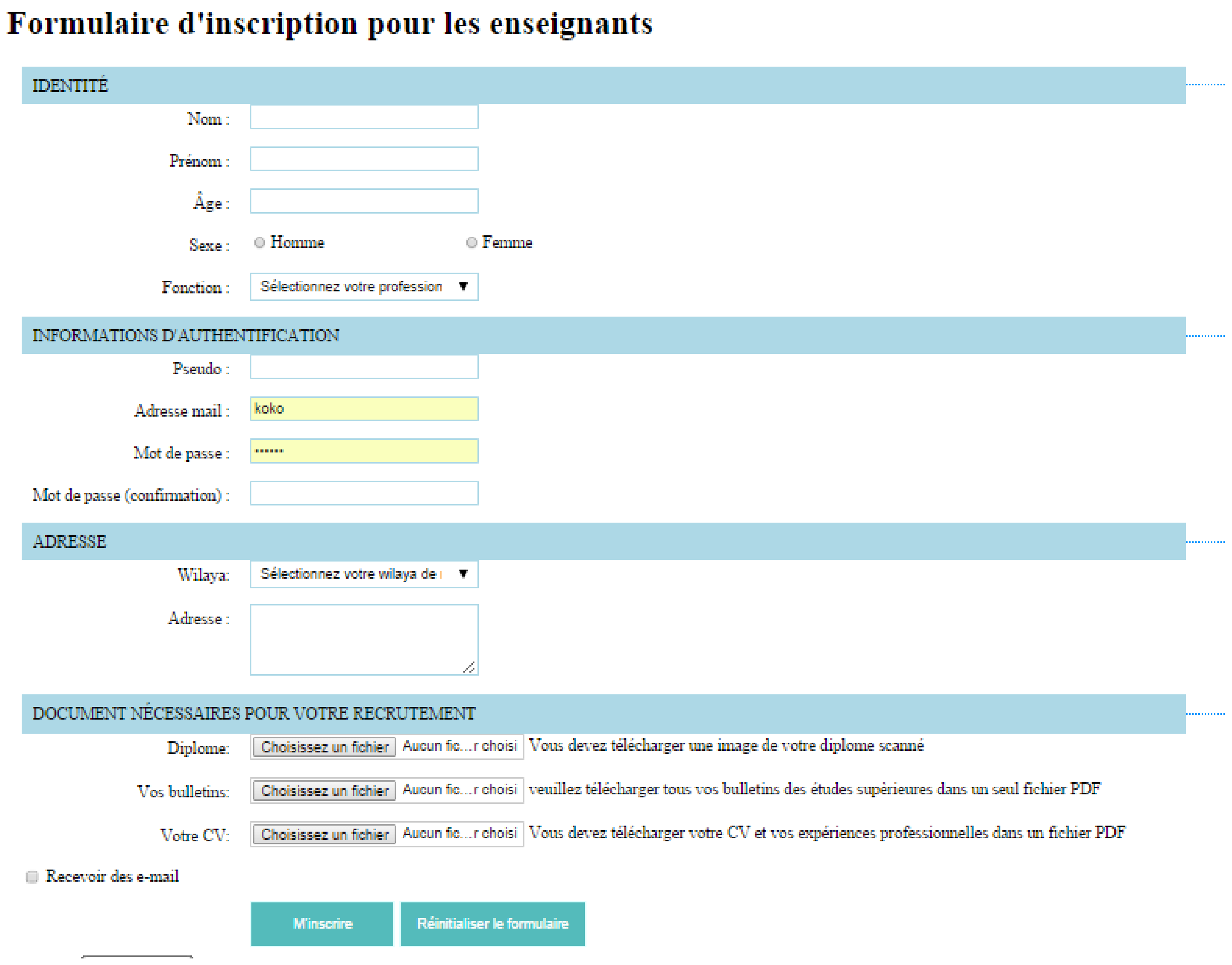

**Figure 54: Formulaire d'inscription des enseignants**

## L'inscription des apprenants :

Conformément au principe de la limitation de collection (minimisation des données), l'apprenant est demandé de fournir un profil réduit qui ne contient que les informations nécessaires à son authentification. La figure suivante (Figure 56) montre le formulaire d'inscription des apprenants :

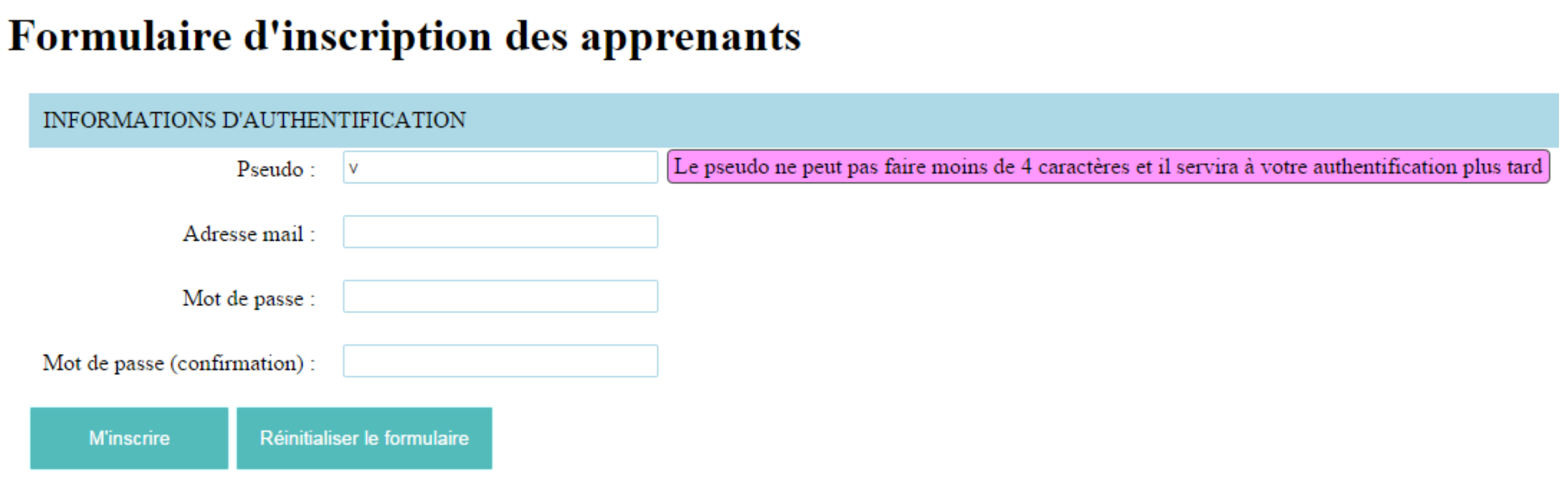

**Figure 55: Formulaire d'inscription des apprenants**

De plus, Le base du site contient un annuaire des utilisateurs avec leur adresse IP permettant de faciliter la localisation des utilisateurs pour les connexions peer2peer (à partir de l'application desktop). Le serveur est aussi un espace de stockage des contenus partagés à accès public. La figure 57 montre la conception de la base de données et la figure 58 montre la base de données réalisée avec un aperçu de l'annuaire des adresses IP.

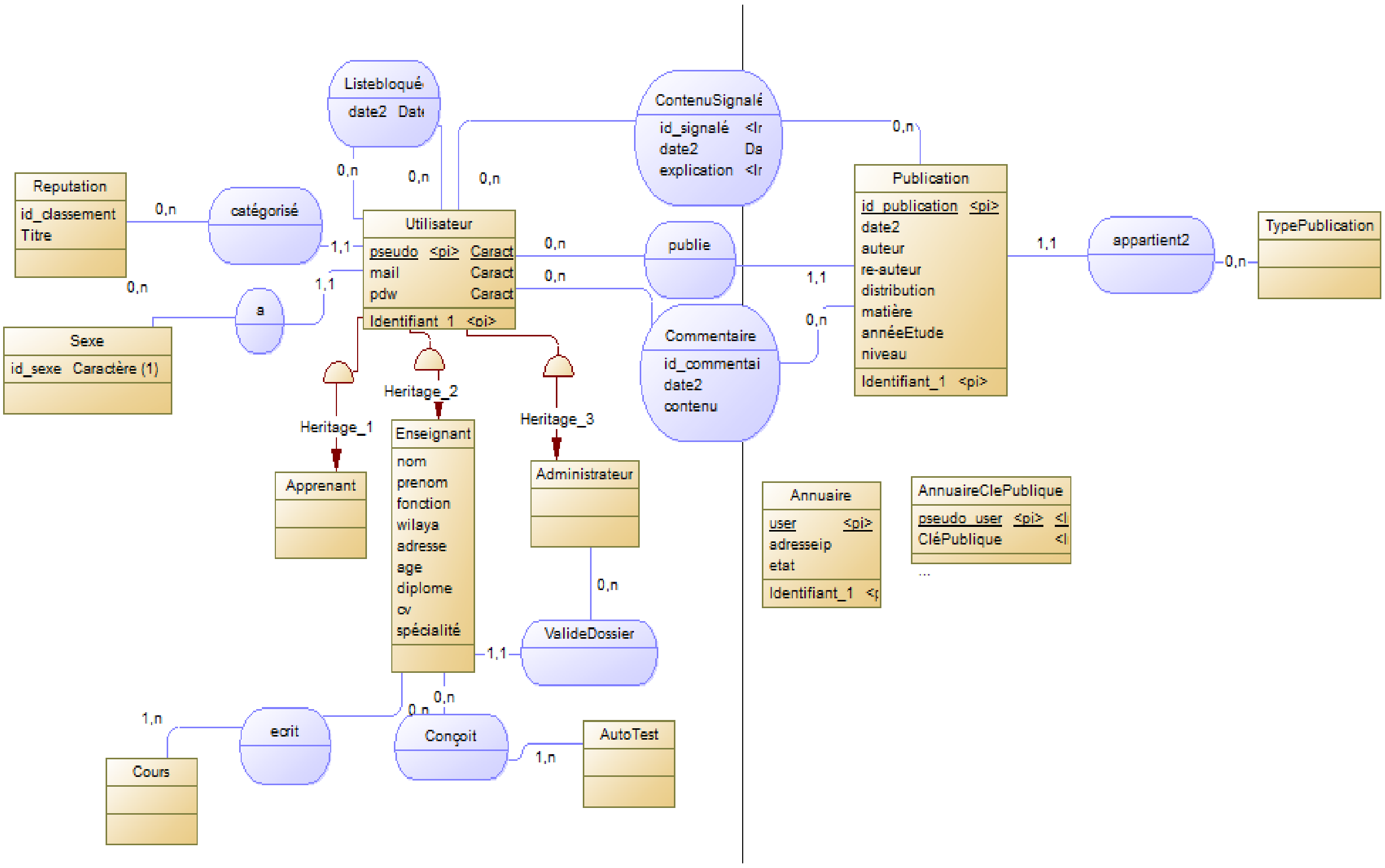

Figure 56: schéma conceptuel de la base de données

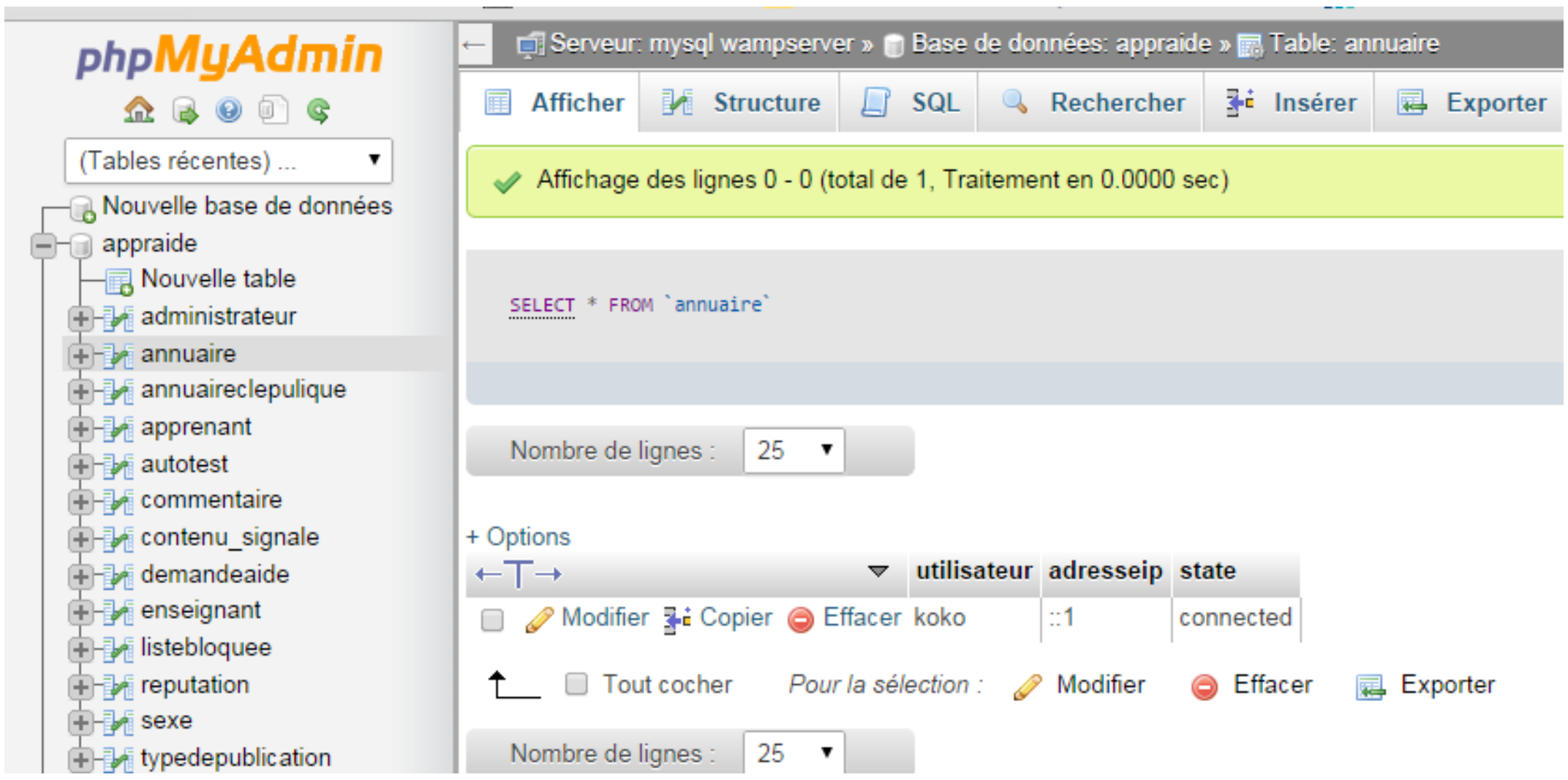

**Figure 57: La base de données avec un aperçu de la table annuaire**

L'adresse IP de l'utilisateur permet d'établir les connexions peer2peer de nos utilisateurs. L'utilisation des adresses IP peut donner l'impression de violer la vie privée des utilisateurs car elle permet de le géo-localiser. Cependant, l'adresse IP n'est pas fixée de façon permanente à l'utilisateur et en cas de sa déconnexion, elle peut être attribuée à d'autres utilisateurs. De plus, les techniques de localisation des utilisateurs permettent seulement de donner une estimation probabiliste de son lieu. Le serveur garde aussi l'information sur les états de connexion des utilisateurs afin de gérer la recherche des pairs aidants non connectés (qui ne sont pas des amis).

### 3.2 L'application desktop :

L'application desktop « ApprAide » devrait contenir 5 outils : un réseau social d'apprentissage, un tableau virtuel, un espace pour les annonces, un outil d'autotest et un outil de consultation des cours conçus par les enseignants. Dans le cadre de ce travail nous avons implémenté les deux premiers outils. La figure suivante montre la fenêtre d'authentification (Figure 59)

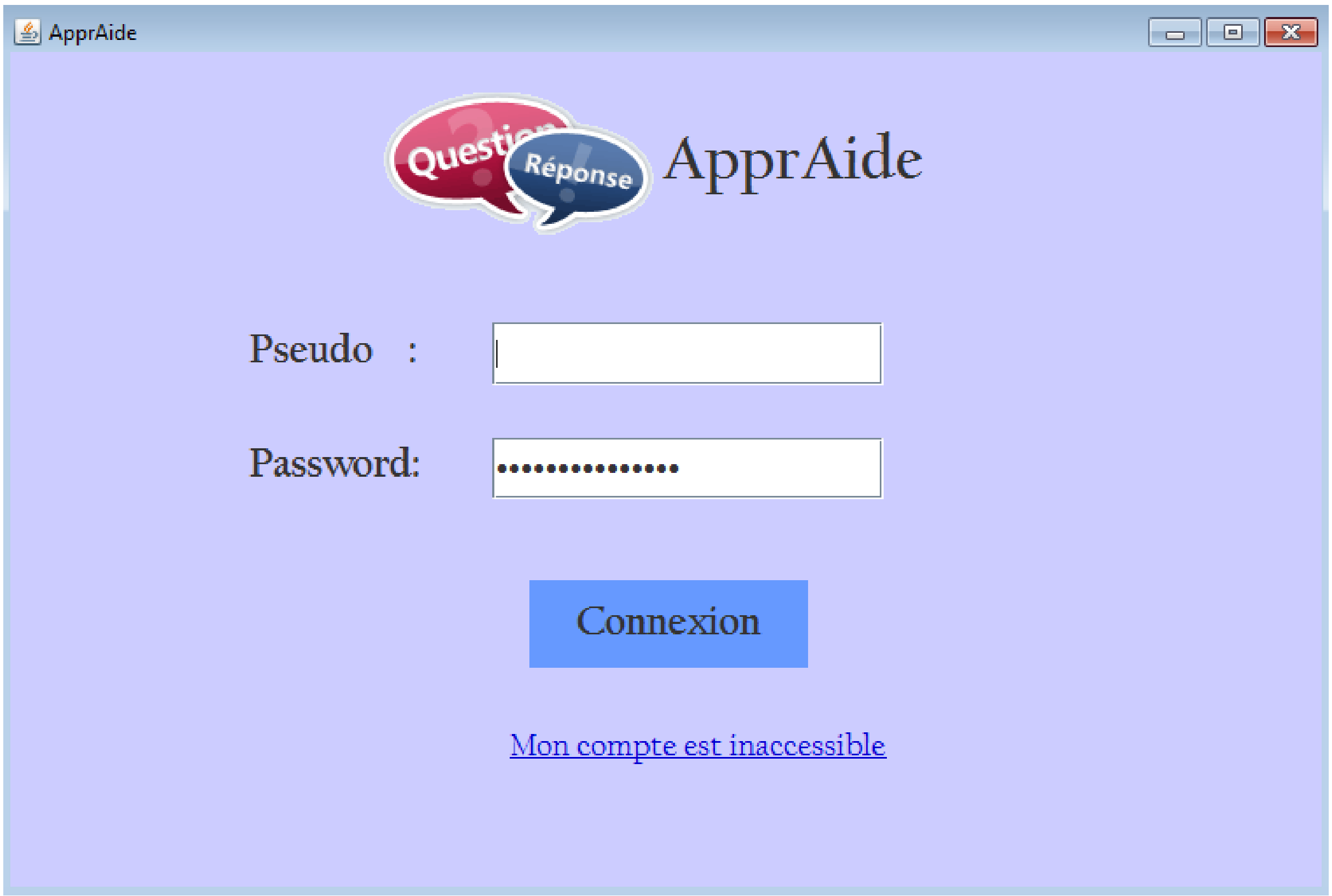

**Figure 58: Fenêtre d'authentification**

Après l'authentification de l'utilisateur par le serveur, l'utilisateur est dirigé vers le réseau social. Par faute de temps l'implémentation du réseau social n'est pas encore terminée, La figure 60 montre comment le réseau devrait être :

- **Le réseau social d'apprentissage :**

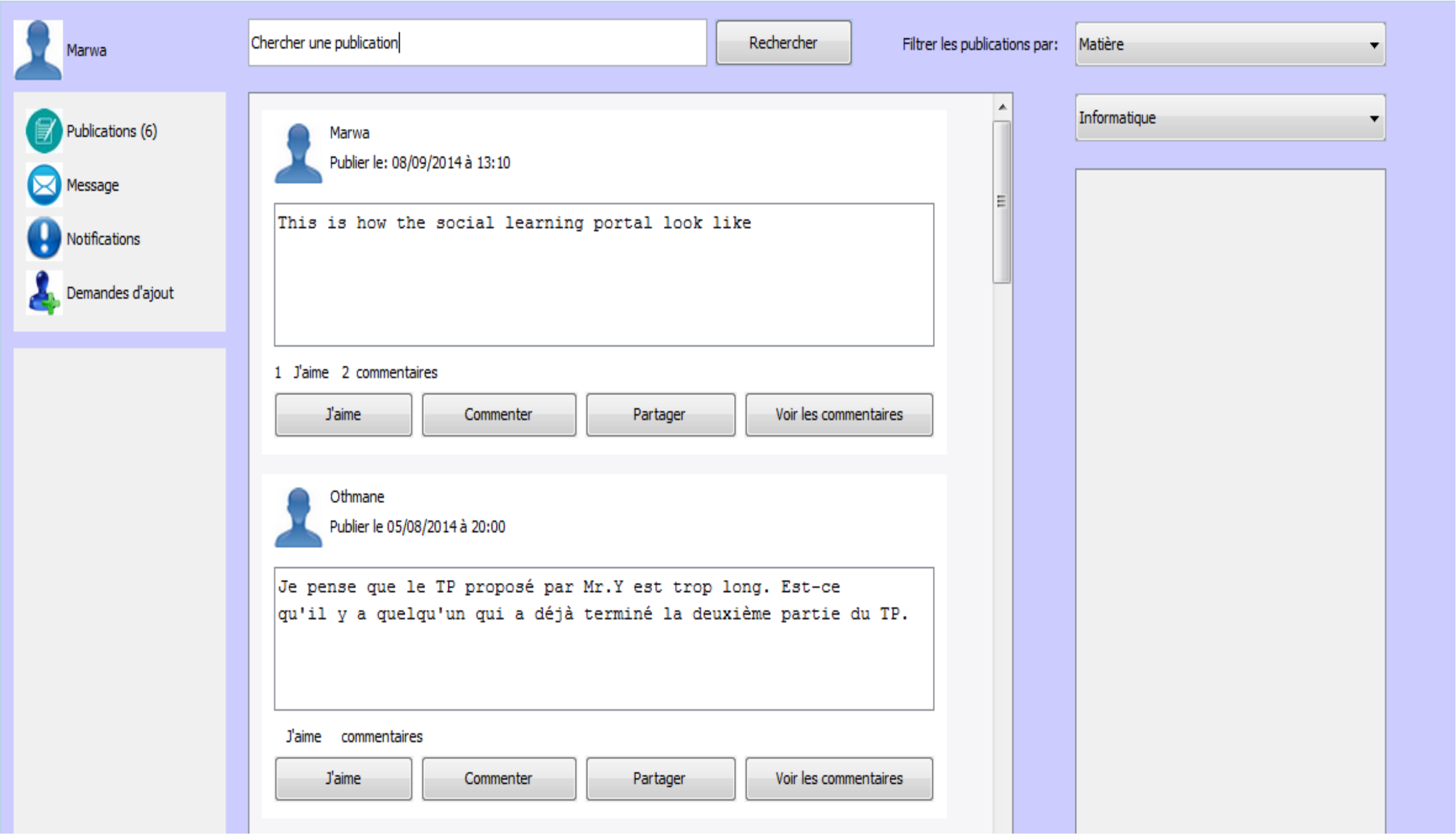

**Figure 59: Réseau social d'apprentissage "ApprAide"**

L'utilisateur peut publier, commenter, aimer ou partager une publication. Avant de publier l'utilisateur doit choisir le type de la publication, la matière, le niveau, l'année d'étude ainsi que la visibilité de sa publication et le droit de distribution. La figure 61 montre la fenêtre « publier » :

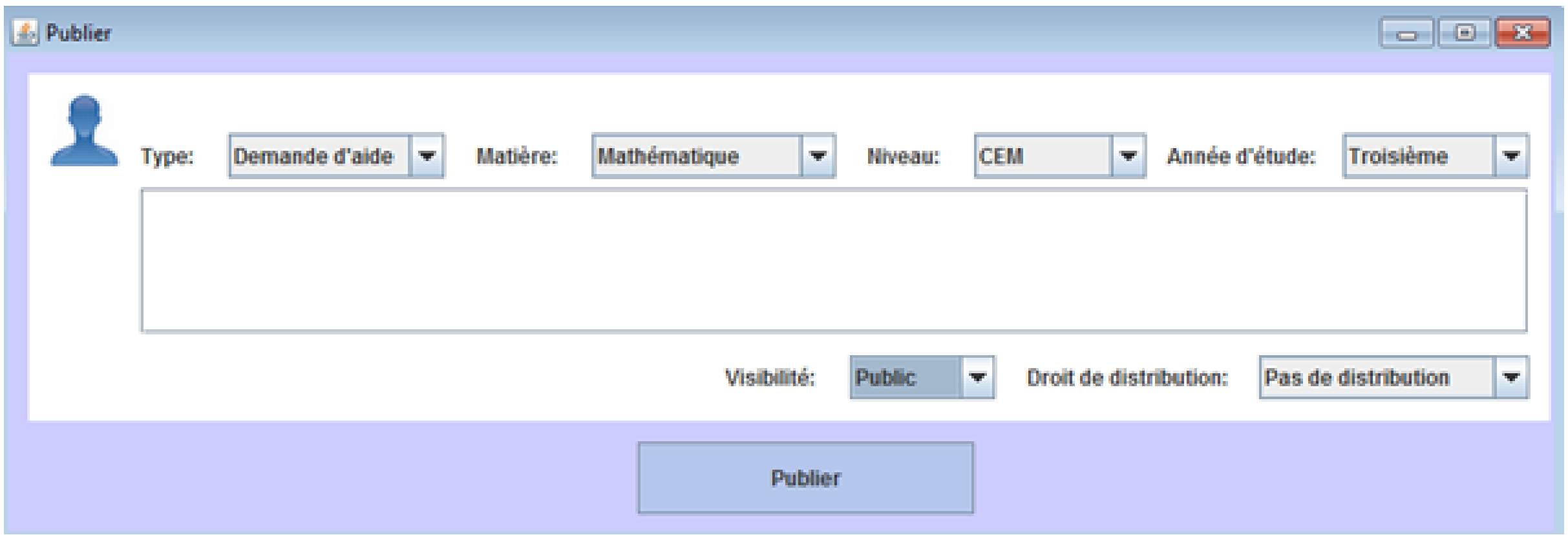

**Figure 60: la fenêtre qui permet à l'utilisateur de publier**

- **Le tableau virtuel pour les cours de soutien**

L'application permet à l'utilisateur d'éditer et de recevoir des demandes d'aide synchrone. La figure suivante montre la page d'édition des demandes d'aide

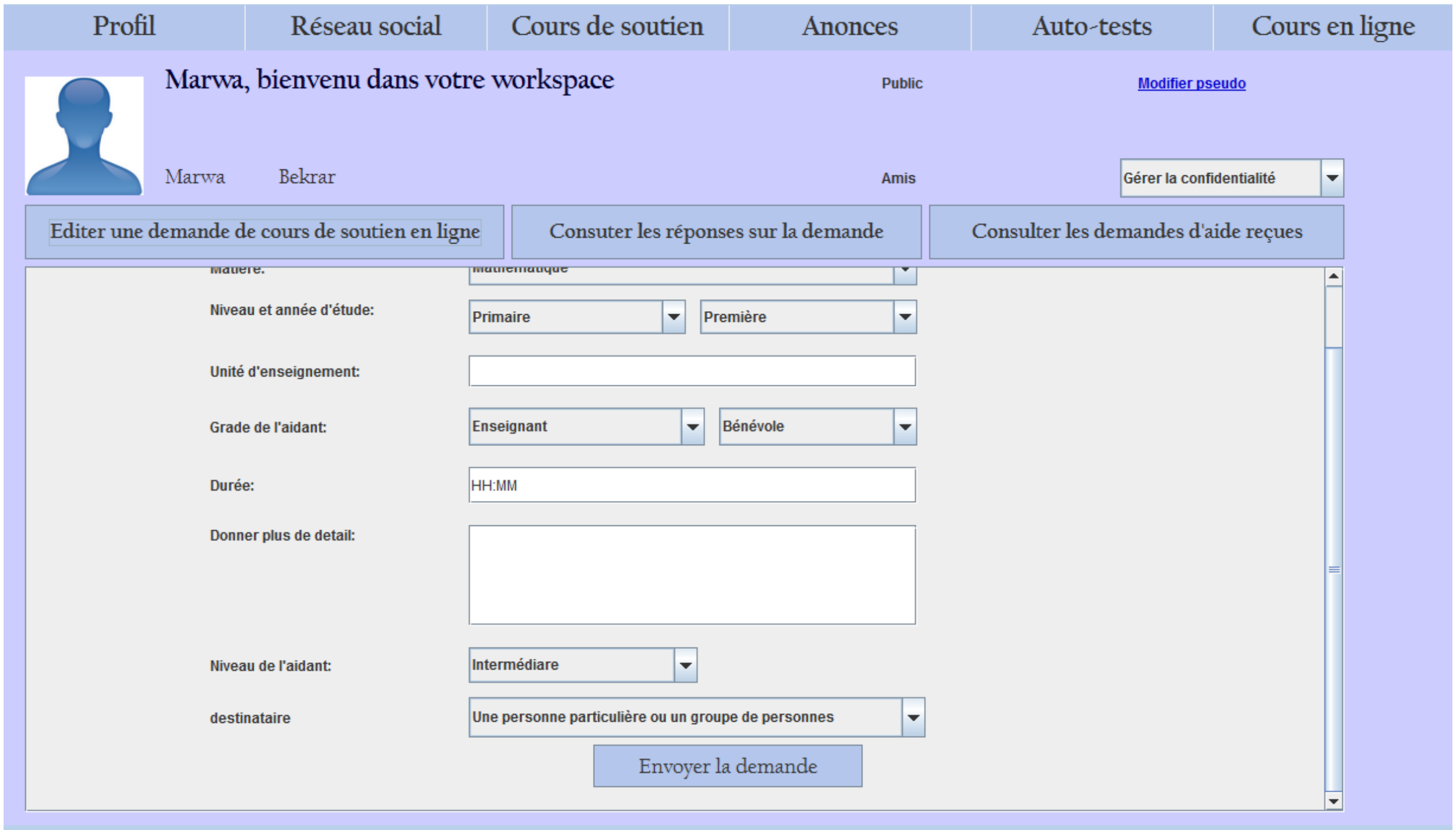

**Figure 61: Editer une demande de cours de soutien**

L'utilisateur peut consulter les réponses sur sa demande d'aide sur la page « Consulter les demandes d'aide reçues » (Figure 63).

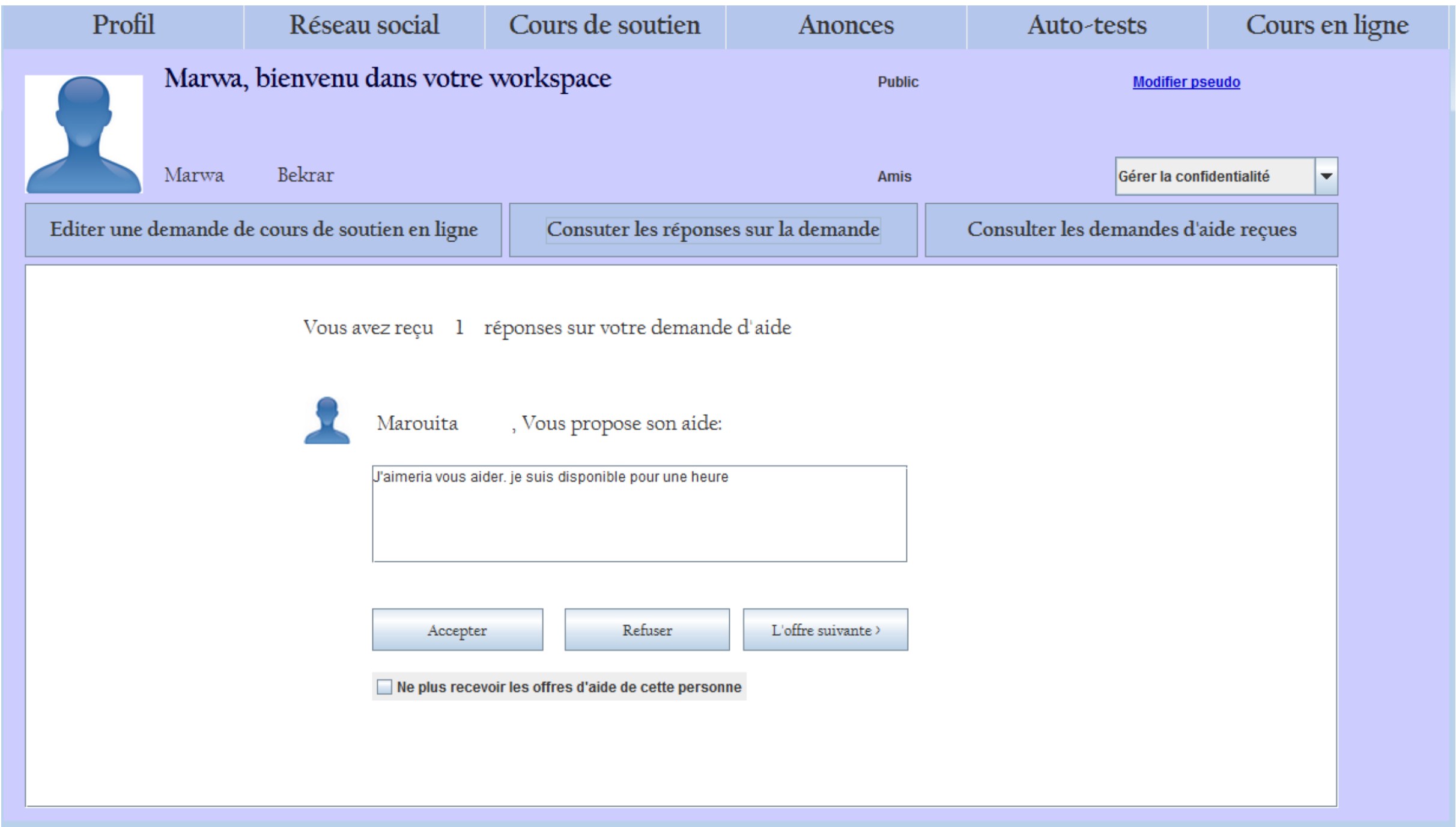

**Figure 62: Consulter les demandes d'aide reçues**

Si l’apprenant accepte l’offre d’aide, il peut travailler avec l’autre personne en utilisant le tableau virtuel. La figure 64 montre un apprenant qui pose une question à son enseignant (en bleu) et la figure 65 montre la réponse de son enseignant (en rouge).

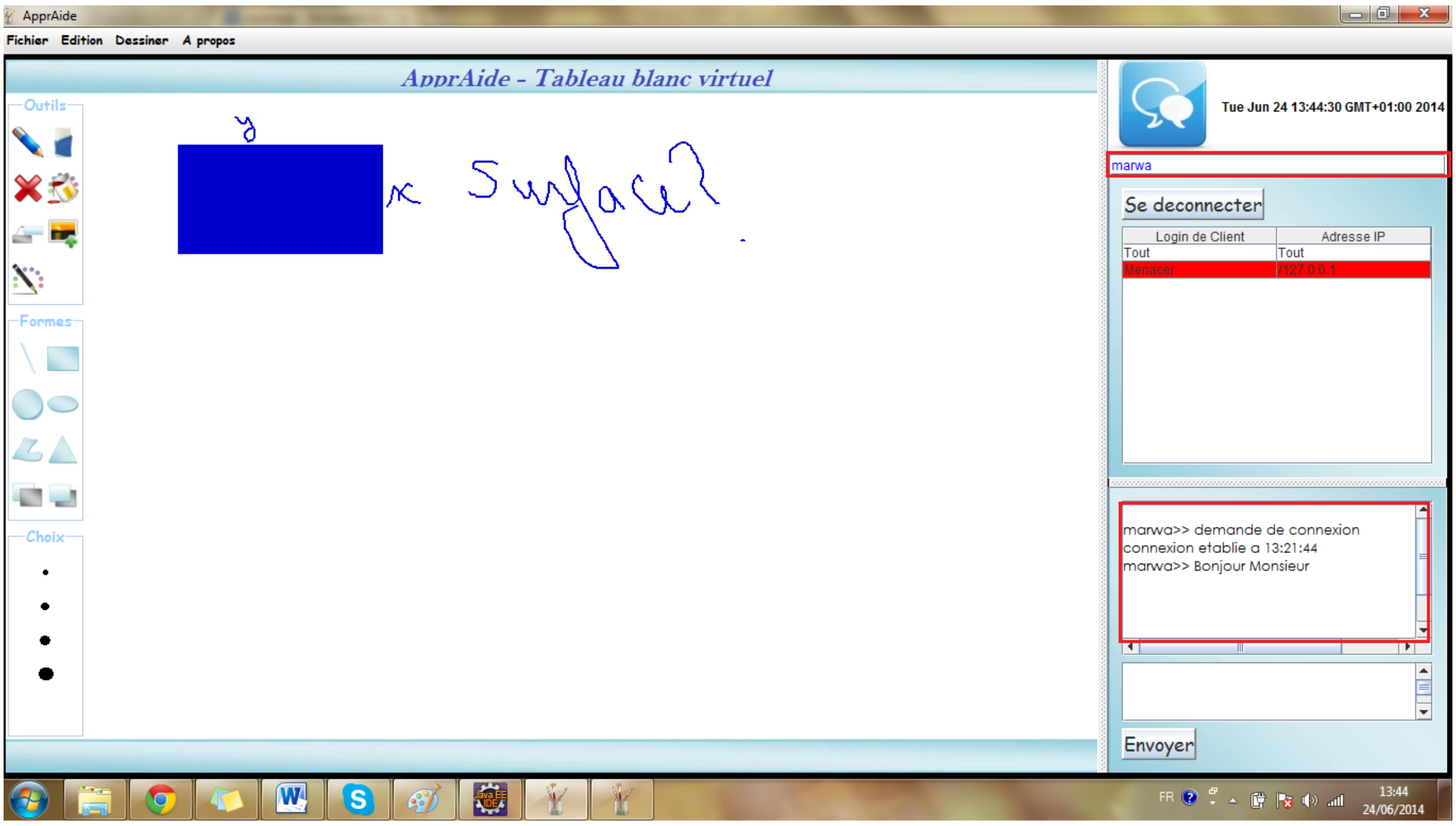

**Figure 63: Le tableau blanc virtuel (avec la question de l’apprenant)**

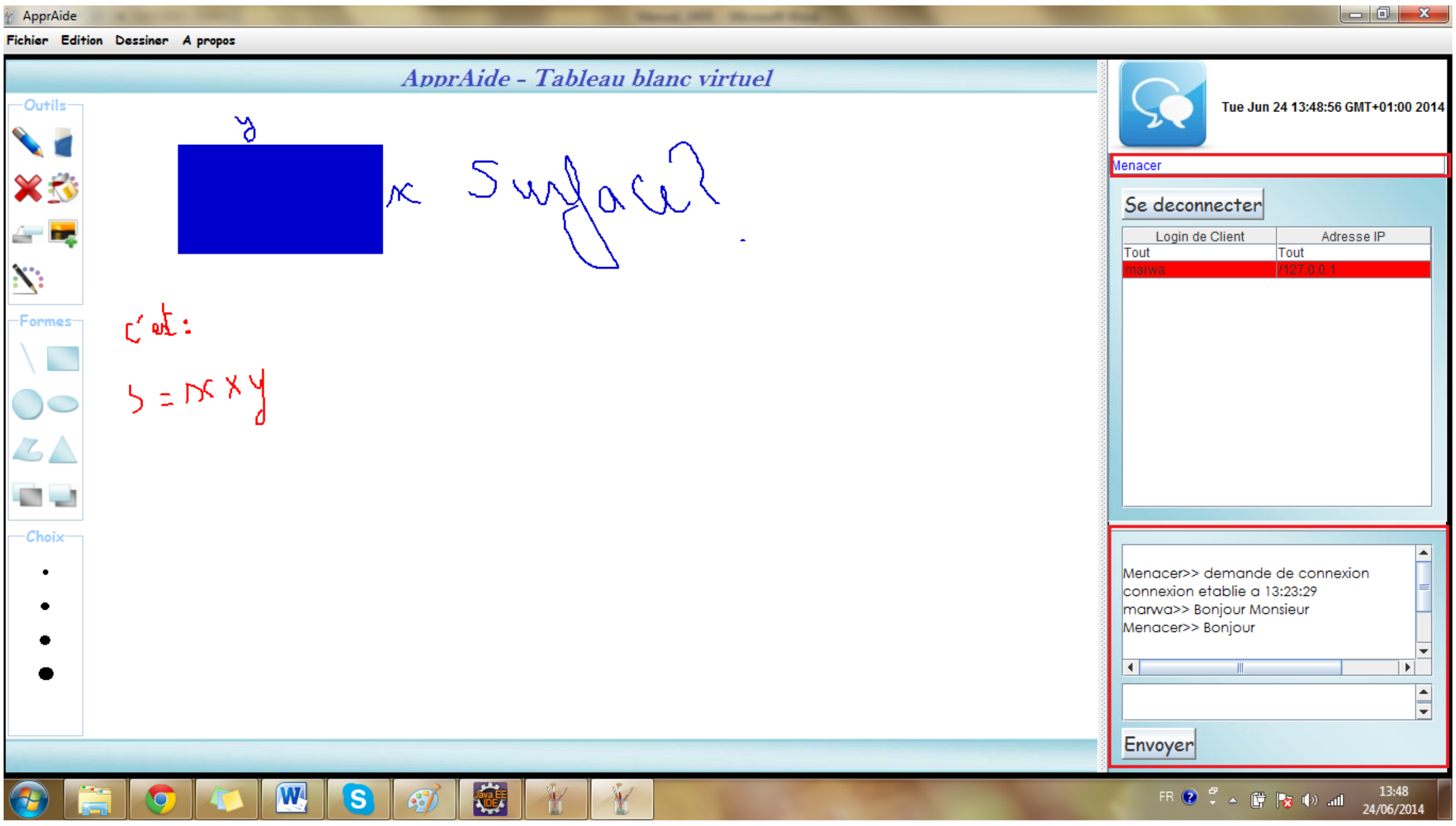

**Figure 64: Le tableau blanc virtuel (avec la réponse de l'enseignant)**

Le tableau blanc contient aussi un outil de chat pour faciliter la communication entre les utilisateurs. L'utilisateur peut aussi paramétrer la réception des demandes d'aide afin de filtrer les demandes indésirables (Figure 66).

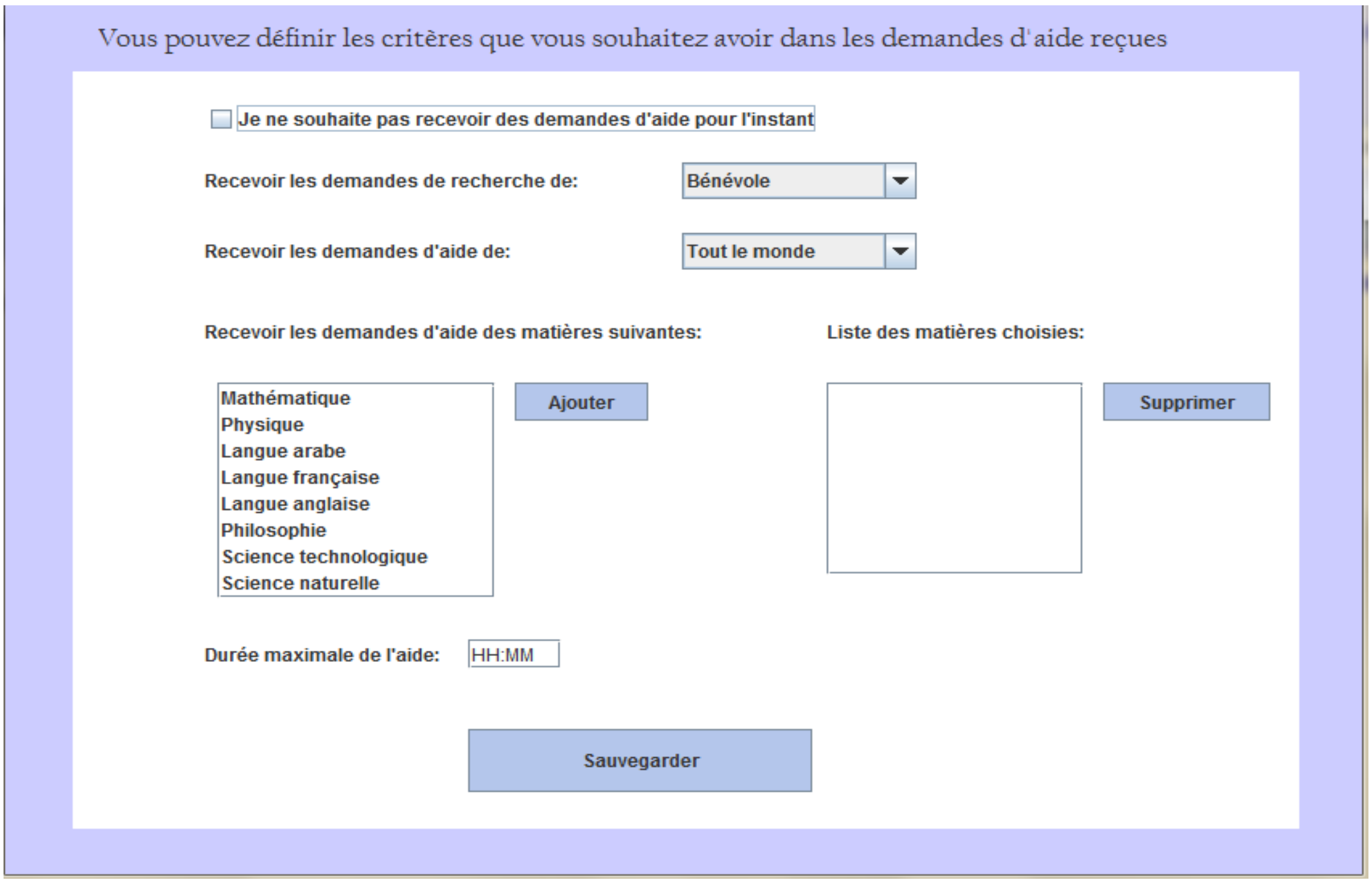

**Figure 65: Paramétrage de la réception des demandes d'aide**

## 4. L'aspect sécurité du système

Pour la sécurité des données échangées nous avons utilisé le RSA pour le cryptage des données. Pour ce faire, nous avons utilisé les packages « javax.crypto » et « java.security» ainsi que le package « java.math.* » de java, pour implémenter le chiffrement RSA. Nous également avons utilisé les Sockets pour l'envoi des objets.

## Conclusion

Le développement du système « ApprAide » nécéssite la maitrise de plusieurs outils et technologies de développement. C'est un système complexe qui engendre plusieurs difficultés en développement ; la première difficulté rencontrée est la non adéquation des langages de programmation des applications Desktop (tels que C# et java) au développement des réseaux sociaux, qui présente des environnements très dynamique. La conception des interfaces a pris la majorité du temps consacré pour cette partie. La deuxième difficulté est le manque de documentations pour le développement d'un tel système en java. Malgré ces difficultés, nous avons pu réaliser plusieurs aspects du système, notamment le site Web, les demandes d'aides synchrones et le tableau virtuel, la publication sur le réseau social ainsi que quelques aspects de sécurité (Cryptographie).

# Conclusion générale

Chercher à protéger la vie privée des apprenants dans les systèmes d'apprentissage est une tâche extrêmement difficile. Sa difficulté réside en un ensemble de contradictions entre les principes qu'impose le respect de la vie privée et les objectifs d'e-Learning. A vrai dire, trouver un compromis entre ces deux représente la difficulté majeure de notre travail notamment en l'absence de toute solution pouvant résoudre ce problème. Pour cette raison, nous avons proposé une approche de conception d'outils pour l'apprentissage social intégrés dans les systèmes d'e-Learning.

Nous tenons à signaler que nous sommes les premiers à avoir proposé une solution à problème. En premier lieu, nous avons utilisé les agents comme des entités autonomes capables de prendre des décisions pour protéger la vie privée des propriétaires des contenus partagés sur le réseau social. Sachant que la technologie d'agents, jusqu'à présent, n'est pas utilisée dans la résolution de ce problème. En second lieu, nous avons proposé une nouvelle méthode de localisation des meilleurs aidants sans divulguer les informations des pairs interagissant. En troisième lieu, nous avons facilité la gestion des clés de déchiffrement pour le cas des utilisateurs sélectionnés, un par un, de la liste d'audience. En quatrième lieu, nous avons assuré le droit à l'oubli en confiant au serveur la diffusion des demandes de suppression des contenus et l'application de l'ordre de suppression par les agents de chaque application concernée lors de la connexion. Actuellement en littérature, le droit à l'oubli n'est pas assuré dans les réseaux sociaux décentralisés.

Malgré les contradictions qui existent entre les objectifs d'e Learning et les principes du respect de la vie privée, notre solution ne s'y oppose pas. Mais ça n'empêche pas de dire qu'elle a à coté de ses nombreux avantages des inconvénients qu'elle a hérités des systèmes décentralisés notamment l'indisponibilité des données et la difficulté de la gestion des mises à jour. Cependant toutes les données que nous avons jugées nécessaires à l'apprentissage (demandes d'aides, test, cours…etc.). et qui n'ont pas vraiment un impact sur la vie privée des utilisateurs, sont, par défaut, dupliqués sur le serveur. Mais, nous avons laissé le choix final à l'utilisateur de changer ou non les paramètres définis par défaut. Toutefois, nous avons décentralisé les données à caractère personnel représentant un danger sur la vie privée.

En tant que fournisseur du service, nous avons pris de notre côté les précautions nécessaires pour protéger l'utilisateur. Par conséquent celui-ci est responsable de tout comportement inapproprié venant de sa part.

Au terme de ce présent mémoire, nous considérons notre travail comme un pas en avant. Il n'est pas parfait, mais perfectible. Les futures recherches pourraient en apporter plus de solutions avec plus de précision.

**Références**

**WEBOGRAPHIE:**